\documentclass{article}

\usepackage{diagrams}
\usepackage{latexsym}
\usepackage{epsfig}
\usepackage{amsfonts}
\let\sect=\section
\def\section{\newpage\sect}

\def\sectionitem{\hbox to 1.1cm{$\diamondsuit$ \hfil}}

\def\text#1{\mbox{\rm #1\ }}

\def\tr{{\rm Tr}}
\def\diag{{\it Diag}}
\def\algebraicspan{{\it Span}}
\def\ie{{\rm i.e.,\/}\ }
\def\etc{{\rm etc.\/}\ }
\def\cf{{\rm cf.\/}\ }
\def\id{\mbox{\it id\,}}
\def\rplus{{\subset\mkern-17mu \hbox{\small + }}}

\def\one{\mbox{\rm 1}\hskip-2.8pt \mbox{\rm l}}
\newcommand{\ZZ}{\mathbb{Z}}
\newcommand{\RR}{\mathbb{R}}
\newcommand{\CC}{\mathbb{C}}
\title{Differential calculus and connections on a quantum plane at a cubic
       root of unity
   \vspace{0.7cm} \\
}

\author{R. Coquereaux${}^1$\thanks{~Email: coque@cpt.univ-mrs.fr}$\;$,
        A. O. Garc\'{\i}a${}^2$\thanks{~Email: ariel@cab.cnea.edu.ar}$\;$,
        R. Trinchero${}^2$\thanks{~Email: trincher@cab.cnea.edu.ar}   \\
\\
${}^1$ {\it Centre de Physique Th\'eorique - CNRS - Luminy, Case 907} \\
       {\it F-13288 Marseille Cedex 9 - France}                       \\
\\
${}^2$ {\it Instituto Balseiro and Centro At\'omico Bariloche}        \\
       {\it CC 439 - 8400 - Bariloche - R\'{\i}o Negro - Argentina}   \\
\\
}

\date{}

\begin{document}

\begin{titlepage}
\thispagestyle{empty}

\maketitle

\vfill

\abstract{
We consider the algebra of $N \times N$ matrices as a reduced quantum plane
on which a finite-dimensional quantum group $\mathcal H$ acts. This quantum
group is a quotient of $U_q(sl(2,\CC))$, $q$ being an $N$-th root of unity.
Most of the time we shall take $N=3$; in that case $\dim(\mathcal H) = 27$.
We recall the properties of this action and introduce a differential
calculus for this algebra: it is a quotient of the Wess-Zumino complex. The
quantum group $\mathcal H$ also acts on the corresponding differential
algebra and we study its decomposition in terms of the representation theory
of $\mathcal H$. We also investigate the properties of connections, in the
sense of non commutative geometry, that are taken as $1$-forms belonging to
this differential algebra. By tensoring this differential calculus with
usual forms over space-time, one can construct generalized connections with
covariance properties with respect to the usual Lorentz group and with
respect to a finite-dimensional quantum group.
}

\vspace{1.0 cm}


\vspace{0.7cm}

\noindent Anonymous ftp or gopher: cpt.univ-mrs.fr

\vspace{0.5 cm}

\noindent {\tt math-ph/9807012}\\
\noindent CPT-98/P.3632 \\
\noindent IT-CNEA-CAB/2906098

\vspace*{0.3 cm}

\end{titlepage}

%%%%%%%%%%%%%%%%%%%%%%%%%%%%%%%%%%%%%%%%%%%%%%%%%%%%%%%%%%%%%%%%%%%%%%%%%%
%%%%%%%%%%%%%%%%%%%%%%%%%%%%%%%%%%%%%%%%%%%%%%%%%%%%%%%%%%%%%%%%%%%%%%%%%%

\section{Introduction}
\label{sec:introduction}

The algebra $\mathcal M$ of $N \times N$ complex matrices can be considered
as a finite quantum space (indeed, a finite-dimensional vector space, as
will be used in what follows). As such, it is a representation and
corepresentation space for finite-dimensional (dual) quantum groups. If some
physical problem involves such an algebra, it could therefore be tempting
to think that this system has a (kind of) quantum symmetry. The role of
fundamental object of the symmetry would now be played by a Hopf algebra,
instead of a usual Lie group. In the case at hand the relevant quantum
groups are particularly interesting, in view of the fact that, unlike the
generic-$q$ case, they have a non-trivial radical and their representation
theory involves indecomposable non-irreducible representations. Being
finite-dimensional, they are also much simpler to analyse explicitly.
Moreover, our knowledge about them is still much less complete than that
of the generic-$q$ case.

Although our paper discusses several topics that can already be found in
the literature (but we incorporate them to improve the reading, to set
notations and for completeness' sake) we develop several aspects that, up
to our knowledge, cannot be found elsewhere. Mainly, these include the
construction of the (finite-dimensional) differential algebra over
$\mathcal M$, the decomposition of this algebra and of its differential
algebra as representations of the quantum group that acts on them, and the
definition of a scalar product on $\mathcal M$ that is invariant under the
quantum group action.

The algebra $\mathcal M$ can indeed be generated by two Heisenberg
generators $x$ and $y$ satisfying the commutation relation $xy = q \, yx$,
where $q$ is an $N$-root of unity. These basic facts are recalled in
Section~\ref{sec:red-q-plane}. As a consequence, there is on this
algebra a coaction of a quotient $\mathcal F$ of the quantum group
$Fun(SL_q(2,\CC))$; this is reminded in Section~\ref{sec:q-group-F}.
Consequently, the dual $\mathcal H$ of $\mathcal F$, also a
finite-dimensional quantum group, acts on the algebra of $N \times N$
matrices. Explicit formulae for the pairing between $\mathcal H$ and
$\mathcal F$ and for the action of $\mathcal H$ on $\mathcal M$ have been
given by \cite{Gluschenkov, Dabrowski}. The structure of $\mathcal H$ had
been studied before by \cite{Alekseev} and its representation theory had
been given by \cite{Coquereaux}. In Section~\ref{sec:q-group-H} we recall
these duality properties and examine this action from the point of view of
the reducible indecomposable representations of the quantum group
$\mathcal H$, which is not a semisimple algebra. The unitary group of the
semi-simple part of $\mathcal H$ turns out to be isomorphic with
$U(3)\times U(2) \times U(1)$; for this reason, this Hopf algebra was
conjectured (see \cite{Connes-2}) to encode a quantum group of ``hidden
symmetries'' in the standard model of electroweak interactions. However,
the task of implementing this observation at the level of the lagrangian
describing the theory has not been completed, and it seems that this idea
would require some non trivial modifications of the model itself;
nevertheless this remark provided one of the motivations for the present
work.

In Section~\ref{sec:stars} we introduce a (unique) real structure on
$\mathcal M$, $\mathcal F$ and $\mathcal H$ and construct a compatible
hermitian scalar product on the space of matrices (the star operation on
$\mathcal M$ does not coincide with the usual hermitian conjugacy on
matrices). Next, in Section~\ref{sec:manin-dual} we introduce the Manin-dual
of our reduced quantum plane, and show how $\mathcal F$ coacts and
$\mathcal H$ acts on it. In Section~\ref{sec:diff-calculus} we define
differential forms on the algebra $\mathcal M$ and study their properties.
The differential algebra $\Omega_{WZ}(\mathcal M)$ that we introduce is a
quotient of the Wess-Zumino complex \cite{Wess-Zumino} constructed
originally for the $2$-dimensional quantum plane. When $q$ is a third
root of unity, $\mathcal M$ is the algebra of $3 \times 3$ matrices and
$\mathcal H$ (or $\mathcal F$) are of dimension $27$. The differential
algebra $\Omega_{WZ}(\mathcal M)$ is of dimension $36$. Since $\mathcal H$
acts on $\mathcal M$ and on its Manin dual, it also acts on
$\Omega_{WZ}(\mathcal M)$ and we study this algebra in terms of the
representation theory of $\mathcal H$. Given an associative algebra ---not
necessarily commutative--- one can define connections and covariant
derivatives, for any choice of a differential calculus over the algebra of
interest. The definition and properties of such connections are studied in
Section~\ref{sec:connections}; we also consider, as a particular case,
connections that are hermitian for the star operation introduced before. In
Section~\ref{sec:space-time}, we show how to couple these differential forms
to usual space-time by constructing a differential algebra equal to the
tensor product of $\Omega_{WZ}(\mathcal M)$ with the usual differential
forms (antisymmetric tensors on space-time). Generalized connections can
then be defined. They incorporate a usual one-form valued in the space of
$3 \times 3$ matrices and two matrix-valued scalar fields. These connections
transform covariantly under a simultaneous action of the usual Lorentz group
and the finite-dimensional quantum group $\mathcal H$. It would certainly be
interesting to build a classical Lagrangian field theory along these lines,
but this involves some deeper problems that are mentioned in the concluding
section.

The paper ends with a number of short appendices. In them we first describe
a set of $3 \times 3$ generalized Gell-Mann matrices with entries in the
quantum group $\mathcal F$. Then we give the structure of the principal
indecomposable modules and the most general covariant metrics on the
representation spaces of the algebra $\mathcal H$, we study the space of
differential operators on $\mathcal M$ and finally write down the universal
$R$-matrix of $\mathcal H$.

%%%%%%%%%%%%%%%%%%%%%%%%%%%%%%%%%%%%%%%%%%%%%%%%%%%%%%%%%%%%%%%%%%%%%%%%%%
%%%%%%%%%%%%%%%%%%%%%%%%%%%%%%%%%%%%%%%%%%%%%%%%%%%%%%%%%%%%%%%%%%%%%%%%%%

\section{The space $\mathcal M$ of $3 \times 3$ complex matrices as a
         reduced quantum plane}
\label{sec:red-q-plane}

%%%%%%%%%%%%%%%%%%%%%%%%%%%%%%%%%%%%%%%%%%%%%%%%%%%%%%%%%%%%%%%%%%%%%%%%%%

\subsection{From elementary $N \times N$ matrices to the $x,y$ generators}

It has been known for a long time \cite{Weyl} that the algebra of $N\times N$
matrices can be generated by two elements $x$ and $y$ with the relations
\begin{equation}
   xy = q yx \ , \qquad \qquad x^N = y^N = \one \ ,
\label{M-relations}
\end{equation}
where $q$ denotes an $N$-th root of unity ($q \neq 1$) and $\one$ is the
unit matrix.

Let us make explicit, in the particular case $N = 3$, this well known (but
sometimes forgotten$\ldots$) property.

Let $q$ be a cubic root of unity ($q \neq 1$) and take
$$
x = \pmatrix{1 & 0 & 0       \cr
             0 & q^{-1} & 0  \cr
             0 & 0 & q^{-2}}
\hspace{2.0cm}
y = \pmatrix{0 & 1 & 0       \cr
             0 & 0 & 1       \cr
             1 & 0 & 0}      \ .
$$
It is then easy to check that the above relations between $x$ and $y$ are
indeed satisfied (use $q^{-1} = q^2$), and that they generate the whole
algebra. Many formulae that we shall write in the following can be easily
generalized when $N$ is an arbitrary integer, but we shall stick to the
case $N=3$.

Any $3 \times 3$ matrix can obviously be expanded on the base made of the
nine elementary matrices $E_{ij}$ (such a matrix has a single non-zero entry
$1$ in position $(i,j)$ and is filled with zeros elsewhere). One can express
the elementary matrices themselves in terms of $x$ and $y$. Calculations
are straightforward, and give
$$
\begin{tabular}{ll}
$ E_{11} = (\one + x + x^2)/3 \qquad \qquad $ &
$ E_{12} = (y + x y + x^2 y)/3 $              \\
$ E_{13} = (y^2 + x y^2 + x^2 y^2)/3 $        &
$ E_{21} = (y^2 + q x y^2 + q^2 x^2 y^2)/3 $  \\
$ E_{22} = (\one + q x + q^2 x^2)/3 $         &
$ E_{23} = (y + q x y + q^2 x^2 y)/3 $        \\
$ E_{31} = (y + q^2 x y + q x^2 y)/3 $        &
$ E_{32} = (y^2 + q^2 x y^2 + q x^2 y^2)/3 $  \\
$ E_{33} = (\one + q^2 x + q x^2)/3 $         & {} \\
\end{tabular}
$$
The unit matrix ($\one$) itself can be written in terms of the generators
$x$ and $y$ (since $x^{3}=y^{3}=\one$). Therefore, one can also express the
usual Gell-Mann matrices that generate $Lie(SU(3))$ in terms of the
generators $x$ and $y$. These expressions are shown in
\ref{app:lambda-matrices}

Warning: the set of $3 \times 3$ matrices is endowed with a usual star
operation (that we denote $\dag$): hermitian conjugacy. It is clear that
$x$ and $y$ are unitary elements (with respect to $\dag$):
$x^\dag = x^{-1}=x^2$ and $y^\dag = y^{-1}=y^2$.
As a matter of fact, this star operation does not have good properties with
respect to a quantum group action that we shall introduce later. We shall
return to this important problem in a forthcoming section.

%%%%%%%%%%%%%%%%%%%%%%%%%%%%%%%%%%%%%%%%%%%%%%%%%%%%%%%%%%%%%%%%%%%%%%%%%%

\subsection{${\mathcal M}=M_3(\CC)$ as a reduced quantum plane}

The associative algebra generated, over the complex numbers, by $x$ and $y$
with the single (quadratic) relation $yx = q^{-1} xy$ is known as ``the
algebra of polynomials over the quantum plane'' and is often denoted by
$Fun_q(\CC^2)$ or by $\CC_q[x,y]$. We shall just call it $\CC_q$.
When $q=1$, this algebra is commutative and can be considered as the algebra
of polynomials $\CC[x,y]$ over the usual plane, $x$ and $y$ being the two
coordinate functions. The dimension of $\CC_q$ ---as a vector space--- is
infinite, since powers of the generators do not satisfy any particular new
relation. On the contrary, in the algebra $M_3(\CC)$ of $3\times 3$
matrices over complex numbers, the generators $x$ and $y$, on top of the
above quadratic relation, satisfy also the cubic relations $x^3=\one$ and
$y^3=\one$. The dimension is then clearly equal to $9$, as it should, since
one can choose the following base of generators:
$\{\one,x,y,x^2,y^2,xy,x^2y,xy^2,x^2y^2\}$.
One can therefore consider $M_3(\CC)$ as the quotient of the associative
algebra $\hat{\CC}_q$, when $q^3=1$, by the bilateral ideal generated by
the relations $x^3-\one = 0$ and $y^3-\one = 0$. Here $\hat{\CC}_q$
denotes the unital extension of $\CC_q$ (the former is obtained by adding
a unit, namely $\one$ to the later). For this reason we can consider the
space of $3 \times 3$ matrices over $\CC$ as a {\sl reduced quantum
plane\/}.

Warning: $\mathcal M \doteq M_3(\CC)$ is {\sl not \/} a quantum group.

%%%%%%%%%%%%%%%%%%%%%%%%%%%%%%%%%%%%%%%%%%%%%%%%%%%%%%%%%%%%%%%%%%%%%%%%%%
%%%%%%%%%%%%%%%%%%%%%%%%%%%%%%%%%%%%%%%%%%%%%%%%%%%%%%%%%%%%%%%%%%%%%%%%%%

\section{The finite-dimensional quantum group $\mathcal F$}
\label{sec:q-group-F}

%%%%%%%%%%%%%%%%%%%%%%%%%%%%%%%%%%%%%%%%%%%%%%%%%%%%%%%%%%%%%%%%%%%%%%%%%%

\subsection{Construction of $\mathcal F$ as an algebra}
\label{subsec:F-as-algebra}

Let us start with a classical analogy and call $x$ and $y$ the coordinate
functions on the plane. One can make a linear change of coordinates and call
$x'$ and $y'$ the new coordinate functions:
\begin{equation}
\pmatrix{x' \cr y'} =
    \pmatrix{a & b \cr c & d} \otimes \pmatrix{x \cr y} \ .
\label{c-L-coaction}
\end{equation}
We can assume the transformation to be unimodular ($ad-bc=1$). Rather than
considering $x$ and $y$ as numbers, we think of them as coordinate
functions. In the same way, we do not take the matrix elements $a,b,c$ and
$d$ as numbers but as {\sl functions\/} on the group of coordinate
transformations: when $g$ denotes such a transformation then
$a(g),b(g),c(g)$ and $d(g)$ are numbers, namely, the matrix elements of $g$.
This change in the perspective explains why we write a tensor product sign
in the previous formula\ldots evaluation on points of the group and of the
space gives the transformed coordinate functions.

One could also introduce line vectors, with coordinate functions $\tilde x$,
$\tilde y$. The same change of coordinates would read
\begin{equation}
\pmatrix{\tilde x & \tilde y} =
    \pmatrix{ x & y} \otimes \pmatrix{a & b \cr c & d} \ .
\label{c-R-coaction}
\end{equation}

From now on, we shall no longer assume that symbols $x$ and $y$ commute but
that they should satisfy the relations discussed in the previous section,
namely $xy = qyx$, $x^3 = y^3 = \one$. Symbols $x$ and $y$ can therefore
be represented by the $3\times 3$ matrices already given. As before,
$\mathcal M$ shall denote the algebra generated by $x$ and $y$.

One then introduces {\sl non-commuting symbols\/} $a,b,c$ and $d$ and
imposes that quantities $x',y'$ (and $\tilde x, \tilde y$) obtained by the
previous matrix equalities should satisfy the same relations as $x$ and $y$.
We call ${\mathcal F}'$ the algebra generated by the elements $a,b,c$ and
$d$. The product in $\mathcal {F' \otimes A}$ is defined by
$(f \otimes z)(g \otimes w) \doteq fg \otimes zw$, in other words,
$a,b,c,d$ commute with $x,y$. We have a similar definition for the
multiplication in $\mathcal {A \otimes F'}$. The two constraints
$x' y' = q y' x' $ and $\tilde x \tilde y = q \tilde y \tilde x$
lead to the six quadratic relations \cite{Manin}
\begin{equation}
\begin{tabular}{ll}
   $ac = qca$ \qquad \qquad & $bd = qdb$             \\
   $ab = qba$               & $cd = qdc$             \\
   $bc = cb$                & $ad-da = (q-q^{-1})bc$ \ .
\end{tabular}
\label{F-products}
\end{equation}

The algebra generated by $a,b,c$ and $d$ (take products and sums) together
with the six above relations is usually denoted $Fun(GL_q(2,\CC))$ and is
the {\sl algebra of would-be functions over the quantum group
$GL_q(2,\CC)$\/}. Calling such elements ``functions'' is, of course, a
misnomer, since they are not valued in any field of numbers and\ldots\/ do
not commute.

The element ${\mathcal D} \doteq da -q^{-1}bc = ad - qbc $ is central
(it commutes with all the elements of $Fun(GL_q(2))$); it is called the
$q$-determinant and we set it equal to\footnote{
We will use $\one$ to denote the unit element of $\mathcal{M, F}$ and the
---to be defined--- quantum group $\mathcal H$, indistinctly. Which one this
symbol refers to, should be easily understood from context. In case this
were not obvious we will use $\one_{\mathcal M}$, $\one_{\mathcal F}$,
$\one_{\mathcal H}$.
}
$\one$. Adding this extra relation
defines the algebra $Fun(SL_q(2,\CC))$.

Now, we should remember that $x^3 = y^3 = \one$. Imposing
$x'^3 = y'^3 = \one$ (and also $\tilde x^3 = \tilde y^3 = \one$) implies
again new relations.

{\small
\noindent For instance,
$$
\begin{tabular}{lll}
$x'^{3}$ & $=$ & $ (a \otimes x + b \otimes y)^{3} $ \\
    ${}$ & $=$ & $ a^{3} \otimes x^{3} + a^{2}b \otimes x^{2}y +
                   aba \otimes xyx + ba^{2} \otimes yx^{2} + $ \\
    ${}$ & ${}$ & $ ab^{2} \otimes xy^{2} + bab \otimes yxy +
                   b^{2}a \otimes y^{2}x + b^{3} \otimes y^{3} $ \\
    ${}$ & $=$ & $ a^{3} \otimes x^{3} +
                   (1+q+q^{2}) a^{2}b \otimes x^{2}y +
                   (1+q+q^{2}) ab^{2} \otimes xy^{2} +
                   b^{3} \otimes y^{3} $ \\
    ${}$ & $=$ & $ a^{3} \otimes x^{3} + b^{3}\otimes y^{3} $ \\
\end{tabular}
$$
where we used $1+q+q^2 = 0$ since $q$ is a third root of unity.
}

\noindent This, together with the analogous constraints on
$y',\tilde x, \tilde y$, imply:
\begin{equation}
\begin{tabular}{ll}
   $a^3 = \one \ , $ & $ b^3 = 0    \ , $ \\
   $c^3 = 0    \ , $ & $ d^3 = \one \ . $
\end{tabular}
\label{F-products-quotient}
\end{equation}

Imposing these cubic relations on $Fun(SL_{q}(2,\CC))$ defines a new algebra
that we denote with $\mathcal F$. We call it the {\sl reduced quantum
unimodular group\/} associated with a cubic root of unity. Actually one
should better call it the {\sl algebra of would-be functions over the
reduced quantum unimodular group\/} $SL_{q}(2,\CC)$ but we shall shorten
the terminology.

Since $a^3 = \one$, multiplying the relation $ad = \one + qbc$ from the
left by $a^2$ leads to
\begin{equation}
   d = a^2 (\one + qbc)
\label{d-element-in-F}
\end{equation}
so $d$ is not needed and can be eliminated. The algebra $\mathcal F$ can
therefore be {\sl linearly \/} generated ---as a vector space--- by the
elements $a^\alpha b^\beta c^\gamma $ where indices $\alpha, \beta, \gamma$
run in the set $\{0,1,2\}$. We see that $\mathcal F$ is a
{\sl finite-dimensional\/} associative algebra, whose dimension is
$$
   \dim({\mathcal F}) = 27 \ .
$$

Let us stress the fact that this very particular algebra $\mathcal F$ (which
turns out to be a quantum group) emerges naturally in relation with the
study of the algebra $\mathcal M$ of $3 \times 3$ matrices.

%%%%%%%%%%%%%%%%%%%%%%%%%%%%%%%%%%%%%%%%%%%%%%%%%%%%%%%%%%%%%%%%%%%%%%%%%%

\subsection{$\mathcal F$ as a quantum group}

$\mathcal F$ is not only an algebra but also a quantum group, in other
words, a Hopf algebra. This amounts to say that besides its algebra
structure (a product), it has a coalgebra structure (a coproduct) and that
the two structures are compatible (hence it is a bialgebra); moreover one
can also define maps called antipode and counit obeying the appropriate
axioms.

There are now several more or less elementary textbooks on the subject of
quantum groups and the interested reader should refer to them for general
properties. We shall simply give the definitions for the coproduct,
antipode and counit, in the present case. The reader will easily show that
all required properties are indeed satisfied. These three maps being algebra
morphisms (or antimorphisms) it is actually enough to define them on the
generators.

\begin{description}
\item[Coproduct:] It is an algebra morphism $\Delta$ from the algebra
   $\mathcal F$ to the algebra $\mathcal{F \otimes F}$,
   \ie $\Delta (uv) = \Delta u \, \Delta v$.
   It is given by $\Delta a = a \otimes a + b \otimes c$,
   $\Delta b = a \otimes b + b \otimes d$,
   $\Delta c = c \otimes a + d \otimes c$,
   $\Delta d = c \otimes b + d \otimes d$.

\item[Antipode:] It is a (linear\footnote{
   We stress that the antipode is a {\em linear antimorphism}, in contrast
   with the star operation that will be introduced in
   Section~\ref{sec:stars} which is an {\em antilinear antimorphism}.
   })
   antimorphism $S$ from $\mathcal F$ to $\mathcal F$,
   \ie $S(uv) = Sv \, Su$. It is given by
   $Sa = d$, $Sb = -q^{-1} b$, $Sc = -q c$, $Sd = a$.

\item[Counit:] It is a morphism $\epsilon$ from $\mathcal F$ to the
   complex numbers $\CC$ and is given by
   $\epsilon(a)=1$, $\epsilon(b)=0$, $\epsilon(c)=0$,
   $\epsilon(d)=1$.
\end{description}

{\small
Actually, one can give a rather short proof of the fact that $\mathcal F$
is a Hopf algebra. It is enough to prove that the two-sided ideal
$\mathcal I$ defining the quotient
${\mathcal F} = Fun(SL_q(2,\CC))/\mathcal I$ is a Hopf ideal. $\mathcal I$
is the ideal generated by the relations $a^3 - \one = 0$, $d^3 - \one = 0$,
$b^3 = 0$ and $c^3 = 0$. Being ${\mathcal I}$ a Hopf ideal means:
$\Delta {\mathcal I} \subset
   {\mathcal I} \otimes Fun(SL_q(2,\CC)) \oplus
      Fun(SL_q(2,\CC)) \otimes {\mathcal I}$,
$\epsilon({\mathcal I})=0$ and $S({\mathcal I}) \subset I$.
This is easy to show. For instance, we see that
\begin{eqnarray*}
\Delta a^3 &=& (\Delta a)^3 = (a \otimes a + b \otimes c)^3 \\
        {} &=& a^3 \otimes a^3 + (1+q+q^2) a^2 b \otimes a^2 c +
                 (1+q+q^2) ab^2 \otimes ac^2 +
                 b^3 \otimes c^3                            \\
        {} &=& a^3 \otimes a^3 + b^3 \otimes c^3
\end{eqnarray*}
but $\Delta \one = \one \otimes \one$, so that
$$
2 \Delta(a^3-\one) = a^3 \otimes (a^3-\one) +
        (a^3-\one) \otimes a^3 + (a^3-\one) \otimes \one +
        \one \otimes (a^3-\one) + 2 b^3 \otimes c^3
$$
which is indeed in
${\mathcal I} \otimes Fun(SL_q(2,\CC)) \oplus
   Fun(SL_q(2,\CC)) \otimes {\mathcal I}$.
}

$\mathcal F$ is, by construction, an associative algebra. However, it
is not semisimple. This can be seen easily from the faithful realization
given in \ref{app:F-faithful-rep} (this realization, in terms of
matrices with entries in an algebra generated by two {\sl commuting\/}
symbols $\xi_1,\xi_2$ whose cube power vanishes is due to Ogievetsky,
\cite{Ogievetsky}). $\mathcal F$ is therefore not a matrix quantum group
in the sense of Woronowicz \cite{Woronowicz}: whatever the involution and
the scalar product we choose, it will never be a $C^*$-algebra.

%%%%%%%%%%%%%%%%%%%%%%%%%%%%%%%%%%%%%%%%%%%%%%%%%%%%%%%%%%%%%%%%%%%%%%%%%%

\subsection{$\mathcal F$ coacting on $\mathcal M$}

What is called coaction of $\mathcal F$ on $\mathcal M$ is precisely
what was described, in simple terms, at the beginning of this section.
Actually, we have two coactions: a left coaction and a right one.
We have
\begin{eqnarray}
   x' \doteq \Delta_L x & = & a \otimes x + b \otimes y \\
   y' \doteq \Delta_L y & = & c \otimes x + d \otimes y \nonumber
\label{L-coaction}
\end{eqnarray}
and
\begin{eqnarray}
   \tilde x \doteq \Delta_R x & = & x \otimes a + y \otimes c \\
   \tilde y \doteq \Delta_R y & = & x \otimes b + y \otimes d \nonumber
\label{R-coaction}
\end{eqnarray}

\noindent Both $\Delta_L$ and $\Delta_R$ can be extended to arbitrary
elements of $\mathcal M$ by imposing the property\footnote{
   Taking into account this condition, we see that the equations
   $$
      \Delta_{R,L} (xy - q yx) = 0
   $$
   are completely equivalent to the ones we used in
   Section~\ref{subsec:F-as-algebra} to obtain the relations
   (\ref{F-products}) for the quantum group $\mathcal F$. Therefore they
   also imply (\ref{F-products}).
}
$$
   \Delta_{L,R}(zw) = \Delta_{L,R}(z) \, \Delta_{L,R}(w)
$$
for any $z,w \in {\mathcal M}$.

$\mathcal M$ itself, endowed with the two coactions $\Delta_L$ and
$\Delta_R$ is {\sl not} a quantum group but a left and right {\em comodule},
\ie a corepresentation space of the quantum group $\mathcal F$. This means
that the right coaction (for instance) maps
${\mathcal M} \mapsto {\mathcal M}\otimes{\mathcal F}$, in such a way that
\begin{equation}
   (\Delta_R \otimes \id)\Delta_R(z) = (\id \otimes \Delta)\Delta_R(z)
\label{comodule-condition-coproduct}
\end{equation}
and
\begin{equation}
   (\id \otimes \epsilon)\Delta_R(z) = z \ ,
\label{comodule-condition-counit}
\end{equation}
for any $z \in {\mathcal M}$. Here one should not confuse $\Delta$ (the
coproduct on the quantum group) with $\Delta_{R,L}$ (the $R,L$-coaction on
$\mathcal M$)! These conditions are indeed satisfied\ldots the interested
reader may check that.

Moreover, $\mathcal M$ is a left and right comodule {\em algebra} over
$\mathcal F$. There are two extra axioms that have to be satisfied in order
to have a comodule algebra structure, in particular a compatibility axiom
between the coaction and the product in $\mathcal M$.
These extra conditions read:
\begin{equation}
   \Delta_{L,R}(zw) = \Delta_{L,R}(z) \, \Delta_{L,R}(w)
\label{comodule-algebra-condition-product}
\end{equation}
for $z,w$ any elements of the algebra. This is just the equation we have
used to extend the coaction to the whole $\mathcal M$, thus is trivially
satisfied. If the algebra also has a unit, the coaction should verify
\begin{equation}
   \Delta_{L,R}(\one) = \one_{\otimes} \ .
\label{comodule-algebra-condition-unit}
\end{equation}

Using this coaction, and recalling that $\mathcal M$ is the algebra of
$3 \times 3$ matrices, one can build a set of generalized Gell-Mann
matrices with entries in the quantum group $\mathcal F$, as it is done
in \ref{app:F-lambda-matrices}

Notice that although $\mathcal F$ {\sl coacts\/} on $\mathcal M$, it does
not {\sl act} on it. This is exactly what happens in the usual (commutative)
situation: the algebra of functions on the group of unimodular
transformations of the plane coacts on the algebra of functions on
the plane (there are indeed tensor product signs in the formula
$x \rightarrow a \otimes x + b \otimes y$, even if it is sometimes
convenient not to write them\ldots). In the present situation, we have no
``points'' and it makes {\it a priori\/} no sense to speak about the action
of a would-be group on would-be points. The coaction of $\mathcal F$ on
$\mathcal M$ is however well defined and allows such a terminological abuse.

As a matter of fact, there is an algebra that {\sl acts} on $\mathcal M$,
it is not $\mathcal F$ but its dual, that we call $\mathcal H$. We shall
return to this in the next section.

%%%%%%%%%%%%%%%%%%%%%%%%%%%%%%%%%%%%%%%%%%%%%%%%%%%%%%%%%%%%%%%%%%%%%%%%%%
%%%%%%%%%%%%%%%%%%%%%%%%%%%%%%%%%%%%%%%%%%%%%%%%%%%%%%%%%%%%%%%%%%%%%%%%%%

\section{The dual $\mathcal H$ of $\mathcal F$}
\label{sec:q-group-H}

%%%%%%%%%%%%%%%%%%%%%%%%%%%%%%%%%%%%%%%%%%%%%%%%%%%%%%%%%%%%%%%%%%%%%%%%%%

\subsection{Classical analogies}

One should think of $\mathcal F$ as an analogue of the space of functions
$Fun(G)$ over a Lie group $G$ (in our case, $G$ itself does not exist) and
$\mathcal H$ ---that we shall introduce now--- as a finite-dimensional
analogue of the universal enveloping algebra of the Lie algebra of $G$.
Another fruitful analogy is to take $\mathcal F$ as the analogue of the
space of functions over a finite group $G$ and $\mathcal H$ as an analogue
of the group algebra $\CC G$.

As in the classical case, arbitrary elements $X$ of $\mathcal H$ can be
understood in several ways:
{\small
\begin{itemize}
\item As distributions on $\mathcal F$: we evaluate $X$ on the ``function''
      $u \in \mathcal F$. The result is a number called
      $\langle X, u \rangle$.

\item As invariant differential operators acting on the ``functions'' $u$.
      The result is another element of $\mathcal F$, that we call $X[u]$.
      Actually, as in the classical case, one can define {\sl two\/} kinds
      of invariant operators: left-invariant ones (coming classically from
      the right action of a group on itself) and right-invariant ones
      (coming from the left action).

\item As a differential operator acting on the space of ``functions''
      $\mathcal M$ (classically functions on a manifold on which the group
      $G$ acts). Let $z$ be such a function. We call $X[z]$ the result.
      This is another element of $\mathcal M$.

\item Abstractly, as an element of the associative algebra $\mathcal H$,
      \ie classically, an element of the enveloping algebra
      ${\mathcal{U}}(Lie(G))$.
\end{itemize}
}
\noindent The remaining part of this section is devoted to a short study
of these various aspects.

%%%%%%%%%%%%%%%%%%%%%%%%%%%%%%%%%%%%%%%%%%%%%%%%%%%%%%%%%%%%%%%%%%%%%%%%%%

\subsection{The finite-dimensional quantum group $\mathcal H$}
\label{subsec:finite-dim-H}

Being the dual of $\mathcal F$, it is clear that $\mathcal H$ is a vector
space of dimension $27$. It can be generated, as a complex algebra,
by elements $X_\pm$, $K$ and $K^{-1}$ obeying a number of relations.
Multiplication and comultiplication in $\mathcal H$ can be obtained from the
corresponding ones in its dual $\mathcal F$, but here we gather most of the
relevant formulae abstractly defining $\mathcal H$ as a Hopf algebra, with
generators $X_{\pm}$ and $K$, without using its duality with $\mathcal F$.

\begin{description}

\item[Product:]
\begin{eqnarray}
   K X_{\pm}     &=& q^{\pm 2} X_{\pm} K                 \nonumber \\
   \left[ X_+ , X_- \right]
                 &=& {1 \over (q - q^{-1})} (K - K^{-1})           \\
   K^3           &=& \one                                \nonumber \\
   X_+^3 = X_-^3 &=& 0                                   \nonumber
\label{H-products}
\end{eqnarray}

\item[Coproduct:]
The comultiplication is an algebra morphism, \ie
$\Delta(XY) = \Delta X \, \Delta Y$. It is given by
\begin{eqnarray}
   \Delta X_+ & \doteq & X_+ \otimes \one + K \otimes X_+      \nonumber \\
   \Delta X_- & \doteq & X_- \otimes K^{-1} + \one \otimes X_-           \\
   \Delta K & \doteq & K \otimes K                             \nonumber \\
   \Delta K^{-1} & \doteq & K^{-1} \otimes K^{-1}              \nonumber
\label{H-coproducts}
\end{eqnarray}

\item[Antipode:]
The antipode $S$ is an anti-automorphism, \ie $S(XY) = SY \, SX$.
It acts as follows:
   $ S \one = \one $,
   $ S K = K^{-1} $,
   $ S K^{-1} = K $,
   $ S X_+ = - K^{-1} X_+ $,
   $ S X_- = - X_- K $.
As usual, the square of the antipode is an automorphism (and it is a
conjugacy given ---up to multiplication by a central element--- by
$K^{-1}$, hence we can write $S^2 u = K^{-1} u K$).

\item[Counit:]
The counit $\epsilon$ is defined by
   $ \epsilon \one = 1 $,
   $ \epsilon K = 1 $,
   $ \epsilon K^{-1} = 1 $,
   $ \epsilon X_+ = 0 $,
   $ \epsilon X_- = 0 $.

\end{description}

The previous multiplication relations allow to order any monomial of the
algebra freely generated by $X_\pm$ and $K$ as
$X_+^\alpha K^\beta X_-^\gamma$. Moreover, as $X_\pm^3 = 0$ and
$K^3 = \one$, it is evident that the $27 = 3^3$ elements
$\{ (X_+^\alpha K^\beta X_-^\gamma) \}_{\alpha,\beta,\gamma \in \{0,1,2\}}$
span ${\cal H}$ as a vector space over $\CC$.

Note that the element $X_+ X_- - (q^{-1} K + q K^{-1})/3$ commutes with all
elements of $\mathcal H$. Therefore its plays the role of a usual Casimir
operator.

{\small
Remark: In \cite{Gluschenkov} the authors study the pairing between a
(reduced) universal enveloping algebra and the algebra of functions on
quantum $SL(2,\CC)$, in the case $q^N = 1$. However, they factorise the
universal algebra over the relations $k^6 = \one$, $I_{\pm}^3 = 0$, rather
than $K^3 = \one$, $X_{\pm}^3 = 0$. Their choice for generators $I_{\pm}$
and $k$ differs from our $X_{\pm}$ and $K$ (actually, $k = K^{1/2}$,
$I_+ = qX_+ K$ and $I_- = qX_- K^2$). The obtained algebra
$\tilde \mathcal H$ is twice as big as ours ($k^3$ is a central element but
is not equal to $\one$).
}

%%%%%%%%%%%%%%%%%%%%%%%%%%%%%%%%%%%%%%%%%%%%%%%%%%%%%%%%%%%%%%%%%%%%%%%%%%

\subsection{$\mathcal H$ as the dual of $\mathcal F$ (distributions)}

Being $\mathcal F$ a quantum group (a Hopf algebra), its dual
${\mathcal H} \doteq {\mathcal F}^*$ is a quantum group as well. Let
$u \in \mathcal F$ and $X \in \mathcal H$. We call $< X, u >$ the
evaluation of $X$ on $u$ (a complex number).

\begin{itemize}
\item
   Using the coproduct $\Delta$ in $\mathcal F$, one defines a product in
   $\mathcal H$ (this is the so-called convolution product of distributions
   and it is often denoted by $\star$, but we shall omit this symbol in the
   sequel):
   $$
      < X_1 X_2, u > \doteq < X_1 \otimes X_2, \Delta u >
   $$
\item
   Using the product in $\mathcal F$, one defines a coproduct (that we
   again denote $\Delta$) in $\mathcal H$\footnote{
      Some authors define
      $<\Delta X, u_1 \otimes u_2> \doteq <X, u_2 u_1>$,
      in order to increase the correspondence with the classical case.
      This alternative definition corresponds to use our
      $\Delta^{op} = \tau\circ\Delta$ as a coproduct, where $\tau$ produces
      a permutation of factors in the tensor product.
   }:
   $$
      <\Delta X, u_1 \otimes u_2 > \doteq < X, u_1 u_2 >
   $$
\item
   The interplay between unit and counit is described by the two relations:
   $< \one_{\mathcal H}, u> = \epsilon_{\mathcal F}(u)$
   and $< X, \one_{\mathcal F}> = \epsilon_{\mathcal H}(X)$.
\end{itemize}

\noindent The two structures of algebra and coalgebra are clearly
interchanged by duality.

$\mathcal H$ is the linear dual of $\mathcal F$. In other words, if
$\mathcal F$ were a space of smooth (resp. continuous) functions over a
compact space, $\mathcal H$ would be the corresponding space of
distributions (resp. measures).

All possible pairings can be computed from those between the generators
$K, X_\pm$ and $a,b,c$. They are given by:
\begin{equation}
\begin{tabular}{cccc}
   $<K,a> = q$   & $<K,b> = 0$   & $<K,c> = 0$   & $<K,d> = q^2$ \\
   $<X_+,a> = 0$ & $<X_+,b> = 1$ & $<X_+,c> = 0$ & $<X_+,d> = 0$ \\
   $<X_-,a> = 0$ & $<X_-,b> = 0$ & $<X_-,c> = 1$ & $<X_-,d> = 0$ \\
\end{tabular}
\label{H-F-pairing}
\end{equation}
The fourth column of this table can be obtained from the first three
(remember that $d = a^2(\one + qbc)$).

Remark: the previously given multiplication and comultiplication equalities
for the generators of $\mathcal H$ (in particular the appearance of $q$
rather than its inverse) cannot be given arbitrarily; indeed they have to be
compatible with those given for the dual, $\mathcal F$. In other words, once
the duality formulae defining the generators $X_\pm$ have been chosen, one
cannot choose independently ``the $q$ of $\mathcal F$'' and ``the $q$ of
$\mathcal H$''. Conversely, if we start with the formulae defining
multiplication and comultiplication in $\mathcal F$ and $\mathcal H$, the
pairing formulae are essentially unique.

{\small
To see this, let us suppose given, for instance, the comultiplication
relations for $\mathcal H$ and determine the constraints for both the
pairings and the multiplication in $\mathcal H$. Choose the basis
${a^i b^j c^k} $ in the vector space $\mathcal F$. At the same time we
choose a basis ${K^i X_+^j X_-^k}$ in the dual vector space $\mathcal F$,
with pairings
$$
\begin{tabular}{cccc}
   $<K,a> = \alpha$ \qquad & $<K,b> = <K,c> = 0$     \\
   $<X_+,b> = \beta$       & $<X_+,a> = <X_+,c> = 0$ \\
   $<X_-,c> = \gamma$      & $<X_-,a> = <X_-,b> = 0$ \\
\end{tabular}
$$
\etc where $\alpha, \beta, \gamma$ are numbers that we are going to
determine.

First of all, let us calculate $<X_+, ab>$ and $<X_+, ba>$.
Known commutation relations in $\mathcal F$ (or comultiplication in
$\mathcal H$) will imply that $\alpha = q$. Indeed,
\begin{eqnarray*}
<X_+,ab> &=& <\Delta X_+, a \otimes b > =
   <X_+ \otimes \one + K \otimes X_+, a \otimes b> = 0 + \alpha \beta \\
<X_+,ba> &=& <\Delta X_+, b \otimes a > =
   <X_+ \otimes \one + K \otimes X_+, b \otimes a> = \beta + 0 \ .
\end{eqnarray*}
But $ab = q ba$, therefore $\alpha = q$.

Then, we shall calculate $<K X_+, b>$ and $<X_+ K, b>$. This will imply
$KX_+ = q^2 X_+ K$. To do that, we need to find $<K, d>$. We know that
$d = a^2(\one + qbc)$, therefore
\begin{eqnarray*}
<K,d> &=& <\Delta K, a^2 \otimes (\one+qbc)> = <K, a^2><K,(\one+qbc)> \\
   {} &=& <\Delta K, a \otimes a>(<K,\one> + q<K,bc>) = q^2(1+0) = q^2 \ .
\end{eqnarray*}
Now,
\begin{eqnarray*}
<K X_+,b> &=& <K \otimes X_+, \Delta b> =
              <K \otimes X_+, a \otimes b + b \otimes d> \\
       {} &=& <K,a><X_+,b> + <K,b><X_+,d> = q \beta + 0 \ .
\end{eqnarray*}
But
\begin{eqnarray*}
<X_+ K,b> &=& <X_+ \otimes K, \Delta b> =
              <X_+ \otimes K, a \otimes b + b \otimes d> \\
       {} &=& <X_+,a><K,b> + <X_+,b><K,d> = 0 + \beta q^2 \ .
\end{eqnarray*}
Therefore
\begin{eqnarray*}
<X_+ K,b> = q <KX_+, b> \quad
      &\longrightarrow& \quad K X_+ = q^2 X_+ K \ .
\end{eqnarray*}

\noindent Finally, there are no conditions on $\beta, \gamma$ and we can
take $\beta = 1$ and $\gamma = 1$.

We conclude that the following conditions are compatible:
$ab = qba$, $<K,a> = q$, $X_+ K = q^2 X_+ K$. Such considerations justify
the defining relations given {\it a priori}, in
Section~\ref{subsec:finite-dim-H}.
}

%%%%%%%%%%%%%%%%%%%%%%%%%%%%%%%%%%%%%%%%%%%%%%%%%%%%%%%%%%%%%%%%%%%%%%%%%%

\subsection{Actions of $\mathcal H$}
\label{subsec:actions-of-H}

One can define several actions of $\mathcal H$ on $\mathcal H$, of
$\mathcal H$ on $\mathcal F$ and of $\mathcal H$ on $\mathcal M$.
We briefly mention them in the present subsection.

%%%%%%%%%%%%%%%%%%%%%%%%%%%%%%%%%%%%%%%%%%%%%%%%%%%%%%%%%%%%%%%%%%%%%%%%%%

\subsection*{\sectionitem $\mathcal H$ acting on $\mathcal H$}

There is a natural action of $\mathcal H$ on itself given by the
multiplication, thus we can define:
\begin{itemize}
\item
   the right action $R[X] Y = YX$,
\item
   the right action $R'[X] Y = S(X) Y$,
\item
   the left action $L[X] Y = XY$,
\item
   the left action $L'[X] Y = Y S(X)$.
\end{itemize}
However, none of these actions is compatible with the algebra structure in
$\mathcal H$, \ie none of these make $\mathcal H$ an $\mathcal H$-module
algebra.

%%%%%%%%%%%%%%%%%%%%%%%%%%%%%%%%%%%%%%%%%%%%%%%%%%%%%%%%%%%%%%%%%%%%%%%%%%

\subsection*{\sectionitem $\mathcal H$ acting on $\mathcal F$}

Let $X,Y$ be elements of $\mathcal H$, \ie linear forms on $\mathcal F$, or
even distributions on the would-be group $G$. Take $u \in \mathcal F$.
Using the pairing $<,>$ between these dual Hopf-algebras, one can define
actions of $\mathcal H$ on $\mathcal F$, that are dual to the ones on
$\mathcal H$ we have just defined:

\begin{itemize}
\item
   the left action defined by
   $$
      <Y, X[u]> = <YX, u> \ .
   $$
   It comes from the right action of $\mathcal H$ on itself given by
   multiplication, $R[X] Y = YX$. This is a left action of $\mathcal H$
   on $\mathcal F$, since $(XY)[u] = X[Y[u]]$.
   It is also compatible with the algebra structure of $\mathcal F$ because
   $X[uv] = X_1[u] X_2[v]$, where $\Delta X = X_1 \otimes X_2$
   (implicit summation). Thus we also say that $\mathcal F$ is a left
   $\mathcal H$-module algebra.
   Notice that $<Y, X[u]> = <YX, u> =$ \linebreak
   $<Y \otimes X, \Delta u> = <Y \otimes X, u_1 \otimes u_2> =
                              <Y, u_1><X,u_2>$.
   Therefore, we have explicitly
   $$
      X[u] = u_1 <X, u_2> \ .
   $$
\item
   the left action defined by $<Y, X[u]> = <S(X)Y, u>$,
   which is not compatible with the multiplication of $\mathcal F$: more
   precisely, it involves a twist because $X[uv] = X_2[u] X_1[v]$.
\item
   the right action defined by
   $$
      <Y, X[u]> = <XY, u> \ .
   $$
   It is obviously a right action on $\mathcal F$, since
   $(XY)[u] = Y[X[u]]$.
\item
   the right action defined by $<Y, X[u]> = < Y S(X), u>$.
\end{itemize}
Again, the first right action is compatible with the algebra structure of
$\mathcal F$ whereas the second one involves a twist.

From now on, we shall choose consistently the left action of $\mathcal H$
on $\mathcal F$ defined by $<Y, X[u]> =$ \linebreak[2] $<YX, u>$.
Explicitly, we find for the left action of the generators $X_\pm$ and $K$
on $\mathcal F$
\begin{equation}
\begin{tabular}{cccccc}
   $ X_+[a] = 0 $ & $ X_+[b] = a $ &
                    $ X_-[a] = b $ & $ X_-[b] = 0 $ &
                    $ K[a] = qa $  & $ K[b] = q^2 b $ \\
   $ X_+[c] = 0 $ & $ X_+[d] = c $ &
                    $ X_-[c] = d $ & $ X_-[d] = 0 $ &
                    $ K[c] = q c $ & $ K[d] = q^2 d $ \\
\end{tabular}
\label{H-acting-on-F}
\end{equation}

For every $X$ in $\mathcal H$, one can associate a linear operator $X[\,]$
from $\mathcal F$ to $\mathcal F$. This operator is sometimes called a
``left invariant operator for the coproduct $\Delta$'' because
$\Delta \circ X[\,] = (\id \otimes X[\,]) \circ \Delta$, where $\circ$
denotes the composition of maps and $\id$ is the identity map.

{\small
\noindent Indeed,
   $\Delta(X[u]) = \Delta(u_1) <X,u_2> =
        (u_{11} \otimes u_{12}) <X, u_2> =
        (\id \otimes \id \otimes <X,\cdot>)
             (u_{11} \otimes u_{12} \otimes u_2)$.
But
$(u_{11} \otimes u_{12} \otimes u_2) = (u_1 \otimes u_{21} \otimes u_{22})$,
because of the coassociativity of $\Delta$ (recall that
$(\Delta \otimes \id) \circ \Delta = (\id \otimes \Delta) \circ \Delta$).
So
$\Delta(X[u]) =
   (\id \otimes \id \otimes <X,\cdot>)(u_1 \otimes u_{21} \otimes u_{22}) =
   u_1 \otimes u_{21} <X, u_{22}> =
   (\id \otimes X[\,])(u_1 \otimes u_2) = (\id \otimes X[\,]) \Delta(u)$.
}

The space of left invariant operators on $\mathcal F$ ---the dual of
$\mathcal H$--- is isomorphic with $\mathcal H$ itself since
$(XY)[u] = X[Y[u]]$. Explicitly, the isomorphism $X \leadsto X[\,]$ and
$X[\,] \leadsto X$ is given by $X[u] =$ \linebreak[2]
$(\id \otimes <X,\cdot>) \circ \Delta$ and $<X, u> = \epsilon \circ X[u]$
where $\epsilon$ is the counit of $\mathcal F$. In a classical
(\ie group-like) situation, it would have been equivalent to evaluate
$X[u]$ at the identity of the group to get $<X, u>$.

On the contrary, the space of right invariant operators (for the coproduct
$\Delta$) is anti-isomorphic with $\mathcal H$, since with a right action
$(XY)[u] = Y[X[u]]$.

%%%%%%%%%%%%%%%%%%%%%%%%%%%%%%%%%%%%%%%%%%%%%%%%%%%%%%%%%%%%%%%%%%%%%%%%%%

\subsubsection*{Remark about the classical case}

An easy way to remember the previous results is the following. First we
introduce classical generators of $Lie(SL(2))$ in the fundamental
representation, \ie the usual matrices $\underline{X}_\pm$,
$\underline{X}_3$ and define $\underline{K} = q^{\underline{X}_3}$:
$$
\underline{X}_+ = \pmatrix{0 & 1 \cr 0 & 0}  \ , \quad
\underline{X}_- = \pmatrix{0 & 0 \cr 1 & 0}  \ , \quad
\underline{X}_3 = \pmatrix{1 & 0 \cr 0 & -1} \ , \quad
\underline{K}   = q^{\underline{X}_3}
                = \pmatrix{q & 0 \cr 0 & q^{-1}} \ .
$$

\noindent Now we have a left action on ${\mathcal F}^{class}$,
which for a generic $\underline{Y} \in Lie(SL(2))$ reads\footnote
{
   Note that the following equations should be interpreted as equalities
   amongst entries of the matrices, \ie
   $$
      \underline{X}_+^L \left[ \pmatrix{a & b \cr c & d} \right] \equiv
         \pmatrix{\underline{X}_+^L[a] & \underline{X}_+^L[b] \cr
                  \underline{X}_+^L[c] & \underline{X}_+^L[d]}   \doteq
         \pmatrix{a & b \cr c & d}\underline{X}_+ \ ,
   $$
   \etc  This explains why this action is a left one, in spite of being
   given by a right multiplication; for example:
   $$ \underline{X}_+^L \left[ \underline{X}_-^L \left[
                        \pmatrix{a & b \cr c & d} \right] \right]     =
      \underline{X}_+^L \left[ \pmatrix{b & 0 \cr d & 0} \right]      =
      \pmatrix{\underline{X}_+^L[b] & 0 \cr \underline{X}_+^L[d] & 0} =
      \pmatrix{a & 0 \cr c & 0}                                       =
      \pmatrix{a & b \cr c & d} \underline{X}_+ \underline{X}_-       =
      (\underline{X}_+ \underline{X}_-)^L \left[
                       \pmatrix{a & b \cr c & d} \right]
   $$

   This action is nothing more than the left action of the classical
   enveloping algebra of a Lie algebra on functions $U \in C(G)$ of the
   corresponding classical group $G$. Being dual to the right-product
   action $R[X]Y = YX$, it is given infinitesimally for the Lie
   generators $x$ by
   $\left. x[U] \right|_g \doteq \left. U \right|_{g(\one + x)} -
                                 \left. U \right|_g$.
}

$$
   \underline{Y}^L \left[ \pmatrix{a & b \cr c & d} \right] \doteq
      \pmatrix{a & b \cr c & d}\underline{Y} \ ,
$$
and a right one which is defined analogously. For the generators, this
action reads explicitly
$$
\pmatrix{a & b \cr c & d}\underline{X}_+ = \pmatrix{0 & a \cr 0 & c}
                                           \ , \quad
\pmatrix{a & b \cr c & d}\underline{X}_- = \pmatrix{b & 0 \cr d & 0}
                                           \ , \quad
\pmatrix{a & b \cr c & d}\underline{X}_3 = \pmatrix{a & -b \cr c & -d} \ .
$$

\noindent Observe that $\underline{X}_\pm$ and $\underline{X}_3$ act by
derivation on ${\mathcal F}^{class}$, and $\underline{K}$ as an
automorphism. Indeed, the above left {\sl classical} action (and the
corresponding right one) of $Lie(SL(2))$ on ${\mathcal F}^{class}$ may
also be written in terms of operators (classical derivations)
$$
\begin{tabular}{ll}
$\underline{X}_+^L = a {\partial\over\partial b}
                   + c {\partial\over\partial d}$ &
$\underline{X}_+^R = c {\partial\over\partial a}
                   + d {\partial\over\partial b}$ \\
$\underline{X}_-^L = b {\partial\over\partial a}
                   + d {\partial\over\partial c}$ &
$\underline{X}_-^R = a {\partial\over\partial c}
                   + b {\partial\over\partial d}$ \\
$\underline{X}_3^L = a {\partial\over\partial a}
                   - b {\partial\over\partial b}
                   + c {\partial\over\partial c}
                   - d {\partial\over\partial d}$ \quad &
$\underline{X}_3^R = a {\partial\over\partial a}
                   + b {\partial\over\partial b}
                   - c {\partial\over\partial c}
                   - d {\partial\over\partial d}$
\end{tabular}
$$

\noindent It is easy to check that, for instance (and as it should!)
$$
   [\underline{X}_3,\underline{X}_+]     = +2\underline{X}_+   \ , \qquad
   [\underline{X}_3^L,\underline{X}_+^L] = +2\underline{X}_+^L \ , \qquad
   [\underline{X}_3^R,\underline{X}_+^R] = -2\underline{X}_+^R \ .
$$

We recover the fact that left fundamental vector fields build up, in the
classical case, the Lie algebra itself.

As we see, it turns out that the formulae (\ref{H-acting-on-F}) expressing
the action of $\mathcal H$ on the quantum {\em generators} $a,b,c,d$ are
the same as the classical ones. However, the quantum action {\sl cannot\/}
be extended by derivations, when the same quantities act on arbitrary
elements of $\mathcal F$: the action of invariant linear operators is
then carried out by {\sl twisted derivations\/} (derivations twisted by
automorphisms). This can be easily seen remembering that one should
use the coproduct of $\mathcal H$ to act on products of $\mathcal F$.

%%%%%%%%%%%%%%%%%%%%%%%%%%%%%%%%%%%%%%%%%%%%%%%%%%%%%%%%%%%%%%%%%%%%%%%%%%

\subsection*{\sectionitem $\mathcal H$ acting on $\mathcal M$}

Using the fact that $\mathcal F$ coacts on $\mathcal M$ (an algebra of
matrices) in two possible ways, and that elements of $\mathcal H$ can be
interpreted as distributions on $\mathcal F$, we obtain two actions of
$\mathcal H$ on the quantum space $\mathcal M$. We shall describe them for
arbitrary elements and give explicit results for the generators.

Let $z \in \mathcal M$. We know that $\mathcal F$ coacts on $\mathcal M$
from the left and from the right. For the left coaction, we have
$\Delta_L z = u_\alpha \otimes z_\alpha$ with $u_\alpha \in \mathcal F$
and $z_\alpha \in \mathcal M$ (implied summation). For instance
$\Delta_L x = a \otimes x + b \otimes y$. Let $X$ be an element of
$\mathcal H$. Using the pairing between $\mathcal H$ and $\mathcal F$,
we set
\begin{equation}
   X^R[z] \doteq (<X,\cdot> \otimes \id) \Delta_L z
               = < X, u_\alpha> z_\alpha \ ,
\label{R-H-action-on-M}
\end{equation}
which makes it into a {\em right} action of $\mathcal H$ on $\mathcal M$.
Moreover, it makes $\mathcal M$ a right-$\mathcal H$-module algebra. For
instance, $X_+^R[x] = <X_+, a> x + <X_+, b> y = 0 x + 1 y = y$.

$\mathcal F$ also coacts on $\mathcal M$ from the right, so that
$\Delta_R z = z_\alpha \otimes v_\alpha$ with $z_\alpha \in \mathcal M$
and $v_\alpha \in \mathcal F$ (implied summation). For instance
$\Delta_R x = x \otimes a + y \otimes c$. Taking $X \in \mathcal H$, and
using the $\mathcal H$-$\mathcal F$ pairing we can define the {\em left}
action
\begin{equation}
   X^L[z] \doteq (\id \otimes <X,\cdot>) \Delta_R z
               = <X, v_\alpha> z_\alpha \ .
\label{L-H-action-on-M}
\end{equation}
With this $L$-action we can check that $\mathcal M$ a
left-$\mathcal H$-module algebra. In particular,
$X_+^L[x] =$ \linebreak[3] $x <X_+, a> + y <X_+, c> = 0 x + 0 y = 0$.

We stress again that we have denoted with $X^L$ the operators on
$\mathcal M$ corresponding to the left-action of an element
$X \in \mathcal H$ on $\mathcal M$, but this action is indeed dual to the
{\em right}-coaction of $\mathcal F$ on $\mathcal M$. The same happens for
the right-action. One should therefore use the notation involving upper
indices $L$ or $R$ with some care.

To obtain the action of an arbitrary element $X \in \mathcal H$ on a
product $zw$ of elements of $\mathcal M$, one may use the definition (above)
or the following property (notation should be clear):
$$
   X^L[zw] = m \left[ (\Delta X)^L [z \otimes w] \right] \ .
$$
It is therefore enough to know the action of the generators $X_{\pm}, K$
on $\mathcal M$ (complete $27 \times 27$ tables, with other conventions,
can be found in \cite{Dabrowski}). One finds:

\begin{table}[h]
$$
\begin{tabular}{l||ccc}
Right     & $K^R$      & $X_+^R$   & $X_-^R$    \\
\hline
\hline
$\one$    & $\one$     & $0$       & $0$        \\
$x^2 y$   & $q x^2 y$  & $-xy^2$   & $\one$     \\
$xy^2$    & $q^2 xy^2$ & $\one$    & $-x^2 y$   \\
\hline
$x$       & $qx$       & $y$       & $0$        \\
$y$       & $q^2 y$    & $0$       & $x$        \\
$x^2 y^2$ & $x^2 y^2$  & $-x$      & $-y$       \\
\hline
$x^2$     & $q^2 x^2$  & $-xy$     & $0$        \\
$xy$      & $xy$       & $y^2$     & $x^2$      \\
$y^2$     & $qy^2$     & $0$       & $-xy$      \\
\end{tabular}
\hspace{1.0cm}
\begin{tabular}{l||ccc}
Left      & $K^L$      & $X_+^L$   & $X_-^L$    \\
\hline
\hline
$\one$    & $\one$     & $0$       & $0$        \\
$x^2 y$   & $q x^2 y$  & $q^2\one$ & $-xy^2$    \\
$xy^2$    & $q^2 xy^2$ & $-x^2 y$  & $q^2 \one$ \\
\hline
$x$       & $q x$      & $0$       & $y$        \\
$y$       & $q^2 y$    & $x$       & $0$        \\
$x^2 y^2$ & $x^2 y^2$  & $-qy$     & $-qx$      \\
\hline
$x^2$     & $q^2 x^2 $ & $0$       & $-q^2 xy$  \\
$xy$      & $xy$       & $qx^2$    & $q y^2$    \\
$y^2$     & $q y^2$    & $-q^2 xy$ & $0$        \\
\end{tabular}
$$
\caption{Right and left actions of $\mathcal H$ on $\mathcal M$.}
\label{table:H-acting-on-M}
\end{table}

\noindent Notice that, up to multiplicative factors,
\nopagebreak
\vspace{0.5cm}
\begin{equation}
\begin{diagram}
  y^2           & \rTo 0      \\
  \uTo{X_+^R} \dDotsto{X_-^R} \\
  xy            &             \\
  \uTo{X_+^R} \dDotsto{X_-^R} \\
  x^2           & \rDotsto 0
\end{diagram}
\quad\hbox{,}\quad
 \begin{diagram}
  y             & \rTo      & 0       \\
                & \luDotsto &         \\
  \uTo{X_+^R} \dDotsto{X_-^R}
                &           & x^2 y^2 \\
                & \ldTo     &         \\
  x             & \rDotsto  & 0
\end{diagram}
\quad\hbox{and}\quad
\begin{diagram}
  xy^2          &           &               \\
                & \rdTo     &               \\
  \uTo{X_+^R} \dDotsto{X_-^R}
                &           & \one          \\
                & \ruDotsto & \dTo \dDotsto \\
  x^2 y         &           & 0
\end{diagram}
\label{graph:right-H-action-on-M}
\end{equation}
\vspace{0.5cm}
\begin{equation}
\begin{diagram}
  y^2           & \rDotsto 0  \\
  \uDotsto{X_-^L} \dTo{X_+^L} \\
  xy            &             \\
  \uDotsto{X_-^L} \dTo{X_+^L} \\
  x^2           & \rTo 0
\end{diagram}
\quad\hbox{,}\quad
\begin{diagram}
  y             & \rDotsto  & 0       \\
                & \luTo     &         \\
  \uDotsto{X_-^L} \dTo{X_+^L}
                &           & x^2 y^2 \\
                & \ldDotsto &         \\
  x             & \rTo      & 0
\end{diagram}
\quad\hbox{and}\quad
\begin{diagram}
  xy^2          &           &               \\
                & \rdDotsto &               \\
  \uDotsto{X_-^L} \dTo{X_+^L}
                &           & \one          \\
                & \ruTo     & \dDotsto \dTo \\
  x^2 y         &           & 0
\end{diagram}
\label{graph:left-H-action-on-M}
\end{equation}
\vspace{0.3cm}

\noindent We see clearly on these diagrams that (up to numerical factors)
the right action is obtained from the left one by interchanging arrows for
$X_+$ and $X_-$.

%%%%%%%%%%%%%%%%%%%%%%%%%%%%%%%%%%%%%%%%%%%%%%%%%%%%%%%%%%%%%%%%%%%%%%%%%%

\subsubsection*{Differential operators on $\mathcal M$ associated with the
                $\mathcal H$ action}

Here, we are only concerned with those differential operators on
$\mathcal M$ associated with the quantum group action of $\mathcal H$.
A more complete analysis can be found in \ref{app:diff-operators-on-M}
They are obtained by considering products and powers of $X_{\pm}^{L,R}$,
$K^{L,R}$ acting from the left or from the right as before. From now on we
will stick with the left action.

We also consider elements of $\mathcal M$ acting by multiplication as
differential operators of order zero: let $z \in \mathcal M$, then, for
example, $x[z] \doteq xz$. Let $X \in \mathcal H$; the reader should
distinguish between $X x$ ---which is a differential operator on
$\mathcal M$ ($Xx[z] \doteq X[x[z]] = X[xz]$)--- from $X[x]$ which is an
element of $\mathcal M$. It makes therefore perfect sense to study the
commutation relations, say, between $X$ and $x$. We shall use such
relations later. Here, we establish the commutation relations between
$x,y$ ---taken as differential operators of degree $0$--- and $X_{\pm},K$.

Let $z \in \mathcal M$. Consider, for instance,
$$
   X_+^L[xz] = X_+^L[x] \, \one[z] + K^L[x] \, X_+^L[z] =
               0 + qx X_+^L[z] \ .
$$
Therefore $X_+^L \, x = qx \, X_+^L$. Similarly, we obtain:
\begin{eqnarray}
   X_+^L \, x &=& q \, x \, X_+^L                \nonumber \\
   X_-^L \, x &=& y \, {{K^L}^{-1}} + x \, X_-^L \nonumber \\
   K^L   \, x &=& q \, x \, K^L                  \nonumber \\
          {} &{}& {}                                       \\
   X_+^L \, y &=& x + q^2 y \, X_+^L             \nonumber \\
   X_-^L \, y &=& y \, X_-^L                     \nonumber \\
   K^L   \, y &=& q^2 y \, K^L                   \nonumber
\label{H-M-relations}
\end{eqnarray}

\noindent Any element of $\mathcal H$ can also be written explicitly in
terms of differential operators on $\mathcal M$; see
\ref{app:diff-operators-on-M}

It may be useful to notice that it is possible to introduce ``conformal
weights'' so that previous formulae are homogeneous with the following
weight assignments: $(X_+^L, K^L, X_-^L) \mapsto (1,0,-1)$, and
$(x,y) \mapsto (1/2,-1/2)$.

%%%%%%%%%%%%%%%%%%%%%%%%%%%%%%%%%%%%%%%%%%%%%%%%%%%%%%%%%%%%%%%%%%%%%%%%%%

\subsection{The structure of the non-semisimple algebra $\mathcal H$}
\label{subsec:structure-of-H}

The algebra $\mathcal H$ can be explicitly defined ---and many of its
properties understood--- without using anything more sophisticated than
basic multiplication or tensor products of matrices as well as elementary
calculus involving anti-commuting numbers (Grassmann numbers). This
construction, inspired from \cite{Alekseev}, was explicitly performed in
\cite{Coquereaux} for $N = 3$ and used to analyse the representation theory
of $\mathcal H$. The analysis for other values of $N$ can be found in
\cite{Ogievetsky}. We just recall here the results and establish the
relations with the previous notations.

The finite-dimensional quantum group $\mathcal H$ can be defined as the
algebra ${\cal H} \doteq M_3 \oplus (M_{2\vert 1}(\Lambda^2))_0$, where
$M_3$ is the set of $3\times 3$ complex matrices and
$(M_{2\vert 1}(\Lambda^2))_0$ is the Grassmann envelope of the associative
$\ZZ_2$-graded algebra $M_{2\vert 1}(\CC^2)$, \ie the even part of its
graded tensor product with a Grassmann algebra $\Lambda^2$ of two
generators.

To see this explicitly, let $\Lambda^2$ be the Grassmann algebra over
$\CC$ with two generators, \ie the linear span of
$\{ 1, \theta_1, \theta_2, \theta_1 \theta_2 \}$ with arbitrary complex
coefficients, where the relations $\theta_1^2 = \theta_2^2 = 0$ and
$\theta_1 \theta_2 = - \theta_2 \theta_1$ hold. This algebra has an even
part, generated by $1$ and $\theta_1 \theta_2$, and an odd part, generated
by $\theta_1$ and $\theta_2$. We call $M_3$ the algebra of $3 \times 3$
matrices over the complex numbers and $M_{2\vert 1}$ another copy of this
algebra that we grade as follows: a matrix $V \in M_{2\vert 1}$ is called
even if it is of the type

$$
   V = \pmatrix{V_{11} & V_{12} & 0 \cr
                V_{21} & V_{22} & 0 \cr
                0      & 0      & V_{33}} \ ,
$$
and odd if it is of the type
$$
   V = \pmatrix{0      & 0      & V_{13} \cr
                0      & 0      & V_{23} \cr
                V_{31} & V_{32} & 0}     \ .
$$

\noindent We call $(M_{2\vert 1}(\Lambda^2))_0$ the Grassmann envelope of
$M_{2\vert 1}$ which is defined as the even part of the tensor product of
$M_{2\vert 1}$ and $\Lambda^2$, \ie the space of $3 \times 3$ matrices $V$
with entries $V_{11}$, $V_{12}$, $V_{21}$, $V_{22}$, $V_{33}$ that are
even Grassmann elements (of the kind $\alpha + \beta \theta_1 \theta_2$)
and entries $V_{13}$, $V_{23}$, $V_{31}$, $V_{32}$ that are odd Grassmann
elements (of the kind $\gamma \theta_1 + \delta \theta_2$). We define
$\mathcal H$ as

$$
   \mathcal H \doteq M_3 \oplus (M_{2\vert 1}(\Lambda^2))_0 \ .
$$
Explicitly,
$$
\mathcal H = \pmatrix{ * & * & * \cr
                       * & * & * \cr
                       * & * & * } \oplus
   \pmatrix{\alpha_{11} + \beta_{11} \theta_1 \theta_2  &
            \alpha_{12} + \beta_{12} \theta_1 \theta_2  &
            \gamma_{13} \theta_1 + \delta_{13} \theta_2 \cr
            \alpha_{21} + \beta_{21} \theta_1 \theta_2  &
            \alpha_{22} + \beta_{22} \theta_1 \theta_2  &
            \gamma_{23} \theta_1 + \delta_{23} \theta_2 \cr
            \gamma_{31} \theta_1 + \delta_{31} \theta_2 &
            \gamma_{32} \theta_1 + \delta_{32} \theta_2 &
            \alpha_{33} + \beta_{33} \theta_1 \theta_2} \ .
$$

\noindent All entries besides the $\theta$'s are complex numbers (the above
$\oplus$ sign is a direct sum sign: these matrices are $6\times 6$ matrices
written as a direct sum of two blocks of size $3\times 3$).

It is obvious that this is an associative algebra, with usual matrix
multiplication, of dimension $27$ (just count the number of arbitrary
parameters). $\mathcal H$ is not semisimple, because of the appearance
of Grassmann numbers in the entries of the matrices. Its semisimple part
$\overline{\mathcal H}$, given by the direct sum of its block-diagonal
$\theta$-independent parts is equal to the $9+4+1 = 14$-dimensional algebra
$\overline{\mathcal H} = M_3(\CC) \oplus M_2(\CC) \oplus \CC$. The radical
(more precisely the Jacobson radical) $J$ of $\mathcal H$ is the left-over
piece that contains all the Grassmann entries, and only the Grassmann
entries, so $\overline{\mathcal H} ={\mathcal H}/J$. The radical has
therefore dimension $13$.

In order to write generators for $\mathcal H$, we need to consider
$6 \times 6$ matrices that have a
$(3\times 3) \oplus ((2\vert 1)\times (2\vert 1))$
block diagonal structure. We introduce elementary matrices $E_{ij}$ for the
$M_3(\CC)$ part and elementary matrices $F_{ij}$ for the
$M_{2\vert 1}(\CC)$ part. The associative algebra $\mathcal H$ defined
previously can be generated by the following three matrices:

\begin{eqnarray*}
X_+ &=& E_{12} + E_{23} +
        (1-\theta_1 \theta_2/2) F_{12} + \theta_1 (F_{23} + F_{31}) \\
X_- &=& - E_{21} - E_{32} +
        (1-\theta_1 \theta_2/2) F_{21} + \theta_2 (F_{13} - F_{32}) \\
K   &=& q^2 E_{11} + E_{22} + q^{-2} E_{33} +
        q F_{11} + q^{-1} F_{22} + F_{33}
\end{eqnarray*}
Explicitly, one gets
{\small
\begin{eqnarray*}
X_+ &\doteq& \pmatrix{\pmatrix{0&1&0\cr 0&0&1\cr 0&0&0} & \pmatrix{} \cr
                      \pmatrix{} & \pmatrix{
                           0 & 1-{\theta_1\theta_2 \over 2} & 0 \cr
                           0 & 0 & \theta_1 \cr
                           \theta_1 & 0 & 0}} \\
X_- &\doteq& \pmatrix{\pmatrix{0&0&0\cr -1&0&0\cr 0&-1&0} & \pmatrix{} \cr
                      \pmatrix{} & \pmatrix{
                           0 & 0 & \theta_2 \cr
                           1-{\theta_1\theta_2 \over 2} & 0 & 0 \cr
                           0 & -\theta_2 & 0}} \\
K &\doteq& \pmatrix{\pmatrix{q^2&0&0\cr 0&1&0\cr 0&0&q^{-2}}&\pmatrix{} \cr
                      \pmatrix{} & \pmatrix{
                           q & 0 & 0 \cr
                           0 & q^{-1} & 0 \cr
                           0 & 0 & 1}}
\end{eqnarray*}
}

\noindent Performing explicit matrix multiplications or using the relations
$E_{ij} E_{jk} = E_{ik}$, $F_{ij} F_{jk} = F_{ik}$ and
$E_{ij} F_{jk} = F_{ij} E_{jk} = 0$,
it is easy to see that the defining relations of $\mathcal H$ are
indeed satisfied.

As mentioned in Section~\ref{subsec:finite-dim-H}, the center of
$\mathcal H$ is generated by a Casimir operator $C$. Its explicit expression
reads
{\small
$$
C = \pmatrix{\pmatrix{-2/3&0&0\cr 0&-2/3&0\cr 0&0&-2/3} & \pmatrix{} \cr
             \pmatrix{} & \pmatrix{
                 1/3 - \theta_1 \theta_2 & 0 & 0 \cr
                 0 & 1/3 - \theta_1 \theta_2 & 0 \cr
                 0 & 0 & 1/3 + \theta_1 \theta_2}}
$$
}

The problem of studying representation theory for $\mathcal H$ is solved by
considering separately all the columns defining
$\mathcal H = M_3 \oplus (M_{2\vert 1}(\Lambda^2))_0$ as a matrix algebra
over a ring. We just ``read'' the following three indecomposable
representations from the explicit definition of $\mathcal H$.
First of all we have\footnote{
The following should be read as ``column vectors''.
}
a three-dimensional irreducible representation
$3_{irr} \doteq (c_1, c_2, c_3)$ (where $c_i$ are complex numbers) coming
from the $M_3$ block. Notice that the three columns of the $M_{3}$ factor
give equivalent irreducible representations.  Next we have two reducible
indecomposable representations (also called ``PIM's'' for ``Projective
Indecomposable Modules'') coming from the columns of
$(M_{2\vert 1}(\Lambda^2))_0$. Notice that the first two columns give
equivalent representations ---that we call $P_e \equiv 6_{eve}$--- and the
last one gives the representation $P_o \equiv 6_{odd}$. Each of these two
PIM's is of dimension $6$. The PIM's are also called ``principal modules''.

$P_o$ and $P_e$, although indecomposable, are not irreducible:
submodules (sub-representations) are immediately found by requiring
stability of the representation spaces under the left multiplication by
elements of $\mathcal H$. The lattice of submodules of $\mathcal H$ is
given and discussed in \cite{Coquereaux}. We recall these results in
\ref{app:H-submodules} Here we just want to note (because it will be
useful shortly) that $6_o$ and $6_e$ both contain indecomposable (but
not irreducible) sub-representations of dimension $3$. Indeed, take
$\lambda_1, \lambda_2 \in \CC$, set
$\lambda \doteq {\lambda_1 \over \lambda_2} \in CP^1$, and define
$\theta_\lambda = \lambda_1 \theta_1 + \lambda_2 \theta_2$.
We obtain, for each $\lambda$, a representation $3_{odd}$ spanned by
$(\gamma \theta_{\lambda}, \gamma'\theta_{\lambda},
   \alpha \theta_1\theta_2)$
and a representation $3_{eve}$ spanned by
$(\beta \theta_{1}\theta_{2}, \beta' \theta_1\theta_2,
   \delta \theta_{\lambda})$.
Here the coefficients $\beta, \beta', \gamma, \gamma', \alpha, \delta$
are arbitrary complex numbers.

%%%%%%%%%%%%%%%%%%%%%%%%%%%%%%%%%%%%%%%%%%%%%%%%%%%%%%%%%%%%%%%%%%%%%%%%%%

\subsection{Decomposition of ${\mathcal M} = M_3(\CC)$ into representations
            of $\mathcal H$}
\label{subsec:M-in-reps-H}

Since there is an action of $\mathcal H$ on $\mathcal M$, it is {\it a
priori\/} clear that $\mathcal M$, as a vector space, can be analysed in
terms of representations of $\mathcal H$. It turns out, as it is easy to
see from (\ref{graph:right-H-action-on-M}) or
(\ref{graph:left-H-action-on-M}), that
$$
   {\mathcal M} \sim 3_{irr} \oplus 3_{eve} \oplus 3_{odd} \ .
$$
More precisely,
\begin{itemize}
\item
   $3_{irr}$ is spanned by $x^2, xy, y^2$,
\item
   $3_{eve}$ is spanned by $x, y, x^2 y^2$,
\item
   $3_{odd}$ is spanned by $\one, x^2 y, xy^2$.
\end{itemize}
If we attribute a $\ZZ_3$-grading $1$ to $x$ and $y$, these three spaces
carry gradings respectively equal to $2,1,0$. This same analysis can be
done in a more general $N$ case, the interested reader can find it in
\cite{Coquereaux-Schieber}.

The algebra $\mathcal M$, as a vector space, carries therefore a
$9$-dimensional reducible representation of $\mathcal H$ and splits into
the direct sum of an irreducible representation of dimension $3$ and two
inequivalent representations of dimension $3$ that are reducible but not
fully reducible. As described previously, these last two representations
$3_{eve}$ and $3_{odd}$ appear as subrepresentations of two projective
indecomposable modules of dimension $6$. It is easy to see (\cf diagrams
below) that $3_{eve} = 2 \rplus 1$ where $2$ denotes a two-dimensional
invariant subspace (but $1$ is not) whereas $3_{odd} = 1 \rplus 2$ where
$1$ denotes a one-dimensional invariant subspace (but $2$ is not).

One possible identification between $\mathcal H$ acting from the left on
$\mathcal M$ and $3_{irr} \oplus 3_{eve} \oplus 3_{odd}$ as elements of
the left regular representation of $\mathcal H$, is as follows:

$$
\begin{tabular}{ccc}
$x^2 \sim E_{12}$ & $x \sim \theta_1\theta_2 F_{11}$            &
                    $x^2 y \sim (\theta_1 - \theta_2) F_{13}$   \\
$xy \sim E_{22}$  & $y \sim \theta_1\theta_2 F_{21}$            &
                    $xy^2 \sim (\theta_1 - \theta_2) F_{23}$    \\
$y^2 \sim E_{32}$ & $x^2 y^2 \sim (\theta_1 - \theta_2) F_{31}$ &
                    $\one \sim \theta_1\theta_2 F_{33}$         \\
\end{tabular}
$$

{\small
\noindent In order to establish this identification, it is enough to
compare the values of $X^L[z]$, for the generators $X$ of $\mathcal H$ and
$z \in \mathcal M$, with those obtained by left matrix multiplication using
the above realisation of $\mathcal H$ as
$M_3 \oplus (M_{2\vert 1}(\Lambda^2))_0$ (of course, we could use right
matrix multiplication and compare with the values of $X^R[z]$ given
before). For instance, in the case of the generator $X_+$ we obtain

$$
\begin{tabular}{ccc}
   $X_+E_{12} = 0$      & $X_+ \theta_1\theta_2 F_{11} = 0$         &
      $X_+ (\theta_1 - \theta_2) F_{13} = -\theta_1\theta_2 F_{33}$   \\
   $X_+E_{22} = E_{12}$ &
      $X_+ \theta_1\theta_2 F_{21} = \theta_1\theta_2 F_{11}$       &
      $X_+ (\theta_1 - \theta_2) F_{23} = (\theta_1-\theta_2) F_{13}$ \\
   $X_+E_{32} = E_{22}$ &
      $X_+ (\theta_1 - \theta_2) F_{31} = -\theta_1\theta_2 F_{21}$ &
      $X_+ \theta_1\theta_2 F_{33} = 0$                               \\
\end{tabular}
$$
to be compared with
$$
\begin{tabular}{ccc}
   $X_+^L[x^2] = 0$       & $X_+^L[x]      = 0$     &
                            $X_+^L[\one] = 0$       \\
   $X_+^L[xy]  = q\,x^2$  & $X_+^L[y]      = x$     &
                            $X_+^L[x^2y] = q^2\one$ \\
   $X_+^L[y^2] = -q^2 xy$ & $X_+^L[x^2y^2] = -q\,y$ &
                            $X_+^L[xy^2] = -x^2y$   \\
\end{tabular}
$$
One obtains in this way two identical $3 \times 9$ tables.
}

This can also be seen by comparing the following diagrams with those
obtained previously:
$$
\begin{diagram}
   E_{12}                  & \rTo     & 0 \\
   \uTo{X_+} \dDotsto{X_-}                \\
   E_{22}                                 \\
   \uTo{X_+} \dDotsto{X_-}                \\
   E_{32}                  & \rDotsto & 0 \\
\end{diagram}
\quad\hbox{,}\quad
\begin{diagram}
   \theta_1\theta_2 F_{11} & \rTo      & 0                            \\
                           & \luDotsto &                              \\
   \dDotsto{X_-} \uTo{X_+} &           & (\theta_1 - \theta_2) F_{31} \\
                           & \ldTo     &                              \\
   \theta_1\theta_2 F_{21} & \rDotsto  & 0
\end{diagram}
\quad\hbox{and}\quad
\begin{diagram}
   (\theta_1 - \theta_2) F_{13} &           &                          \\
                                & \rdTo     &                          \\
   \dDotsto{X_-} \uTo{X_+}      &           & \theta_1 \theta_2 F_{33} \\
                                & \ruDotsto & \dDotsto \dTo            \\
   (\theta_1 - \theta_2) F_{23} &           & 0
\end{diagram}
$$

%%%%%%%%%%%%%%%%%%%%%%%%%%%%%%%%%%%%%%%%%%%%%%%%%%%%%%%%%%%%%%%%%%%%%%%%%%
%%%%%%%%%%%%%%%%%%%%%%%%%%%%%%%%%%%%%%%%%%%%%%%%%%%%%%%%%%%%%%%%%%%%%%%%%%

\section{Reality structures}
\label{sec:stars}

We now want to obtain a reality structure (a star) and a scalar product
on the space of matrices $\mathcal M$, with covariance properties under
the quantum group transformations.

%%%%%%%%%%%%%%%%%%%%%%%%%%%%%%%%%%%%%%%%%%%%%%%%%%%%%%%%%%%%%%%%%%%%%%%%%%

\subsection{Real forms and stars on quantum groups}

A $*$-Hopf algebra $\mathcal F'$ is an associative algebra that satisfies
the following properties (for all elements $a, b$ in $\mathcal F'$):

\begin{itemize}
\item
   $\mathcal F'$ is a Hopf algebra (a quantum group), with coproduct
   $\Delta$, antipode $S$ and counit $\epsilon$.

\item
   $\mathcal F'$ is an involutive algebra, \ie it has an involution $*$
   (a `star' operation). This operation is involutive ($**a = a$),
   antilinear ($*(\lambda a) = \overline{\lambda} *a$, where $\lambda$ is
   a complex number), and anti-multiplicative ($*(ab) = (*b)(*a)$).

\item
   The involution is compatible with the coproduct, in the following sense:
   if $\Delta a = a_1 \otimes a_2$, then $\Delta *a = *a_1 \otimes *a_2$.

\item
   The involution is also compatible with the counit:
   $\epsilon(*a) = \overline{\epsilon(a)}$.

\item
   The following relation with the antipode holds: $S * S * a = a$.
\end{itemize}

\noindent Actually, the last relation is a consequence of the others.
It can also be written
$$
   S \, * = * \, S^{-1} \ .
$$
It may happen that $S^2 = \id$, in which case $S \, * = * \, S$, but it
is not generally so (usually the square of the antipode is only a
conjugacy).

If one wishes, using the $*$ on $\mathcal F'$, one can define a star
operation on the tensor product ${\mathcal F}' \otimes {\mathcal F}'$, by
$*(a \otimes b) = *a \otimes *b$. The third condition reads then:
$$
   \Delta \, * = * \, \Delta \ ,
$$
so one can say $\Delta$ is a $*$-homomorphism between $\mathcal F'$ and
${\mathcal F}' \otimes {\mathcal F}'$ (each with its respective star). It
can also be said that $\epsilon$ is a $*$-homomorphism between
$\mathcal F'$ and $\CC$ with the star given by complex conjugation.

A star operation as above, making the Hopf-algebra a $*$-Hopf algebra, is
also called a {\sl real form\/} for the given algebra. Very often, it is
convenient to write $a^* \doteq *a$. A real element is, by definition, an
element such that $a^* = a$. Notice that a product of real elements $a$ and
$b$ is not real in general ($(ab)^* = b^* a^* = ba$), but the anticommutator
of two real elements, \ie the Jordan product, is real
($(ab+ba)^* = b^* a^* + a^* b^* = ab+ba$).

%%%%%%%%%%%%%%%%%%%%%%%%%%%%%%%%%%%%%%%%%%%%%%%%%%%%%%%%%%%%%%%%%%%%%%%%%%

\subsubsection{Twisted stars}

It may happen that one finds an involution $*$ on a Hopf algebra for
which the third axiom fails in a special way, namely, the case where
$\Delta a = a_1 \otimes a_2$ but $\Delta *a = *a_2 \otimes *a_1$. In
this case $S \, * = * \, S$. Such an involution is called a {\sl twisted
star\/} operation. Remember that, whenever one has a coproduct $\Delta$
on an algebra, it is possible to construct another coproduct $\Delta^{op}$
by composing the first one with the tensorial flip: if
$\Delta a = a_1 \otimes a_2$, then $\Delta^{op} a \doteq a_2 \otimes a_1$.
If one defines a star operation on the tensor product (as before) by
$*(a \otimes b) \doteq *a \otimes *b$, the property defining a twisted
star reads
$$
   \Delta \, * = * \, \Delta^{op} \ .
$$

One should be cautious: some authors introduce a different star operation
on the tensor product, namely $*'(a \otimes b) \doteq *b \otimes *a$. In
that case, a twisted star operation obeys $\Delta \, * = *' \, \Delta$ !

$*$-Hopf algebras seem to be ``better'' than Hopf algebras endowed with
twisted stars (their representation theory is more interesting). In any
case we shall only be interested in genuine ``true'' $*$-Hopf algebras and
the only reason why we mention these twisted stars here is that we want
to rule out some of the possibilities that we shall meet later.

%%%%%%%%%%%%%%%%%%%%%%%%%%%%%%%%%%%%%%%%%%%%%%%%%%%%%%%%%%%%%%%%%%%%%%%%%%

\subsection{General results concerning real forms for the classical group
            $SL(2,\CC)$ and the quantum group $Fun(SL_q(2,\CC))$}

The following results are known and can be found easily in the literature.
Classically, one can define several real forms for $SL(2,\CC)$, by selecting
the elements which are unitary with respect to a given scalar product
$G$, which defines a conjugacy on the group. This last one, in turn, may be
thought as coming from a real form on the universal enveloping algebra of
the Lie algebra of the group. If
$u = \mbox{\scriptsize$\pmatrix{\alpha & \beta \cr \gamma & \delta}$}
   \in SL(2,\CC)$,
one may impose $\overline{u^t} G u = G$. There are three cases:

\begin{enumerate}
\item
   Here $G = \mbox{\scriptsize$\pmatrix{1 & 0 \cr 0 & 1}$}$. The
   corresponding real form is called $SU(2)$. The matrix elements obey
   $\alpha^* = \delta$, $\beta^* = -\gamma$, $\gamma^* = -\beta$ and
   $\delta^* = \alpha$.

\item
   Here $G = \mbox{\scriptsize$\pmatrix{1 & 0 \cr 0 & -1}$}$. The
   corresponding real form is called $SU(1,1)$. The matrix elements obey
   $\alpha^* = \delta$, $\beta^* = \gamma$, $\gamma^* = \beta$ and
   $\delta^* = \alpha$.

\item
   Here $G = \mbox{\scriptsize$\pmatrix{0 & 1 \cr 1 & 0}$}$. The
   corresponding real form is called $SL(2,\RR)$. The matrix elements obey
   $\alpha^* = \alpha$, $\beta^* = \beta$, $\gamma^* = \gamma$ and
   $\delta^* = \delta$.

\end{enumerate}

\noindent In the classical case, $SU(2)$ and $SL(2,\RR)$ have the same
{\sl finite\/}-dimensional representations (they do not have the same
infinite-dimensional ones). For this reason there is a one to one
correspondence between the Hopf algebras of these two groups (indeed, the
Hopf algebra of a group can be considered as the space of polynomials on
matrix elements of finite-dimensional representations). The star operation
allows one to distinguish between the two groups. Moreover, in the classical
case, the two groups $SU(1,1)$ and $SL(2,\RR)$ are isomorphic. One could
also consider the complex group $SL(2,\CC)$ itself as a real group (the spin
covering of the Lorentz group) but the dimensionality is twice bigger and
we are not going to follow this route in the quantum case ---at least not
in the present paper.

In the quantum case, one has also three possibilities for the star
operations on $Fun(SL_q(2,\CC))$ (up to Hopf automorphisms):

\begin{enumerate}
\item
   The real form $Fun(SU_q(2))$. The matrix elements obey
   $a^* = d$, $b^* = -qc$, $c^* = -q^{-1} b$ and $d^* = a$. Moreover,
   $q$ should be real.

\item
   The real form $Fun(SU_q(1,1))$. The matrix elements obey
   $a^* = d$, $b^* = qc$, $c^* = q^{-1} b$ and $d^* = a$. Moreover,
   $q$ should be real.

\item
   The real form $Fun(SL_q(2,\RR))$. The matrix elements obey
   $a^* = a$, $b^* = b$, $c^* = b$ and $d^* = d$.
   Here $q$ can be complex but it should be a phase.
\end{enumerate}

\noindent In the quantum case, it is already clear from these results
that taking $q$ a root of unity is incompatible with the $SU_q$ real forms,
and that the only possibility if we assume $q^N = 1$ is to choose the star
corresponding to $Fun(SL_q(2,\RR))$. Notice also that, in the quantum case,
the two real forms $Fun(SU_q(1,1))$ and $Fun(SL_q(2,\RR))$ are different
(already the range of values where $q$ can be chosen do not coincide).

%%%%%%%%%%%%%%%%%%%%%%%%%%%%%%%%%%%%%%%%%%%%%%%%%%%%%%%%%%%%%%%%%%%%%%%%%%

\subsection{Real forms on $\mathcal F$}

Having only one real form compatible with a complex $q$ on
$Fun(SL_q(2,\CC))$ already tells us that there is at most one real
form on its quotient $\mathcal F$. We only have to check that the
star operation preserves the ideal and coideal defined by
$a^3 = d^3 = \one$, $b^3 = c^3 = 0$. This is trivial because
$a^* = a$, $b^* = b$, $c^* = c$ and $d^* = d$. Hence, this star
operation can be restricted to $\mathcal F$.

This real form can be considered as a reduced quantum group associated
with the real form $Fun(SL_q(2,\RR))$ of $Fun(SL_q(2,\CC))$.

%%%%%%%%%%%%%%%%%%%%%%%%%%%%%%%%%%%%%%%%%%%%%%%%%%%%%%%%%%%%%%%%%%%%%%%%%%

\subsection{Real structures and star operations on $\mathcal{M, H}$}

Now we would like to introduce an involution (a star operation) on the
comodule algebra $\mathcal M$. This involution should be compatible with the
coaction of $\mathcal F$. That is, we are asking for covariance of the star
under the (right,left) $\mathcal F$-coaction,
\begin{equation}
   (\Delta_{R,L} \, z)^* = \Delta_{R,L} (z^*) \ , \quad
                        \text{for any} z \in {\mathcal M} \ .
\label{*-delta-condition}
\end{equation}

\noindent Here we have also used the notation $*$ for the star on the
tensor products, which is defined as $(A \otimes B)^* = A^* \otimes B^*$.
Using, for instance, the left coaction
{\scriptsize
$\pmatrix{x' \cr y'} = \pmatrix{a & b \cr c & d} \otimes \pmatrix{x \cr y}$
}
in (\ref{*-delta-condition}), we see immediately that $Fun(SU_q(2))$
corresponds to the choice $x^* = y, y^* = -q^{-1} x$. Writing $xy = qyx$,
one also recovers the fact that, in this case, $q$ should be real. In the
same way, we find that the real form $Fun(SU_q(1,1))$ corresponds to the
choice $x^* = y, y^* = q^{-1} x$. Finally, the real form $Fun(SL_q(2,\RR))$
corresponds to $x^* = x$ and $y^* = y$.

As $q^3 = 1$, we stick to the $Fun(SL_q(2,\RR))$ case since it is the only
one compatible with a complex $q$. We therefore define the star on
${\mathcal M} = M_3(\CC)$ by
\begin{equation}
   x^* = x \ , \qquad y^* = y \ .
\label{*-on-M}
\end{equation}

We now want to find a compatible $*$ on the algebra $\mathcal H$. Of course,
one could proceed by using duality with $\mathcal F$, but it is both handy
and instructive to proceed otherwise.

We let $\mathcal M$ {\sl act\/} on itself by multiplication and we want
to promote this star operation to an involution defined for all operators
acting on $\mathcal M$. In particular, on the operators determined by the
action of $\mathcal H$, given in Section~\ref{subsec:actions-of-H}. To this
end, we apply the $*$ operation to the commutation relations between $x,y$
and the differential operators $X^L$ (or $X^R$), and read off the results.
For instance, $X_+^L \, y = x + q^2 y \, X_+^L$ implies
$y \, (X_+^L)^* = x + \overline{q^2} (X_+^L)^* \, y$ and this suggests
$(X_+^L)^* = - q^2 (X_+^L)$. In this way we obtain:

\begin{eqnarray*}
   (X_+^L)^* &=& -q^2 X_+^L \\
   (X_-^L)^* &=& -q X_-^L      \\
   (K^L)^*   &=& K^L
\end{eqnarray*}
and
\begin{eqnarray*}
   (X_+^R)^* &=& -q X_+^R      \\
   (X_-^R)^* &=& -q^2 X_-^R \\
   (K^R)^*   &=& K^R
\end{eqnarray*}

\noindent Notice the change of $q$ into $q^{-1}$ when going from left to
right. As the space of operators coming from a left action is isomorphic
with $\mathcal H$ itself, the star operation in $\mathcal H$ is the same
as the one obtained for the operators $X_{\pm}^L, K^L$, namely
\begin{equation}
   X_+^* = -q^2 X_+ \ , \quad X_-^* = -q X_- \ , \quad K^* = K \ .
\label{*-on-H}
\end{equation}

{\small
Now, of course, one has to check that this involution satisfies all the
axioms given at the beginning of this section. This is rather
straightforward. For illustration, let us compute $\Delta (X_-^*)$
and $S*S*X_-$:
\begin{eqnarray*}
   \Delta (X_-^*) &=& -q \Delta X_- =
                      -q (X_-\otimes K^{-1} + \one \otimes X_-)       \\
               {} &=& X_-^* \otimes (K^{-1})^* + \one^* \otimes X_-^* \\
			{} &=& (\Delta X_-)^*
\end{eqnarray*}
and
\begin{eqnarray*}
   S * S * X_-    &=& S*S(-q X_{-}) = S*(-qSX_-) = S(*(q X_- K)) =
                      S(q^2 K^* X_-^*)                                   \\
               {} &=& q^2 S(K(-q X_-) = - S(X_-) S(K) = -(-X_- K K^{-1}) \\
                  &=& X_-
\end{eqnarray*}
}

Adding the cubic relations $x^3 = \one, y^3 = \one$ in $\mathcal M$, and
the corresponding ones in the algebra $\mathcal H$, does not change anything
to the determination of the star structures. This is because the (co)ideals
are preserved by the involutions, and thus the quotients can still be done.

Anyway, as $\mathcal H$ is dual to $\mathcal F$ (or $U_q(sl(2))$ dual to
$Fun(SL_q(2,\CC))$), we should have dual $*$-structures. This means the
relation
\begin{equation}
   <h^*, u> = \overline{<h, (Su)^*>} \ , \quad
                 h \in{\mathcal H}, u \in{\mathcal F}
\label{dual-*-structures}
\end{equation}
holds. Moreover, the covariance condition for the star, equation
(\ref{*-delta-condition}), may also be written dually as a condition for
$*$ under the action of $\mathcal H$. This can be done pairing the
$\mathcal F$ component of equation (\ref{*-delta-condition}) with some
$h \in {\mathcal H}$. One should next recall the definition of the left
action (\ref{L-H-action-on-M}) (the right action if we are using the left
coaction in (\ref{*-delta-condition})) and the duality
(\ref{dual-*-structures}) that relates both $*$-Hopf stars. One gets
finally the constraint on $*_{\mathcal M}$ to be $\mathcal H$ covariant,
\begin{equation}
   h(z^*) = \left[ (S h )^* z \right]^* \ , \quad
                 h \in{\mathcal H}, z \in{\mathcal M} \ .
\label{*-action-condition}
\end{equation}

Note that the $*$-operation that we obtain on the algebra
${\mathcal M} = M_3(\CC)$ does not coincide with the usual hermitian
conjugacy ($\dag$-operation) on matrices, at least when $x$ and $y$
are represented with the $3 \times 3$ matrices given in
Section~\ref{sec:red-q-plane} (they were such that $x^\dag = x^{-1} = x^2$
and $y^\dag = y^{-1} = y^2$). Moreover, it can be seen that it is not
possible to find a star operation on the quantum group $\mathcal F$ that
would respect the $\dag$ involution on $\mathcal M$.

The situation here is reminiscent of the case where one introduces a
``bar''-operation (Dirac conjugacy) on Dirac matrices (Clifford algebra
for usual space-time) that is distinct from usual hermitian conjugacy. One
remembers that, using this involution, one can build a scalar product on
the space of spinors which is spin-invariant (the scalar product built using
the usual hermitian conjugacy being positive but not invariant). The
situation here is somewhat similar.

{\bf Remark:}
Note finally that one could be tempted to chose the involution defined by
$K^\star = K^{-1}$, $X_+^\star = \pm X_-$ and $X_-^\star = \pm X_+$.
However, as it is easy to check (look for instance at the compatibility
with the coproduct), this is a {\em twisted} star operation. This last star
operation is the one one would need to interpret the unitary group of
$\mathcal H$ as $U(3) \times U(2) \times U(1)$, which could be related to
the gauge group of the Standard Model \cite{Connes-2}. Instead of using the
standard Hopf-structure on $U_q(sl(2,\CC))$ with this twisted star, a
better sound mathematical alternative would be to look for a non-standard
Hopf-structure for this quantum group (\ie a different coproduct, but
preserving the multiplication), such that $\star$ becomes a normal star
operation.

%%%%%%%%%%%%%%%%%%%%%%%%%%%%%%%%%%%%%%%%%%%%%%%%%%%%%%%%%%%%%%%%%%%%%%%%%%

\subsubsection{Matrix representation of the $*$-operation on $\mathcal M$}

The star operation for $\mathcal M$ introduced above ($*x = x$, $*y = y$)
can be written explicitly in terms of $3 \times 3$ matrices. Let us define
the {\sl charge conjugation matrix\/} $C$:
\begin{equation}
   C \doteq \pmatrix{1 & 0 & 0 \cr
                     0 & 0 & 1 \cr
                     0 & 1 & 0}
\label{charge-conjugation-C}
\end{equation}
For every element $m$ of the algebra $\mathcal M$ we set
\begin{equation}
   * \, m \doteq (C \, m \, C^{-1})^\dag \ ,
\label{star-with-C}
\end{equation}
where $\dag$ denotes the usual hermitian conjugation (the transpose of the
complex conjugate matrix). This is clearly a star operation: it is
conjugate-linear, involutive ($*(*m) = m$), anti-multiplicative
($*(mn) = (*n)(*m)$), and unital ($*\one = \one$). Remark also that
$C = C^{-1} = C^\dag$.

Let us represent the generators $x$ and $y$, as before, by
$x = \mbox{\scriptsize
     $\pmatrix{1 & 0 & 0 \cr 0 & q^{-1} & 0 \cr 0 & 0 & q^{-2}}$}$ and
$y = \mbox{\scriptsize
     $\pmatrix{0 & 1 & 0 \cr 0 & 0 & 1 \cr 1 & 0 & 0}$}$. With
(\ref{star-with-C}) it is then easy to check that $*x = x$, $*y = y$.
Therefore the star operation defined explicitly as before in terms of the
charge conjugation matrix $C$ implements the abstract star operation
introduced above. We already know that this star structure on $\mathcal M$
gives rise to star structures on the quantum group $\mathcal F$ and its dual
$\mathcal H$ that are compatible with their Hopf algebra structure.

The most general self-adjoint element should satisfy $*m = m$, hence
the vector space generators
$$
   \one \,,\ x \,,\ y \,,\ x^2 \,,\ q\,xy \,,\ y^2 \,,\
      q^2 x^2 y \,,\ q^2 xy^2 \,,\ q\,x^2 y^2
$$
are ``real'' in the sense that they are self-adjoint for $*$. The same is
obviously true for any linear combination of these generators with real
coefficients. Using this, it is easy to write the most general $3\times 3$
matrix that is self-adjoint for the star $*$. It can be written as:
$$
   \alpha = \alpha_{irr} + \alpha_{eve} + \alpha_{odd} \ ,
$$
with

\begin{eqnarray*}
\alpha_{irr} &=& \pmatrix{a_{20}     & q \, a_{11} & a_{02} \cr
                          a_{02}     & q \, a_{20} & a_{11} \cr
                          q^2 a_{11} & a_{02}      & q^2 a_{20}}
              =  a_{20} \, x^2 + a_{11} \, qxy + a_{02} \, y^2   \ , \\
\alpha_{eve} &=& \pmatrix{a_{10}     & a_{01}      & q \, a_{22} \cr
                          q^2 a_{22} & q^2 a_{10}  & a_{01}      \cr
                          a_{01}     & a_{22}      & q \, a_{10}}
              =  a_{10} \, x + a_{01} \, y + a_{22} \, q x^2 y^2 \ , \\
\alpha_{odd} &=& \pmatrix{a_{00}      & q^2 \, a_{21} & q^2 \, a_{12} \cr
                          q \, a_{12} & a_{00}        & a_{21}        \cr
                          q \, a_{21} & a_{12}        & a_{00}}
              =  a_{00}\,\one + a_{21} \, q^2 x^2 y + a_{12} \, q^2 xy^2 \ ,
\end{eqnarray*}
where all the $a_{rs}$ are arbitrary {\sl real\/} numbers (warning: indices
$r,s$ of $a_{rs}$ refer to the coefficients of $x^r y^s$ and not to the line
or column indices of the corresponding matrix elements).

Notice that starting from the defining relations for the algebra
$\mathcal M$ in terms of $x$ and $y$, and imposing $x^* = x, y^* = y$, it is
not possible to find a matrix representation of this algebra for which these
matrices would be ($\dag$-)hermitian. Indeed such matrices would be in such
a case diagonalisable with real eigenvalues, but the condition
$x^3 = y^3 = \one$ would then imply $x = y = \one$.

%%%%%%%%%%%%%%%%%%%%%%%%%%%%%%%%%%%%%%%%%%%%%%%%%%%%%%%%%%%%%%%%%%%%%%%%%%

\subsubsection{Inverses and unitary elements in $\mathcal M$}

To study the inverses of elements of $\mathcal M$, it is useful to define
\begin{eqnarray*}
   C_0(t) &=& {1 \over 3} (\exp(t) + \exp(qt) + \exp(q^2 t)) \\
   C_1(t) &=& {d \over dt} C_0(t)                            \\
   C_2(t) &=& {d \over dt} C_1(t)
\end{eqnarray*}
These functions appear as generalizations of sine and cosine functions when
one replaces $\ZZ_2$ by $\ZZ_3$. They seem to have been invented several
times in the literature (an old reference is \cite{Humbert}). Call
\begin{eqnarray*}
   D(a,b,c) &=& a^3 + b^3 + c^3 - 3abc \\
   T(a,b,c) &=& a^2 - bc
\end{eqnarray*}
In particular, it is easy to see that $D(C_0,C_1,C_2) = 1$
(a generalization of $\sin^2 + \cos^2 = 1$).

Writing ${\mathcal M} = 3_{irr} \oplus 3_{eve} \oplus 3_{odd}$, it
can be shown that (with obvious notations)
$$
   Inverse(3_{eve}) \subset 3_{irr} \ , \qquad
   Inverse(3_{irr}) \subset 3_{eve} \ , \qquad \text{but} \
   Inverse (3_{odd}) \subset 3_{odd} \ .
$$

\noindent More precisely, given an arbitrary element $a_{odd} \in 3_{odd}$,
we have
\begin{eqnarray*}
   a_{odd}      &=& a_{00} \, \one + a_{12} \, q^2 xy^2 +
                    a_{21} \, q^2 x^2 y \\
   a_{odd}^{-1} &=& {1 \over D(a_{00},a_{12},a_{21})} \left[
                       T(a_{00},a_{12},a_{21}) \, \one+
                       T(a_{21},a_{00},a_{12}) \, q^2 xy^2 +
                       T(a_{12},a_{21},a_{00}) \, q^2 x^2y \right] \ .
\end{eqnarray*}

\noindent Taking $a_{eve} \in 3_{eve}$,
\begin{eqnarray*}
   a_{eve}      &=& a_{10} \, x + a_{01} \, y + a_{22} \, q x^2 y^2 \\
   a_{eve}^{-1} &=& {1 \over D(a_{10},a_{01},a_{22})} \left[
                       T(a_{10},a_{01},a_{22}) \, x^2 +
                       T(a_{22},a_{10},a_{01}) \, qxy +
                       T(a_{01},a_{22},a_{10}) \, y^2 \right] \ .
\end{eqnarray*}

\noindent Finally, for a general $a_{irr} \in 3_{irr}$, we see that
\begin{eqnarray*}
   a_{irr}      &=& a_{20} \, x^2 + a_{11} \, qxy + a_{02} \, y^2 \\
   a_{irr}^{-1} &=& {1 \over D(a_{20},a_{11},a_{02})} \left[
                       T(a_{20},a_{11},a_{02}) \, x +
                       T(a_{02},a_{20},a_{11}) \, y +
                       T(a_{11},a_{02},a_{20}) \, q x^2 y^2 \right] \ .
\end{eqnarray*}

\noindent The result $Inverse(3_{odd}) \subset 3_{odd}$ should not be
surprising, since $3_{odd}$ is an abelian subalgebra of $\mathcal M$
(the two other subspaces are not subalgebras); this algebra is actually
isomorphic with the algebra $\CC[\ZZ_3]$ of the abelian group $\ZZ_3$.
Moreover, using this last isomorphism we can write
$$
   {\mathcal M} = \CC[\ZZ_3] \oplus x \, \CC[\ZZ_3] \oplus x^2 \, \CC[\ZZ_3]
                = 3_{odd} \oplus 3_{eve} \oplus 3_{irr} \ .
$$
Remark that when the coefficients $a_{ij}$ are real, the generic elements
$a_{odd}$, $a_{eve}$ and $a_{irr}$ written above are also real.

Since $3_{odd}$ is a subalgebra, it is interesting to consider the
unitary elements within $3_{odd}$ (with respect to our star operation).
A unitary element ($u^* = u^{-1}$) can be written $u = \exp(ia)$, with
$a^* = a$ ($a$ is ``real''). Therefore, working on $3_{odd}$, $u$ can
be expanded as
$u = \exp[i(a_{00}\,\one + a_{12}\,q^2 x y^2 + a_{21}\,q^2 x^2 y)]$,
where $a_{ij} \in \RR$. Since $x^2 y$ commutes with $xy^2$, the previous
exponential can be written as a product of three exponentials, each with
its own real parameter: $u = u_0 u_1 u_2$, with
\begin{eqnarray*}
   u_0 &=& \exp(i a_{00} \, \one) = \exp(i a_{00}) \, \one \\
   u_1 &=& \exp(i a_{12} \, q^2 x y^2 ) =
           C_0(i a_{12}) \, \one + C_1(i a_{12}) \, q^2 x^2 y +
           C_2(i a_{12}) \, q^2 x y^2 \\
   u_2 &=& \exp(i a_{21} \, q^2 x^2 y ) =
           C_0(i a_{21}) \, \one + C_1(i a_{21}) \, q^2 x^2 y +
           C_2(i a_{21}) \, q^2 x y^2
\end{eqnarray*}

%%%%%%%%%%%%%%%%%%%%%%%%%%%%%%%%%%%%%%%%%%%%%%%%%%%%%%%%%%%%%%%%%%%%%%%%%%

\subsection {A quantum group invariant scalar product on the space of
             $3 \times 3$ matrices}
\label{subsec:scalar-product-on-M}

The usual scalar product on the space ${\mathcal M} = M_3(\CC)$ of
$3 \times 3$ matrices is
$m_1, m_2 \rightarrow (m_1, m_2) = \tr(m_1^\dag m_2)$. For every linear
operator $\ell$ acting on $\mathcal M$, we can define the usual adjoint
$\ell^\dag$; however, this adjoint does not coincide with the star operation
introduced previously. Our aim in this section is to find another scalar
product better suited for our purpose.

We call $z, z'$ two generic elements of $\mathcal M$ (linear
combinations of $x^r y^s$), and $h$ a generic element of $\mathcal H$ (a
linear combination of $X_+^\alpha K^\beta X_-^\gamma$). We know that the
first ones act on $\mathcal M$ like multiplication operators and that the
others act on $\mathcal M$ by twisted derivations or automorphisms. We also
know the action of our star operation $*$ on these linear operators. We
shall now obtain a unique scalar product by imposing that $*$ coincides with
the adjoint associated with this scalar product. That is, we are asking for
an inner product on $\mathcal M$ such that the (left) actions of
$\mathcal M$ and $\mathcal H$ (each with its respective star) on that vector
space may be thought as $*$-representations. Hence, for every linear
operator $\ell$ acting on the nine-dimensional {\sl vector space\/}
$\mathcal M$, we impose:
\begin{equation}
   (z, \ell . z') = (\ell^* . z, z') \ .
\label{condition-on-scalar-product}
\end{equation}

Moreover, $(\cdot,\cdot)$ should be hermitian, so that
$$
   (z', z) = \overline{(z, z')} \ .
$$
As usual, $(\cdot,\cdot)$ is taken as antilinear with respect to the first
argument and linear with respect to the second:
$(\alpha z, \beta z') = \overline{\alpha}\beta (z, z')$, with
$\alpha, \beta \in \CC$.

Due to the fact that $(z, z') = (\one, z^* z')$, it is enough to compute
$(\one, z)$ for all the $z$ belonging to $\mathcal M$. We now use the
action of $\mathcal H$ in (\ref{condition-on-scalar-product}) and obtain
$$
   (h^*.\one, z) = (\one, h.z) \ .
$$
Note that we are now calling $h.z$ the left action that was previously
denoted with $h^L[z]$. Taking $h = K$ and $z = x^r y^s$ we see that (use
the results of Section~\ref{subsec:actions-of-H},
Table~\ref{table:H-acting-on-M}):
$$
   (\one, K. x^r y^s) = q^{(r-s)} (\one, x^r y^s)
$$
but
$$
   (\one, K. x^r y^s) = (K^*.\one, x^r y^s) =
                        (K.\one, x^r y^s) = (\one, x^r y^s) \ ,
$$
and therefore we conclude that $(\one, x^r y^s) = 0$ unless $r=s$.
We have then to find $(\one,\one)$, $(\one,xy)$ and $(\one,x^2 y^2)$.

Taking $h = X_+$ we find
\begin{eqnarray*}
   (\one, X_+.z) &=& (X_+^*.\one, z) = (-q^2 X_+.\one, z) \\
              {} &=& -q \, (0, z) = 0 \ .
\end{eqnarray*}
With $z = x^2 y$ we have $X_+.x^2 y = q^2 \one$, so $(\one, \one) = 0$.
Selecting $z = y^2$, we have $X_+.y^2 = -q^2 xy$ and thus $(\one, xy) = 0$.

Hence, the only non vanishing scalar product of the type $(\one, z)$ is
$(\one, x^2 y^2)$. From this quantity we deduce eight other non-zero scalar
products $(z,z')$, where $z$ and $z'$ are basis elements $x^r y^s$. For
instance, $(x, xy^2) = (\one, x^*.xy^2) = (\one, x^2 y^2)$.

Hermiticity with respect to $*$ implies that $(xy,xy)$ should be real,
indeed $(xy,xy) = \overline{(xy,xy)}$. We set
$$
   (xy,xy) = 1 \ .
$$
We should however remember that this normalization condition is arbitrary
and could be set to any other positive real number (this comment may become
physically relevant when one introduces a ``coupling constant'' playing a
role in some physical model).

One can summarise the results in the formula:
\begin{eqnarray}
   (x^p y^t, x^r y^s) = q^2 \delta_{p+r,2} \, \delta_{t+s,2} \,
      [q^2 \delta_{s,0} + q\delta_{s,1} + \delta_{s,2}] \ ,
\label{M-scalar-product}
\end{eqnarray}
where $\delta_{i,j}$ is the Kroneker delta. In the basis
$$
   \{ \{ x^2y,xy^2,x^2y^2,y^2 \}, \{ y,x,\one,x^2 \}, \{ xy \} \}
$$
the $9 \times 9$ matrix of scalar products $(z,z')$ reads
$$
   \pmatrix{0        & U & 0 \cr
            U^{\dag} & 0 & 0 \cr
            0        & 0 & 1} \ ,
$$
where $U$ is the $4 \times 4$ diagonal matrix $\diag(1,q,q,q)$. Notice that
the trace and determinant of this matrix are equal to $1$. The total space
splits (Witt's decomposition) into the sum of two complementary isotropic
spaces of dimension $4$ and a supplementary space of dimension $1$.

The real part of this non degenerate hermitian scalar product is a bilinear
form of signature $(5,4)$. More precisely, if one uses the basis:
$$
\begin{array}{ll}
   u_1 = -y + x^2 y \qquad & u_2 = y + x^2 y   \\
   v_1 = -1 + x^2 y^2      & v_2 = 1 + x^2 y^2 \\
   w_1 = -x + xy^2         & w_2 = x + xy^2    \\
   t_1 = -x^2 + y^2        & t_2 = x^2 + y^2   \\
   s   = xy                & {}
\end{array}
$$
our hermitian form can be written
{\small
$$
   \pmatrix{-2& 0& 0& 0& 0& 0& 0& 0& 0 \cr
             0& 2& 0& 0& 0& 0& 0& 0& 0 \cr
             0& 0& 1& q-q^2& 0& 0& 0& 0& 0 \cr
             0& 0& -q+q^2& -1& 0& 0& 0& 0& 0 \cr
             0& 0& 0& 0& 1& q-q^2& 0& 0& 0 \cr
             0& 0& 0& 0& -q+q^2& -1& 0& 0& 0 \cr
             0& 0& 0& 0& 0& 0& 1& q-q^2& 0 \cr
             0& 0& 0& 0& 0& 0& -q+q^2& -1& 0 \cr
             0& 0& 0& 0& 0& 0& 0& 0& 1}
$$
}
Its real part is diagonal in this base (since $\overline q = q^2$), and
reads
$$
   \diag(-2, 2, 1, -1, 1, -1, 1,- 1, 1) \ .
$$
The signature $(5,4)$ can be read immediately from the above.

As we saw in Section~\ref{subsec:M-in-reps-H}, the algebra $\mathcal M$
can be written as the sum of three vector spaces of dimensionality $3$
corresponding to the decomposition of the underlying vector space as the sum
of three indecomposable representations of the algebra ${\mathcal H}$:
${\mathcal M} = 3_{irr} \oplus 3_{eve} \oplus 3_{odd}$, where
\begin{eqnarray*}
   3_{irr} &=& \algebraicspan(x^2,xy,y^2)      \\
   3_{odd} &=& \algebraicspan(\one,x^2 y,xy^2) \\
   3_{eve} &=& \algebraicspan(x,y,x^2 y^2) \ .
\end{eqnarray*}
It is therefore interesting to write the scalar product in a basis adapted
to this decomposition.

In the basis $\{\{ x^2,xy,y^2 \},\{ x,y,x^2 y^2 \},\{ \one,x^2 y,xy^2 \}\}$,
the scalar product can be written as
$$
   G = \pmatrix{ B & 0 & 0 \cr
                 0 & 0 & B \cr
                 0 & B & 0} \ ,
$$
where $B$ is the $3 \times 3$ block
$$
   B = \pmatrix{ 0 & 0 & q^2 \cr
                 0 & 1 & 0 \cr
                 q & 0 & 0} \ .
$$

Using the charge conjugation matrix (\ref{charge-conjugation-C}) to write
the star as in (\ref{star-with-C}), the scalar product on $\mathcal M$ can
be written as
\begin{equation}
   (m_1,m_2) = {1 \over 3}\,\tr(B^t \, C \, m_1^* m_2) =
               {1 \over 3}\,\tr(B^t \, m_1^\dag \, C \, m_2) \ ,
\label{scalar-product-with-matrices}
\end{equation}
where the trace is a usual trace on the matrices.

%%%%%%%%%%%%%%%%%%%%%%%%%%%%%%%%%%%%%%%%%%%%%%%%%%%%%%%%%%%%%%%%%%%%%%%%%%

\subsubsection{Quantum group invariance of the scalar product}

We should now justify why the above scalar product was called a quantum
group invariant one. Remember we only said the scalar product was such that
the stars coincide with the adjoint-operators, or such that the actions are
given by $*$-representations.

We refer the reader to \cite{CoGaTr}, where it is shown that the
$*$-representation condition on the scalar product,
\begin{equation}
   (h.z,w) = (z,h^* .w) \ , \quad h \in {\mathcal H} \ ,
\label{star_rep}
\end{equation}
automatically fulfils one of the two alternative invariance conditions
that can be imposed on the scalar product, namely
\begin{equation}
   ((Sh_1)^* .z,h_2.w) = \epsilon(h) (z,w) \ , \quad
                         \text{with} \Delta h = h_1 \otimes h_2 \ .
\label{cond_for_scalar_prod_inv}
\end{equation}

{\small
As a side note, we can check the classical limit: let $H$ be the universal
enveloping algebra of some Lie algebra, and $x$ any antihermitian generator
of the Lie algebra (such that $\exp(x)$ is unitary). As the natural Hopf
structure on this particular cocommutative $H$ is given by $\epsilon(x)=0$,
$S(x)=-x$, and $\Delta x = x\otimes 1 + 1\otimes x$,
(\ref{cond_for_scalar_prod_inv}) reduces in this case to
\begin{eqnarray*}
   ((*Sx).z, 1.w) + ((*S1).z, x.w) &=& (x.z, w) + (z, x.w) \\
                                   &=& 0 \ .
\end{eqnarray*}
Note that this last equation is just what is needed to get the scalar
product invariance in the usual sense:
\begin{eqnarray*}
   (z^\prime ,w^\prime) &\doteq & (\exp(x).z,\exp(x).w) \\
                        &=& ((1+x).z,(1+x).w) + {\cal O}(x^2) =
                            (z,w) + (x.z,w) + (z,x.w) + {\cal O}(x^2) \\
                        &=& (z,w)
\end{eqnarray*}
}

The relations dual to (\ref{cond_for_scalar_prod_inv}) and (\ref{star_rep})
are those that apply to the coaction of $\mathcal F$ instead of the action
of $\mathcal H$. These are
\begin{equation}
   (\Delta_R \, z, \Delta_R \, w) = (z,w) \one_{\mathcal F} \ ,
\label{dual_cond_for_scalar_prod_inv}
\end{equation}
where $(\Delta_R \, z, \Delta_R \, w)$ should be understood as
$(z_i, w_j) T_i^* T_j$ if $\Delta_R \, z = z_i \otimes T_i$, \etc, and
\begin{equation}
   (z,\Delta_R w) = ((1\otimes S)\Delta_R z,w) \ ,
\label{dual_star_rep}
\end{equation}
respectively. We have used here the right-coaction, but the formulas for
the left coaction can be trivially deduced from the above ones.

It is worth noting that these equations for the coaction of $\mathcal F$
{\em imply} the previous ones for the action of $\mathcal H$, and are
completely equivalent assuming non-degeneracy of the pairing
$<\cdot,\cdot>$. Moreover, (\ref{dual_cond_for_scalar_prod_inv}) is a
requirement analogous to the condition of classical invariance by a group
element action.

If we select an orthonormal basis $\{z_i\}$ of $\mathcal M$ such that
$\Delta_R z_i = z_j \otimes T_i^j$ with
$\Delta T_i^j = T_k^j \otimes T_i^k$, (\ref {dual_cond_for_scalar_prod_inv})
and (\ref{dual_star_rep}) reduce to
$$
   (T_i^k)^* T_j^k = \delta _{ij} \one_{\mathcal F}
$$
and
$$
   ST_i^j = (T_j^i)^* \ ,
$$
respectively. This last equation is exactly what we use in the classical
(Lie algebra) case: the unitarity condition for matrices (the antipode is
the correct quantum generalization of the group inverse), that warrants the
unitarity of the representation (in the sense of invariance of the scalar
product).

Applying the above discussion to the invariant subspaces of the Hopf-algebra
$\mathcal H$ we obtain on each one the most general invariant scalar
products that we list in \ref{app:metrics-on-H-representations}

%%%%%%%%%%%%%%%%%%%%%%%%%%%%%%%%%%%%%%%%%%%%%%%%%%%%%%%%%%%%%%%%%%%%%%%%%%
%%%%%%%%%%%%%%%%%%%%%%%%%%%%%%%%%%%%%%%%%%%%%%%%%%%%%%%%%%%%%%%%%%%%%%%%%%

\section{The Manin dual ${\mathcal M}^!$ of $\mathcal M$}
\label{sec:manin-dual}

In order to find the Manin dual \cite{Manin} of a given quadratic algebra
$\mathcal A$ given by the relations $E_{ij} x^i x^j = 0$ amongst its
generators $\{ x^i \}$, one has first to determine all the matrices
$\mathcal E$ such that
${\mathcal E}_{ij} E_{ij} = \tr({\mathcal E}^t E) = 0$.
The Manin dual ${\mathcal A}^!$ of $\mathcal A$ is defined as a
quadratic algebra with relations ${\mathcal E}_{ij} \xi^i \xi^j = 0$.
This is equivalent to say that the quadratic relations in ${\mathcal A}^!$
are orthogonal to the relations in $\mathcal A$, with the pairing (or
``scalar product'') $<\xi^i , x^j> = \delta^{ij}$.

The quadratic relations defining $\mathcal M$ ($xy = qyx$) can be written
$E_{ij} x^i x^j = 0$, with $x^1 \doteq x, x^2 \doteq y$ and
$E_{ij} = \mbox{\small $\pmatrix{0 & 1 \cr -q & 0}$}$. Hence we have,
in the present case,
${\mathcal E}^t E = \mbox{\small $
                    \pmatrix{-q{\mathcal E}_{21} & {\mathcal E}_{11} \cr
                             -q{\mathcal E}_{22} & {\mathcal E}_{12}}$}$.
Thus
$$
   \tr({\mathcal E}^t E) = -q{\mathcal E}_{21} + {\mathcal E}_{12}
      \equiv 0 \ \ \leadsto \ \ {\mathcal E}_{12} = q{\mathcal E}_{21} \ .
$$

\noindent The matrices solution of our problem span therefore a vector
space of dimension $3$:
$$
{\mathcal E} = \pmatrix{{\mathcal E}_{11} & q{\mathcal E}_{21} \cr
                        {\mathcal E}_{21} & {\mathcal E}_{22}} =
   {\mathcal E}_{11} {\mathcal E}^{(1)} +
   {\mathcal E}_{21} {\mathcal E}^{(2)} +
   {\mathcal E}_{22} {\mathcal E}^{(3)}
$$
with
$$
   {\mathcal E}^{(1)}=\pmatrix{1 & 0 \cr 0 & 0}, \quad
   {\mathcal E}^{(2)}=\pmatrix{0 & q \cr 1 & 0}, \quad
   {\mathcal E}^{(3)}=\pmatrix{0 & 0 \cr 0 & 1} .
$$

\noindent Our algebra $\mathcal M$ is not quadratic since we impose the
relations $x^3 = y^3 = \one$. Nevertheless, forgetting momentarily those
constraints, we define its Manin dual ${\mathcal M}^!$ as the algebra
generated {\sl over the complex numbers\/} by $\xi^1, \xi^2$, satisfying
the relations
$$
   {\mathcal E}^{(\sigma)}_{ij} \xi^i \xi^j = 0 \, , \quad
      \sigma \in \{1,2,3\}
$$
\ie
$$
   (\xi^1)^2 = 0                    \: , \quad
   q\,\xi^1 \xi^2 + \xi^2 \xi^1 = 0 \: , \quad
   (\xi^2)^2 = 0                    \ .
$$
This means, in particular, that any cubic polynomial in the $\xi$'s is
identically zero. There is no need to introduce additional constraints for
$\xi^1$, $\xi^2$ orthogonal to the cubic relations in $\mathcal M$.

We shall write $dx \doteq \xi^1$ and $dy \doteq \xi^2$, so ${\mathcal M}^!$
is defined by these two generators and the relations
\begin{equation}
   dx^2 = 0                     \: , \quad
   dy^2 = 0                     \: , \quad
   q \, dx \, dy + dy \, dx = 0 \ .
\label{M!-relations}
\end{equation}

%%%%%%%%%%%%%%%%%%%%%%%%%%%%%%%%%%%%%%%%%%%%%%%%%%%%%%%%%%%%%%%%%%%%%%%%%%

\subsection{$\mathcal F$ coacting on ${\mathcal M}^!$}

Once the coaction of $\mathcal F$ on $\mathcal M$ has been defined as
in Section~\ref{sec:q-group-F}, and once the quantum group product relations
have been obtained, it is easy to check that the coaction of $\mathcal F$
on ${\mathcal M}^!$ is given by the same formulae as for $\mathcal M$
itself. Namely, writing
$$
   \pmatrix{dx' \cr dy'} =
      \pmatrix{a & b \cr c & d} \otimes \pmatrix{dx \cr dy}
$$
and
$$
   \pmatrix{\tilde dx & \tilde dy} =
      \pmatrix{dx & dy} \otimes \pmatrix{a & b \cr c & d}
$$
ensures that $q \, dx' \, dy' + dy' \, dx' = 0$ and
$q \, \tilde dx \, \tilde dy + \tilde dy \, \tilde dx = 0$,
given that the relation $q \, dx \, dy + dy \, dx = 0$ is satisfied.
Actually, one can also recover $Fun(SL_q(2,\CC))$ imposing just the
invariance of relations (\ref{M-relations}) and (\ref{M!-relations})
under the quantum group left-coactions.

The coaction can be read from those formulae, for instance
$\Delta_R (dx) = dx \otimes a + dy \otimes c$.

%%%%%%%%%%%%%%%%%%%%%%%%%%%%%%%%%%%%%%%%%%%%%%%%%%%%%%%%%%%%%%%%%%%%%%%%%%

\subsection{$\mathcal H$ acting on ${\mathcal M}^!$}

Since the formulae for the (left or right) coactions are the same, the
formulae for the actions of $\mathcal H$ on $\mathcal M$ and
${\mathcal M}^!$ must also coincide. For instance, using $X_-^L[x] = y$
we find immediately $X_-^L[dx] = dy$. This corresponds to an irreducible
two-dimensional representation of $\mathcal H$. Anyway, we shall return
to this problem in the next section, since we are going to introduce a
differential algebra $\Omega_{WZ}({\mathcal M})$ built over $\mathcal M$
and investigate its contents in terms of representations of $\mathcal H$.

%%%%%%%%%%%%%%%%%%%%%%%%%%%%%%%%%%%%%%%%%%%%%%%%%%%%%%%%%%%%%%%%%%%%%%%%%%
%%%%%%%%%%%%%%%%%%%%%%%%%%%%%%%%%%%%%%%%%%%%%%%%%%%%%%%%%%%%%%%%%%%%%%%%%%

\section{Covariant differential calculus on $\mathcal M$}
\label{sec:diff-calculus}

%%%%%%%%%%%%%%%%%%%%%%%%%%%%%%%%%%%%%%%%%%%%%%%%%%%%%%%%%%%%%%%%%%%%%%%%%%

\subsection{Differential algebras associated with a given algebra}

Given an algebra $\mathcal A$, there is a universal construction that
allows one to build the so-called {\sl algebra of universal differential
forms\/} $\Omega({\mathcal A}) = \sum_{p=0}^\infty \Omega^p({\mathcal A})$
over $\mathcal A$. This differential algebra is universal, in the sense that
any other differential algebra with $\Omega^0({\mathcal A}) = {\mathcal A}$
will be a quotient of the universal one. Schematically, the construction of
$\Omega(\mathcal A)$ goes as follows. Let $m$ be the multiplication map:
$m(a \otimes b) \doteq ab$, for $a$ and $b$ in $\mathcal A$ and let
$\ker(m)$ be the kernel of this map ($a \otimes b \in \ker(m)$ implies that
$ab = 0$). One builds the algebra of universal forms as the graded vector
space:
\begin{eqnarray*}
\Omega^0({\mathcal A}) &\doteq & {\mathcal A}        \\
\Omega^1({\mathcal A}) &\doteq &
   \ker(m) \subset {\mathcal A} \otimes {\mathcal A} \\
\Omega^p({\mathcal A}) &\doteq &
   \Omega^1({\mathcal A}) \otimes_{\mathcal A}
   \Omega^1({\mathcal A}) \otimes_{\mathcal A} \ldots
   \otimes_{\mathcal A} \Omega^1({\mathcal A})
      \qquad \text{($p+1$ times)}
\end{eqnarray*}
The above tensor products are taken over the algebra ${\mathcal A}$.
Multiplication is defined as follows:
$$
(a_0 \otimes a_1 \otimes \ldots \otimes a_n)
   (b_0 \otimes b_1 \otimes \ldots \otimes b_n)
\doteq
   (a_0 \otimes a_1 \otimes \ldots \otimes a_n b_0 \otimes b_1
        \otimes \ldots b_n) \ .
$$
One also writes $a_0 da_1 \doteq a_0 \otimes a_1 - a_0 a_1 \otimes \one$,
\etc Notice that $d\one = 0$. More details about this construction can be
found, for instance, in \cite{Coquereaux-JGP}.

In the present situation, ${\mathcal A} = {\mathcal M}$.
Since $\Omega^1({\mathcal M}) = \ker(m)$, we find that
$\dim(\Omega^1({\mathcal M})) =
   \dim({\mathcal M} \otimes {\mathcal M}) - \dim({\mathcal M}) =
   9^2 - 9 = 72$.
The dimension of $\Omega({\mathcal M})$ itself is infinite (notice that
there are no particular relations between $dx$ and $dy$ so
$\Omega^p({\mathcal M})$ never vanishes).

For practical purposes, it is often not very convenient to work with the
algebra of universal forms. First of all, it is very ``big''. Next, it does
not remember anything of the coaction of ${\mathcal F}$ on the algebra
${\mathcal M}$ (the $0$-forms).

Starting from a given algebra, there are several constructions that allow
one to build ``smaller'' differential calculi. As already mentioned, they
will all be quotients of the algebra of universal forms by some
(differentiable) ideal. One possibility for such a construction was
described by \cite{Connes}, another one by \cite{Dubois-Violette},
and yet another one by \cite{Coquereaux-Haussling-Scheck}. In the present
case, however, we shall use something else, namely the differential calculus
$\Omega_{WZ}$ introduced by Wess and Zumino \cite{Wess-Zumino} (see also
\cite{Pusz}) in the case of the quantum $2$-plane. Its main interest, as we
shall see below, is that it is covariant with respect to the action (or
coaction) of a quantum group. This differential algebra is, as usual, a
graded vector space, and our first step will be to define the differentials
$dx$ and $dy$ together with their commutation relations.

%%%%%%%%%%%%%%%%%%%%%%%%%%%%%%%%%%%%%%%%%%%%%%%%%%%%%%%%%%%%%%%%%%%%%%%%%%

\subsection{The Wess-Zumino complex}

The differential algebra $\Omega_{WZ}$ that we are going to use was
introduced historically by \cite{Wess-Zumino}, in the case of the quantum
plane. It is only one of the many possible differential calculi that one
can associate with this algebra, but it is the only one (up to trivial
modifications) that is both quadratic and compatible with the coaction of
the group $Fun(SL_q(2,\CC))$. Since it will play an important role in what
follows, we shall briefly recall how it can be obtained.

First of all
$\Omega_{WZ} = \Omega_{WZ}^0 \oplus \Omega_{WZ}^1 \oplus \Omega_{WZ}^2 $
is a graded vector space.

\begin{itemize}
\item
   Forms of grade $0$ just coincide with the functions on the quantum plane,
   \ie they are polynomials in $x$ and $y$.

\item
   Forms of grade $1$ are of the type
   $a_{rs} x^r y^s dx + b_{rs} x^r y^s dy$, where $dx$ and $dy$ are those
   introduced in Section~\ref{sec:manin-dual}, devoted to the construction
   of the Manin dual ${\mathcal M}^!$. The Manin dual is the subalgebra
   $\{ \lambda_{00} + \lambda_{10} \, dx + \lambda_{01} \, dy +
       \lambda_{11} \, dx \, dy \}$
   of $\Omega_{WZ}$, where the coefficients $\lambda_{ij}$ belong to the
   field of scalars.

\item
   Forms of grade $2$ are of the type $c_{rs} x^r y^s dx \, dy$.
\end{itemize}

\noindent Next, $\Omega_{WZ}$ is an algebra. We need therefore to find the
relations between $x$, $y$, $dx$ and $dy$. Moreover, we want $\Omega_{WZ}$
to be a {\em differential} algebra, so we set $d(x) = dx$, $d(y) = dy$ and
the Leibniz rule (for $d$) should hold. Also, we set
$d \one = 0$ and $d^2 = 0$.

Assuming quadraticity of the algebra, we expand {\it a priori\/}
$x \, dx$, $x \, dy$, $y \, dx$ and $y \, dy$ in terms of
$dx \, x$, $dy \, x$, $dx \, y$ and $dy \, y$. This involves sixteen
unknown coefficients. Applying $d$ to the above, we get four relations
between these coefficients. We are left with twelve independent parameters.

Differentiating the relation $xy - qyx = 0$ and replacing $x \, dx$,
$x \, dy$, $y \, dx$ and $y \, dy$ by what was postulated before gives one
identity that fixes three of the unknown coefficients (actually one finds
four constraints but one of them is not independent of the others). We are
left with $12-3 = 9$ independent parameters.

Next, one uses compatibility with the (left or right) coaction of
$Fun(SL_q(2,\CC))$, \ie one writes
$x' = a \otimes x + b \otimes y$,
$y' = c \otimes x + d \otimes y$,
$dx' = a \otimes dx + b \otimes dy$,
$dy' = c \otimes dx + d \otimes dy$
and imposes that the relations\footnote{
Observe that asking for $xy=qyx$ and $dx\,dy=-q^{-1}dy\,dx$ to be
preserved under the coaction provides an alternative way of defining the
quantum group. Here we are just refering to relations that mix one of
$\{ x,y \}$ with one of $\{ dx,dy \}$.
}
between $x', y'$ and $dx', dy'$ are the same as the relations between the
corresponding untransformed elements (the unprimed ones). In this way one
obtains four identities that fix $8$ additional parameters. One is left with
just one ($ = 9-8$) free parameter.

The last parameter is fixed by checking associativity of the cubics, for
instance $(x \, dy) \, dx$ should be equal to $x \, (dy \, dx)$. One finds
that this last parameter should be either equal to $q$ or to $1/q$.

The final result is as follows:
\begin{equation}
\begin{tabular}{ll}
   $xy = qyx$                 &                                        \\
                              &                                        \\
   $x\,dx = q^2 dx\,x$ \qquad & $x\,dy = q \, dy\,x + (q^2 - 1) dx\,y$ \\
   $y\,dx = q \, dx\,y$       & $y\,dy = q^2 dy\,y$                    \\
                              &                                        \\
   $dx^2 = 0$                 & $dy^2 = 0$                             \\
   $dx\,dy + q^2 dy\,dx = 0$  &                                        \\
\end{tabular}
\label{WZ-relations}
\end{equation}
It may be useful to notice that $dy \, x = (q-1) y \, dx + q^2 x \, dy$.

%%%%%%%%%%%%%%%%%%%%%%%%%%%%%%%%%%%%%%%%%%%%%%%%%%%%%%%%%%%%%%%%%%%%%%%%%%

\subsection{A reduced Wess-Zumino complex}

In the case $q^3 = 1$, we add the following to the defining relation of
the quantum plane: $x^3 = \one, y^3 = \one$. As we know, this defines the
algebra $\mathcal M$ (the space of $3\times 3$ complex matrices) as a
quotient of the quantum plane. Adding the same two cubic relations to the
differential algebra $\Omega_{WZ}$ defines a differential algebra that we
shall denote $\Omega_{WZ}({\mathcal M})$. The fact that it is well defined
as a differential algebra is not totally obvious and requires some checking.
Technically, one has to verify that we are taking the quotient by a
differential ideal. In plain terms, one has to check that
$d(x^3) = d\one = 0$ and that $d(y^3) = d\one = 0$. This is indeed so:
\begin{eqnarray*}
   d(x^3) &=& d(x^2) x + x^2 dx = (dx)x^2 + x(dx)x + x^2 dx =
              (1+q+q^2)\, x^2 dx = 0 \\
   d(y^3) &=& d(y^2) y + y^2 dy = (dy)y^2 + y(dy)y + y^2 dy =
              (1+q+q^2)\, y^2 dy = 0
\end{eqnarray*}
Note that $\dim(\Omega_{WZ}^0) = 9$, $\dim(\Omega_{WZ}^1) = 9+9 = 18$
and $\dim(\Omega_{WZ}^2) = 9$.

%%%%%%%%%%%%%%%%%%%%%%%%%%%%%%%%%%%%%%%%%%%%%%%%%%%%%%%%%%%%%%%%%%%%%%%%%%

\subsection{The differential of an element of $\mathcal M$}

Let $m$ be an arbitrary $3 \times 3$ matrix. Let us call
$m_{ij}, \: i,j \in \{1,2,3\}$ the matrix elements of $m$. By using
elementary matrices, we know how to decompose $m$ on the basis $x^r y^s$
(see Section~\ref{sec:red-q-plane}). This way we can write
$m = m_{rs} x^r y^s$ (where, of course, the coefficients $m_{rs}$ do not
coincide at all with the $m_{ij}$!). It is then straightforward to compute
the one-form $dm$ belonging to the reduced Wess-Zumino differential algebra
$\Omega_{WZ}({\mathcal M})$ by using the Leibniz rule together with the
commutation relations given previously. If we write
$$
   dm = (dm)_x dx + (dm)_y dy \ ,
$$
we obtain (here the indices are of type $i,j$, \ie refer to the matrix
elements of $m$):
\begin{eqnarray*}
dm &=& {1 \over 3}
       \pmatrix{(m_{11}-m_{33})(1-q^2) & (m_{12}-m_{31})(q^2-q) &
                                         (m_{32} - m_{13})(1-q) \cr
                (m_{21}-m_{13})(q^2-q) & (m_{11}-m_{22})(1-q)   &
                                         (m_{23}-m_{12})(1-q^2) \cr
                (m_{23}-m_{31})(1-q)   & (m_{32}-m_{21})(1-q^2) &
                                         (m_{33}-m_{22})(q^2-q)} dx \\
{} &{}&        + \pmatrix{m_{12}     & -m_{13}q^2 & 0           \cr
                          0          & m_{23}     & -m_{21} q^2 \cr
                          -m_{32}q^2 & 0          & m_{31}} dy
\end{eqnarray*}
Notice that differences of cubic roots of $1$ appear in the matrix elements
of $(dm)_x$.

%%%%%%%%%%%%%%%%%%%%%%%%%%%%%%%%%%%%%%%%%%%%%%%%%%%%%%%%%%%%%%%%%%%%%%%%%%

\subsection{The action of $\mathcal H$ on $\Omega_{WZ}({\mathcal M})$}
\label{subsec:H-action-on-Omega}

%%%%%%%%%%%%%%%%%%%%%%%%%%%%%%%%%%%%%%%%%%%%%%%%%%%%%%%%%%%%%%%%%%%%%%%%%%

\subsubsection{The action of $\mathcal H$ on
               $\Omega_{WZ}^0({\mathcal M}) = {\mathcal M}$}

We already saw that the nine-dimensional space $\mathcal M$ can be
decomposed into three indecomposable representations of $\mathcal H$
called $3_{irr}$, $3_{eve}$ and $3_{odd}$. The $3_{irr}$ is irreducible and
spanned by $x^2, xy, y^2$. The $3_{eve}$ is reducible indecomposable and
spanned by $x, y, x^2 y^2$; it contains an invariant $2$-dimensional
subspace spanned by $x,y$. The $3_{odd}$ is also reducible indecomposable
and spanned by $\one, x^2 y, xy^2$; it contains an invariant
$1$-dimensional subspace spanned by $\one$.

%%%%%%%%%%%%%%%%%%%%%%%%%%%%%%%%%%%%%%%%%%%%%%%%%%%%%%%%%%%%%%%%%%%%%%%%%%

\subsubsection{The action of $\mathcal H$ on $\Omega_{WZ}^1({\mathcal M})$}

Since
$\Omega_{WZ}^1({\mathcal M}) = {\mathcal M}\,dx \oplus {\mathcal M}\,dy$,
and as $\{dx,dy\}$ span the two-dimensional irreducible representation of
${\mathcal H}$, it is {\it a priori \/} clear that we should decompose
$3_{irr} \otimes 2$, $3_{eve} \otimes 2$ and $3_{odd} \otimes 2$ in
indecomposable representations of $\mathcal H$. The action of elements
$X \in \mathcal H$ on one-forms is obtained, as usual, from the coproduct
rule; for instance,
$$
   X_+^L[y\,dy] = X_+^L[y] \one[dy] + K^L[y] X_+^L[dy] =
                  x\,dy + q^2 y\,dx \ .
$$
In this way we obtain the following results (we only give the tables
associated with the left action):

\begin{itemize}
\item
   The case $3_{odd} \otimes 2 = 3_{irr} \oplus 3_{eve}$.

\medskip
$$
\begin{tabular}{l||ccc}
              & $K^L$             & $X_+^L$        & $X_-^L$ \\
\hline
\hline
$\one \, dx$  & $q \, dx$         & $0$            & $dy$    \\
$x^2 y \, dx$ & $q^2 x^2 y \, dx$ & $q^2\, dx$     &
                $-q^2 x y^2 \, dx + x^2 y \, dy$             \\
$xy^2 \, dx$  & $x y^2 \, dx$     & $-x^2 y \, dx$ &
                $q dx + xy^2 \, dy$                          \\
\hline
$\one \, dy$  & $q^2 \, dy$       & $dx$           & $0$     \\
$x^2 y \, dy$ & $x^2 y \, dy$     & $q^2 \, dy + q x^2 y \, dx$      &
                $-q x y^2 \, dy$                             \\
$x y^2 \, dy$ & $q x y^2 \, dy$   & $-x^2 y \, dy + q^2 x y^2 \, dx$ &
                $dy$                                         \\
\end{tabular}
$$
\medskip

This gives $3_{irr} \oplus 3_{eve}$, since, up to multiplicative
factors (dashed arrows stand for the action of $X_-^L$ and continuous ones
for $X_+^L$)

\medskip
$$
\begin{diagram}
  q \, x^2 y \, dx - dy          & \rTo 0     \\
  \uTo \dDotsto                  &            \\
  -x^2 y \, dy + q^2 x y^2 \, dx &            \\
  \uTo \dDotsto                  &            \\
  -q \, dx + q \, x y^2 \, dy    & \rDotsto 0 \\
\end{diagram}
\hspace{1.0cm} \oplus \hspace{1.0cm}
\begin{diagram}
  dy            & \rDotsto  & 0                              \\
                & \luTo     &                                \\
  \uDotsto \dTo &           & x^2 y \, dy + q \, x y^2 \, dx \\
                & \ldDotsto &                                \\
  dx            & \rTo      & 0
\end{diagram}
$$
\medskip

Note that $3_{eve}$ contains an irreducible $2$-dimensional representation
spanned by $\{ dx, dy \}$.

\item
   The case $3_{eve} \otimes 2 = 3_{irr} \oplus 3_{odd}$.

\medskip
$$
\begin{tabular}{l||ccc}
                & $K^L$             & $X_+^L$      & $X_-^L$   \\
\hline
\hline
$x \, dx$       & $q^2 x \, dx$     & $0$          &
                  $q^2 y \, dx + x \, dy$                      \\
$y \, dx$       & $y \, dx$         & $x \, dx$    & $ y\, dy$ \\
$x^2 y^2 \, dx$ & $q x^2 y^2 \, dx$ & $-q y \, dx$ &
                  $-x \, dx + x^2 y^2 \, dy$                   \\
\hline
$x \, dy$      & $x \, dy$           & $q x \, dx$                  &
                 $q y \, dy$                                         \\
$y \, dy$      & $qy \, dy$          & $x \, dy + q^2 y \, dx$      &
                 $0$                                                 \\
$x^2y^2 \, dy$ & $q^2 x^2 y^2 \, dy$ & $-q y \, dy + x^2 y^2 \, dx$ &
                 $-q^2 x \, dy$                                      \\
\end{tabular}
$$
\medskip

This corresponds to $3_{irr} \oplus 3_{odd}$, since, up to multiplicative
factors,

\medskip
$$
\begin{diagram}
  y \, dy                & \rDotsto 0 \\
  \uDotsto \dTo          & \\
  q \, x \, dy + y \, dx & \\
  \uDotsto \dTo          & \\
  x \, dx                & \rTo 0 \\
\end{diagram}
\hspace{1.0cm} \oplus \hspace{1.0cm}
\begin{diagram}
  x^2y^2\, dy - x \, dx &           & \\
                        & \rdDotsto & \\
  \uDotsto \dTo         &           & x\, dy - q \, y \, dx \\
                        & \ruTo     & \dTo \dDotsto \\
  y \, dy - q^2 \, x^2 y^2\, dx &   & 0
\end{diagram}
$$
\medskip

Remark that $3_{odd}$ contains an irreducible $1$-dimensional
representation spanned by $x \, dy - q \, y \, dx$.

\item
   The case $3_{irr} \otimes 2 = 6_{eve}$.

\medskip
$$
\begin{tabular}{l||ccc}
             & $K^L$          & $X_+^L$        & $X_-^L$                  \\
\hline
\hline
$x^2 \, dx$  & $x^2 \, dx$    & $0$            & $-qxy \, dx + x^2 \, dy$ \\
$xy \, dx$   & $qxy \, dx$    & $qx^2 \, dx$   & $ y^2 \, dx + xy \, dy$  \\
$y^2 \, dx$ & $q^2 y^2 \, dx$ & $-q^2xy \, dx$ & $y^2 \, dy$              \\
\hline
$x^2 \, dy$  & $q x^2 \, dy $ & $q^2 x^2 \, dx$         & $-xy \, dy$     \\
$xy \, dy$   & $q^2 xy \, dy$ & $qx^2 \, dy + xy \, dx$ & $q^2 y^2 \, dy$ \\
$y^2 \, dy$  & $y^2 \, dy$    & $-q^2xy \, dy+q y^2 \, dx$ & $0$          \\
\end{tabular}
$$
\medskip

This actually gives the six-dimensional indecomposable representation
$6_{eve}$ (which is projective, \cf \ref{app:H-submodules}). It contains
the four-dimensional indecomposable representation $4_{eve}$, a family of
indecomposables of the type $3_e^\lambda$, and one irreducible of dimension
two, spanned by $\{ -q^2 x^2 \, dy + xy \, dx, -q xy \, dy + y^2 \, dx\}$.

\medskip
$$
\begin{diagram}
  0                         &       & x^2\, dx  & \rTo     & 0            \\
  \uTo                      & \ldDotsto &       & \luTo    &              \\
  -q^2 x^2 \, dy + xy \, dx &       &           &          &
                              xy \, dx + x^2 \, dy                        \\
  \uTo \dDotsto             & & \ldDotsto\luTo  &          & \uTo\dDotsto \\
  -q xy \, dy + y^2 \, dx   &       &           &          &
                              xy \, dy + y^2 \, dx                        \\
  \dDotsto                  & \luTo &          & \ldDotsto &              \\
  0                         &       & y^2 \, dy & \rDotsto & 0            \\
\end{diagram}
$$
\medskip
\end{itemize}

%%%%%%%%%%%%%%%%%%%%%%%%%%%%%%%%%%%%%%%%%%%%%%%%%%%%%%%%%%%%%%%%%%%%%%%%%%

\subsubsection{The action of $\mathcal H$ on $\Omega_{WZ}^2({\mathcal M})$}

The differential $2$-form $dx\, dy$ that generates (over $\mathcal M$)
$\Omega_{WZ}^2({\mathcal M})$ is $\mathcal H$-invariant, since
$X_+^L[dx\, dy] = X_-^L[dx\, dy] = 0$ and $K^L[dx\, dy] = dx \, dy$.
Hence $\Omega_{WZ}^2({\mathcal M}) = {\mathcal M} \, dx\, dy$ has exactly
the same decomposition in representations of $\mathcal H$ as $\mathcal M$.

%%%%%%%%%%%%%%%%%%%%%%%%%%%%%%%%%%%%%%%%%%%%%%%%%%%%%%%%%%%%%%%%%%%%%%%%%%

\subsection{Cohomology of $d$ on $\Omega_{WZ}({\mathcal M})$ and
            $\mathcal H$-representations}
\label{subsec:d-cohomology}

We will now analyse the action of the differential operator $d$ on each
level of the differential complex, as given by a basis adapted to the
action of the quantum group $\mathcal H$. As usual, we shall call
${\mathcal Z}^p \doteq \{ \omega \in \Omega^p \:
                          \text{with} \: d \omega = 0 \}$,
the space of $p$-cocycles and
${\mathcal B}^p \doteq \{ \omega \in \Omega^p \:
                          \text{such that} \: \omega = d \phi \, , \,
                          \text{for}\: \phi \in \Omega^{p-1} \}$,
the space of $p$-coboundaries. Of course,
${\mathcal B}^p \subset {\mathcal Z}^p$, and we set
$H^p \doteq {\mathcal Z}^p / {\mathcal B}^p$.

%%%%%%%%%%%%%%%%%%%%%%%%%%%%%%%%%%%%%%%%%%%%%%%%%%%%%%%%%%%%%%%%%%%%%%%%%%

\subsubsection{Action of $d$ on $\Omega_{WZ}^0({\mathcal M})$}

The differential operator $d$ acts on a basis of $\mathcal M$ as shown
in the following table. The range and image vectors are classified
according to the $\mathcal H$-representation to which they belong.

$$
\begin{tabular}{c|ccc|c}
{}        & {}       & $d$           &                      &           \\
\hline \hline
{}        & $\one$   & $\rightarrow$ & $0$                  &           \\
            \cline{2-4}
$3_{odd}$ & $x^2 y$  & $\rightarrow$ & $-qxy\,dx + x^2 dy$  & $2$       \\
{}        & $xy^2$   & $\rightarrow$ & $qy^2dx - q^2xy\,dy$ &           \\
\hline \hline
{}        & $x$      & $\rightarrow$ & $dx$                             \\
$3_{eve}$ & $y$      & $\rightarrow$ & $dy$                 & $3_{eve}$ \\
            \cline{2-4}
{}        & $x^2y^2$ & $\rightarrow$ & $-xy^2dx - q^2x^2y\,dy$ &        \\
\hline \hline
{}        & $x^2$    & $\rightarrow$ & $-q^2 x\,dx$         &           \\
$3_{irr}$ & $xy$     & $\rightarrow$ & $q^2 y\,dx + x\,dy$  & $3_{irr}$ \\
{}        & $y^2$    & $\rightarrow$ & $-q^2 y\,dy $        &           \\
\end{tabular}
$$

\noindent As it should ($d$ was built as a {\em covariant} differential
operator), $d$ preserves the representations, mapping
$3_{eve} \mapsto 3_{eve}$, $3_{irr} \mapsto 3_{irr}$, and
$3_{odd} \mapsto \text{the quotient} 3_{odd}/1 = 2$.

%%%%%%%%%%%%%%%%%%%%%%%%%%%%%%%%%%%%%%%%%%%%%%%%%%%%%%%%%%%%%%%%%%%%%%%%%%

\subsubsection{Action of $d$ on $\Omega_{WZ}^1({\mathcal M})$}

Dividing the space according to representations of $\mathcal H$, $d$ acts
as follows:
$$
\begin{tabular}{c|cccc|c}
{}                  & {} & {} & $d$ & {} & {} \\
\hline \hline
$3_{odd} \otimes 2$ & {} & {} & {}  & {} & {} \\
\hline
{}        & $x^2 y\,dx - q^2 dy$    & & $\rightarrow$ & $-qx^2dx\,dy$  & \\
$3_{irr}$ & $-x^2y\,dy + q^2xy^2dx$ & & $\rightarrow$ & $-xy\,dx\,dy$  &
                                                        $3_{irr}$ \\
{}        & $-q\,dx + qxy^2 dy$     & & $\rightarrow$ & $q^2y^2dx\,dy$ & \\
\hline
{}        & $dx$ & $\in{\mathcal B}^1 \subset {\mathcal Z}^1$ &
                   $\rightarrow$ & $0$ & \\
$3_{eve}$ & $dy$ & $\in{\mathcal B}^1 \subset {\mathcal Z}^1$ &
                   $\rightarrow$ & $0$ & \\
            \cline{2-5}
{}        & $x^2y\,dy + qxy^2dx$ &
                   $\in{\mathcal B}^1 \subset {\mathcal Z}^1$ &
                   $\rightarrow$ & $0$ & \\
\hline \hline
$3_{eve} \otimes 2$ & {} & {} & {} & {} & {} \\
\hline
{}        & $y\,dy$ & $\in{\mathcal B}^1 \subset {\mathcal Z}^1$ &
                      $\rightarrow$ & $0$ & \\
$3_{irr}$ & $x\,dy + q^2 y\,dx$ &
                      $\in{\mathcal B}^1 \subset {\mathcal Z}^1$ &
                      $\rightarrow$ & $0$ & \\
{}        & $x\,dx$ & $\in{\mathcal B}^1 \subset {\mathcal Z}^1$ &
                      $\rightarrow$ & $0$ & \\
\hline
{}        & $x\,dy - qy\,dx$      & & $\rightarrow$ & $-q\,dx\,dy$     & \\
            \cline{2-5}
$3_{odd}$ & $y\,dy - q^2x^2y^2dx$ & & $\rightarrow$ & $-q^2x^2y\,dx\,dy$ &
                                                      $3_{odd}$          \\
{}        & $x^2 y^2dy - x\,dx$   & & $\rightarrow$ & $-xy^2 dx\,dy$   & \\
\hline \hline
$3_{irr}\otimes 2$  & {} & {} & {} & {} & {} \\
\hline
{}        & $xy\,dx + x^2dy$     &                     &
                                   $\rightarrow$ & $x\,dx\,dy$   &     \\
{}        & $xy\,dy + y^2dx$     &                     &
                                   $\rightarrow$ & $-qy\,dx\,dy$ & $2$ \\
            \cline{2-5}
$6_{eve}$ & $x^2dx$              & $\in{\mathcal Z}^1$ &
                                   $\rightarrow$ & $0$           &     \\
{}        & $y^2dy$              & $\in{\mathcal Z}^1$ &
                                   $\rightarrow$ & $0$           &     \\
            \cline{2-5}
{}        & $x^2dy - qxy\,dx$    & $\in{\mathcal B}^1 \subset{\mathcal Z}^1$
                                 & $\rightarrow$ & $0$           &     \\
{}        & $qy^2dx - q^2xy\,dy$ & $\in{\mathcal B}^1 \subset{\mathcal Z}^1$
                                 & $\rightarrow$ & $0$           &     \\
\end{tabular}
$$

%%%%%%%%%%%%%%%%%%%%%%%%%%%%%%%%%%%%%%%%%%%%%%%%%%%%%%%%%%%%%%%%%%%%%%%%%%

\subsubsection{Action of $d$ on $\Omega_{WZ}^2({\mathcal M})$}

Here $d$ gives trivially $0$ on all the elements, because there is no
$\Omega^3$ subspace in our differential complex; however, one should
distinguish between exact and closed forms:
\begin{eqnarray*}
   3_{odd} &:& \{ dx\, dy, x^2 y\, dx\, dy, xy^2 dx\, dy \}
               \in {\mathcal B}^2 \subset {\mathcal Z}^2 \\
   3_{eve} &:& \{ x\, dx\, dy, y\, dx\, dy \}
               \in {\mathcal B}^2 \subset {\mathcal Z}^2 \\
           & & \{ x^2 y^2 dx\, dy \}
               \in {\mathcal Z}^2                        \\
   3_{irr} &:& \{ x^2 dx\, dy, xy\, dx\, dy, y^2 dx\, dy \}
               \in {\mathcal B}^2 \subset {\mathcal Z}^2 \\
\end{eqnarray*}

%%%%%%%%%%%%%%%%%%%%%%%%%%%%%%%%%%%%%%%%%%%%%%%%%%%%%%%%%%%%%%%%%%%%%%%%%%

\subsubsection{Cohomology of $d$}

From the previous tables, we see that
\begin{eqnarray*}
   \dim{(\mathcal Z}^0)=1~ \ , &\dim({\mathcal B}^0)=0 \ , &\text{hence}
      \dim(H^0)=1 \ , \\
   \dim({\mathcal Z}^1)=10 \ , &\dim({\mathcal B}^1)=8 \ , &\text{hence}
      \dim(H^1)=2 \ , \\
   \dim({\mathcal Z}^2)=9~ \ , &\dim({\mathcal B}^2)=8 \ , &\text{hence}
      \dim(H^2)=1 \ .
\end{eqnarray*}
Remark that $\chi = \dim(H^0) - \dim(H^1) + \dim(H^2) = 1 - 2 + 1 = 0$.

%%%%%%%%%%%%%%%%%%%%%%%%%%%%%%%%%%%%%%%%%%%%%%%%%%%%%%%%%%%%%%%%%%%%%%%%%%

\subsection{Star operations on the differential calculus
            $\Omega_{WZ}({\mathcal M})$}
\label{subsec:*-on-Omega}

Given the $*$ operation on the algebra $\mathcal M$, we want to extend it
to the differential algebra $\Omega_{WZ}({\mathcal M})$. As usual, it has
to be involutive, anti-multiplicative for the algebra structure in
$\Omega_{WZ}({\mathcal M})$, and complex sesquilinear. Moreover,
it should be compatible with the coaction of $\mathcal F$. However,
there is no reason {\it a priori\/} to impose that $*$ should commute
with $d$.

The quantum group covariance condition is, again, just the commutativity of
the $*$, $\Delta_{R,L}$ diagram, or, algebraicaly,
\begin{equation}
   (\Delta_{R,L} \omega)^* = \Delta_{R,L} (\omega^*) \ .
\label{*-coaction-on-omega-condition}
\end{equation}

\noindent In any case, it is enough to determine the action of $*$ on the
generators $dx$ and $dy$, since we already determined the $*$ operation on
$\mathcal M$ ($* x = x$, $* y = y$).

Taking $\Delta_{L} dx = a \otimes dx + b \otimes dy$, we get
$(\Delta_{L} dx)^* = a \otimes dx^* + b \otimes dy^*$, to be compared
with $\Delta_{L} (dx^*)$. It is trivial that the solution $dx^* = dx$
is a permissible one. But it is also the only one, up to complex phases.
To see this one should expand $dx^*$ as a generic element of
$\Omega^1_{WZ}({\mathcal M})$ (we want a grade-preserving $*$) in the
previous condition. Solving for the coefficients, and adding the
requirement $x^* dx^* = q dx^* x^*$ (for instance, it is the $*$ of
$dx\,x = q^2 x\,dx$), we see that the result is:
\begin{equation}
   dx^* = dx \ , \qquad dy^* = dy \ .
\label{star-in-Omega}
\end{equation}

Remark: On $\Omega_{WZ}$, $d$ is a graded derivation. First, $*$ is
defined on $\Omega_{WZ}^0 = {\mathcal M}$. It is also defined on the
$d$ of the generators of ${\mathcal M}$ (in our case $* dx = dx$ and
$* dy = dy$). Then $*$ is extended to the whole of the differential
algebra $\Omega_{WZ}$ by imposing the anti-multiplicative property
$*(\omega_1 \omega_2) = (* \omega_2) (* \omega_1)$. Let
$\phi = a_i \, dx^i$ be an arbitrary element of $\Omega_{WZ}^1$. Then
$d(a_i \, dx^i) = d(a_i) dx^i$, so
$$
   *d(a_i \, dx^i) = (da_i \, dx^i)^* = (dx^i)^* \, (da_i)^* =
                     dx^i \, (da_i)^* \ .
$$
But, at the same time,
$$
   d*(a_i \, dx^i) = d((dx^i)^* \, (a_i)^*) = d(dx^i (a_i)^*) =
                     - dx^i \, d(a_i^*) \ .
$$

\noindent In the case of a (real) manifold, we have an almost trivial star,
with $* a_i = a_i$ and $* da_i = da_i$ for $a_i \in \Omega_{WZ}^0$, so
$d* a_i = da_i$ and one finds\footnote{
This shows in particular that usual differential forms on a real manifold
(forms that are real in the naive sense) cannot be ``real'' for the star
operation. Indeed, for a one-form, $* \phi = \phi$ implies
$* d\phi = - d\phi$. This peculiar aspect of usual differential forms
is a general feature of any differential algebra and can only be
circumvented at the expense of introducing {\sl graded} star operations.
We do not use such graded star operations in the present paper.
}
$* d\phi = - d * \phi$, when $\phi \in \Omega_{WZ}^1$. One recovers the fact
that $*$ is almost trivial, ``almost'' since it is still antimultiplicative:
$*(dx\,dy) = dy\,dx$.

In our case the conclusion is the same, indeed, it can be checked that,
for all elements $a_i$ in ${\mathcal M}$, one has $*(da_i) = d(* a_i)$.
Therefore (with $\phi = a_i \, dx^i \in \Omega_{WZ}^1$),
$d * \phi = - * d\phi$. More generally,
\begin{equation}
   d * \omega = (-1)^p * d \omega \quad \text{when} \quad
                \omega \in \Omega_{WZ}^p \ .
\label{d-*-relation}
\end{equation}
A general expression for the structure of the most general hermitian
one-forms will be given later.

Warning: we are not saying that the above involution is the only one that
one can define on the Wess-Zumino complex. For instance, one could very well
try to extend the involution $\dag$ (the one for which $x^\dag = x^2$) to
this differential algebra. Such an extension may be possible, but it would
not be compatible with the coaction of $\mathcal F$ (it would only be
compatible with dilations of $x$ and $y$ by numerical scaling factors).
Loosing the compatibility with the coaction of $\mathcal F$ is however
clearly inacceptable since the main interest of the differential complex
of Wess-Zumino rests on the fact that it is compatible with the coaction
(indeed, one can construct many other differential algebras over
$\mathcal M$ that are {\sl not\/} compatible with the coaction of
$\mathcal F$!).

%%%%%%%%%%%%%%%%%%%%%%%%%%%%%%%%%%%%%%%%%%%%%%%%%%%%%%%%%%%%%%%%%%%%%%%%%%

\subsection{Multiplicative properties of $\Omega_{WZ}({\mathcal M})$
            and ${\mathcal H}$-representations}
\label{subsec:product-on-Omega}

We now analyze the behaviour ---as representations of $\mathcal H$---
of the product in $\Omega_{WZ}({\mathcal M})$. Some facts are trivially
obtained:
\begin{eqnarray*}
   \Omega^2_{WZ}({\mathcal M}) \cdot \Omega^2_{WZ}({\mathcal M}) &=& 0 \\
   \Omega^1_{WZ}({\mathcal M}) \cdot \Omega^2_{WZ}({\mathcal M}) &=&
   \Omega^2_{WZ}({\mathcal M}) \cdot \Omega^1_{WZ}({\mathcal M})  =  0
\end{eqnarray*}

\noindent Moreover, as $dx \, dy$ is $\mathcal H$-invariant we know that
$\Omega^2_{WZ}({\mathcal M})$ is isomorphic to $\mathcal M$. Using this,
and the fact that $dx \, dy$ is $\mathcal M$-central, we see that the
products $\Omega^2 \cdot \Omega^0$ and $\Omega^0 \cdot \Omega^2$ can be
understood in terms of products in $\mathcal M$.

Anyway, the multiplication table for representations in $\mathcal M$ is
very easy to write down (for instance using the grading of the generators):
$$
\begin{tabular}{c|ccc}
${\mathcal M} \cdot {\mathcal M}$ & $3_{odd}$ & $3_{eve}$ & $3_{irr}$ \\
\hline\hline
$3_{odd}$                         & $3_{odd}$ & $3_{eve}$ & $3_{irr}$ \\
$3_{eve}$                         & $3_{eve}$ & $3_{irr}$ & $3_{odd}$ \\
$3_{irr}$                         & $3_{irr}$ & $3_{odd}$ & $3_{eve}$ \\
\end{tabular}
$$

\noindent Next we summarize the product relations amongst elements of
$\Omega^1_{WZ}({\mathcal M})$:
$$
\begin{tabular}{l|ccc}
$\Omega^1({\mathcal M}) \cdot \Omega^1({\mathcal M})$ &
                                $3_{odd} \otimes 2$ & $3_{eve} \otimes 2$ &
                                $3_{irr} \otimes 2 = 6_{eve}$     \\
\hline\hline
$3_{odd} \otimes 2$           & $3_{odd}$ & $3_{eve}$ & $3_{irr}$ \\
$3_{eve} \otimes 2$           & $3_{eve}$ & $3_{irr}$ & $3_{odd}$ \\
$3_{irr} \otimes 2 = 6_{eve}$ & $3_{irr}$ & $3_{odd}$ & $3_{eve}$ \\
\end{tabular}
$$

\noindent Finally, here is the corresponding table for products
between elements of $\mathcal M$ in a given representation and elements
of $\Omega^1_{WZ}({\mathcal M})$, also with fixed transformation
properties:
$$
\begin{tabular}{c|ccc}
${\mathcal M} \cdot \Omega^1({\mathcal M})$ or
$\Omega^1({\mathcal M}) \cdot {\mathcal M}$ &
            $3_{odd} \otimes 2$ & $3_{eve} \otimes 2$ &
                                  $3_{irr} \otimes 2 = 6_{eve}$ \\
\hline\hline
$3_{odd}$ & $3_{odd} \otimes 2$ & $3_{eve} \otimes 2$ &
                                  $3_{irr} \otimes 2$ \\
$3_{eve}$ & $3_{eve} \otimes 2$ & $3_{irr} \otimes 2$ &
                                  $3_{odd} \otimes 2$ \\
$3_{irr}$ & $3_{irr} \otimes 2$ & $3_{odd} \otimes 2$ &
                                  $3_{eve} \otimes 2$ \\
\end{tabular}
$$

%%%%%%%%%%%%%%%%%%%%%%%%%%%%%%%%%%%%%%%%%%%%%%%%%%%%%%%%%%%%%%%%%%%%%%%%%%
%%%%%%%%%%%%%%%%%%%%%%%%%%%%%%%%%%%%%%%%%%%%%%%%%%%%%%%%%%%%%%%%%%%%%%%%%%

\section{Noncommutative generalized connections on $\mathcal M$ and their
         curvature}
\label{sec:connections}

%%%%%%%%%%%%%%%%%%%%%%%%%%%%%%%%%%%%%%%%%%%%%%%%%%%%%%%%%%%%%%%%%%%%%%%%%%

\subsection{Generalized connections in non commutative geometry}

Let $\Omega$ be a differential calculus over a unital associative
algebra $\mathcal A$, \ie a graded differential algebra with
$\Omega^0 = \mathcal A$. Let $\mathcal A'$ be a right module over
$\mathcal A$. A covariant differential $\nabla$ on $\mathcal A'$ is a map
${\mathcal A'} \otimes_{\mathcal A} \Omega^p \mapsto
               {\mathcal A'} \otimes_{\mathcal A} \Omega^{p+1}$,
such that
$$
   \nabla( \psi \lambda) = (\nabla \psi) \lambda + (-1)^s \psi \, d \lambda
$$
whenever $\psi \in {\mathcal A'} \otimes_{\mathcal A} \Omega^s$ and
$\lambda \in \Omega^t$. $\nabla$ is clearly not linear with respect to the
algebra $\mathcal A$ but it is easy to check that the curvature $\nabla^2$
is a linear operator with respect to $\mathcal A$.

In the particular case where the module $\mathcal A'$ is taken as the
algebra $\mathcal A$ itself, any one-form $\omega$ (any element of
$\Omega^1$) defines a covariant differential. One sets simply
$\nabla \one = \omega$, where $\one$ is the unit of the algebra
$\mathcal A$. When $f \in \mathcal A$, one obtains
$$
   \nabla f = \nabla \one f =
              (\nabla \one) f + \one \, df = df + \omega f \ .
$$
Moreover,
$\nabla^2 f = \nabla (df + \omega f) =
              d^2 f + \omega df + (\nabla \omega) f - \omega df =
              (\nabla \omega) f$.
The curvature, in that case, is
$$
   \rho \doteq \nabla \omega = \nabla \one \, \omega =
                               (\nabla \one) \, \omega + \one \, d \omega =
                               d \omega + \omega^2 \ .
$$

\noindent Take $u$ as an invertible element of $\mathcal A$, and act with
$d$ on the equation $u^{-1} u = \one$. Using the property $d\one = 0$ one
obtains $du^{-1} = - u^{-1} du \, u^{-1}$. Define
$\omega' = u^{-1} \omega u + u^{-1} du$ and compute the new curvature
$\rho' = d\omega' + \omega'^2$. One obtains immediately
$\rho' = u^{-1} ( d \omega + \omega^2) u = u^{-1} \rho u$. This shows
that the usual formulae hold without having to assume commutativity of
the algebra $\mathcal A$.

Remark that here we take the module $\mathcal A'$ (a necessary ingredient
in the construction of any gauge theory) as the algebra $\mathcal A$ itself.
More generally, we could have chosen a free module ${\mathcal A}^p$, or
even a projective module over $\mathcal A$. We shall not consider here
this more general situation. Therefore, in some sense, our connections are
an analogue of usual ``abelian Yang Mills fields'', but the algebra
$\mathcal A$, contrarily to what happens in conventional gauge field
theories, will be non-commutative.

%%%%%%%%%%%%%%%%%%%%%%%%%%%%%%%%%%%%%%%%%%%%%%%%%%%%%%%%%%%%%%%%%%%%%%%%%%

\subsection{Connections on $\mathcal M$ and their curvature}

We now return to the specific case where ${\mathcal A}={\mathcal M}$ is
the algebra of functions over the quantum plane at a cubic root of unity.

The most general connection is defined by an element $\phi$ of
$\Omega_{WZ}^1(\mathcal M)$. Since we have a quantum group action of
$\mathcal H$ on $\Omega_{WZ}$, it is convenient to decompose $\phi$
into representations of this algebra as obtained in
Section~\ref{subsec:H-action-on-Omega}. We set

\begin{eqnarray*}
   \phi &=& \phi_{3i} + \phi'_{3i} + \phi_{3e} + \phi_{3o} + \phi_{6e} \ ,
\end{eqnarray*}
where
\begin{eqnarray*}
   \phi_{3i}  &=& a_{i1}\, (q x^2 y \, dx - dy) +
                  a_{i2}\, (q^2 x y^2 \, dx - x^2 y \, dy) +
                  a_{i3}\, q (dx - x y^2 \, dy)                          \\
   \phi'_{3i} &=& b_{i1}\, q^2 y\, dy + b_{i2}\, (y\, dx + q x\, dy) +
                  b_{i3}\, q^2 x\, dx                                    \\
   \phi_{3e}  &=& a_{e1}\, dy + a_{e2}\, (q x y^2 \, dx + x^2 y \, dy) +
                  a_{e3}\, dx                                            \\
   \phi_{3o}  &=& b_{o1}\, q (x\, dx - x^2 y^2\, dy) +
                  b_{o2}\, q (q y\, dx - x\, dy) +
                  b_{o3}\, q (x^2 y^2\, dx - q y\, dy)                   \\
   \phi_{6e}  &=& c_1\, (xy\, dx - q^2 x^2\, dy) +
                  c_2\, (y^2\, dx - q xy\, dy) +
                  c_3\, q x^2\, dx + c_4\, q y^2\, dy +                  \\
              & & c_5\, q^2 (xy\, dx + x^2\, dy) +
                  c_6\, q (xy\, dy + y^2\, dx)
\end{eqnarray*}

The coefficients $a_i$ and $b_i$ refer to the three-dimensional irreducible
representations, $a_e$ and $b_o$ to the even and odd three-dimensional
indecomposable representations, and $c$ to the six-dimensional
indecomposable representation $6_{eve}$.

The exact expression of the curvature $\rho = d \phi + \phi^2$ is not
very illuminating, but, thanks to the knowledge of the general features of
the multiplication in $\Omega_{WZ}$ (Section~\ref{subsec:product-on-Omega})
and the properties of $d$ (Section~\ref{subsec:d-cohomology}) we can make
the following observations:

\begin{itemize}
   \item
      $\phi \in 3_{irr} \subset 3_{odd} \otimes 2 \quad \Rightarrow \quad
       d\phi \in 3_{irr}$, $\phi^2 \in 3_{odd}$. Actually, direct
      calculation shows that $\phi^2 = 0$, so $\rho = d\phi \in 3_{irr}$.

   \item
      $\phi \in 3_{eve} \subset 3_{odd} \otimes 2 \quad \Rightarrow \quad
       d\phi = 0$, and $\rho = \phi^2 \in 3_{odd}$.

   \item
      $\phi \in 3_{irr} \subset 3_{eve} \otimes 2 \quad \Rightarrow \quad
       d\phi = 0$, and $\rho = \phi^2 \in 3_{irr}$.

   \item
      $\phi \in 3_{odd} \subset 3_{eve} \otimes 2 \quad \Rightarrow \quad
       d\phi \in 3_{odd}$, $\phi^2 \in 3_{irr}$. In fact, it can be shown
      that here also $\phi^2 = 0$, so $\rho = d\phi \in 3_{odd}$.

   \item
      $\phi \in 6_{eve} = 3_{irr} \otimes 2 \quad \Rightarrow \quad
       d\phi \in 2$, $\phi^2 \in 3_{eve}$, and $\rho \in 3_{eve}$.
\end{itemize}

%%%%%%%%%%%%%%%%%%%%%%%%%%%%%%%%%%%%%%%%%%%%%%%%%%%%%%%%%%%%%%%%%%%%%%%%%%

\subsubsection*{Hermitian connections}

As we know, the only star operation compatible with the quantum group action
of $\mathcal H$ on the differential algebra $\Omega_{WZ}$, when $q^3 = 1$,
is the one obtained in Section~\ref{subsec:*-on-Omega} ($dx^* = dx$,
$dy^* = dy$). Imposing the hermiticity property $\phi = \phi^*$ on the
connection implies that all the coefficients $a_i, b_i, a_e, b_o, c$ should
be {\sl real\/}.

Consider, for instance, the most general hermitian connection with
coefficients in the representation $3_{eve}$, namely
$$
   \phi = \phi_{3e} =
          a_{e1}\, dy + a_{e3}\, dx +
          a_{e2}\, (x^2 y \, dy + qxy^2 \, dx) \ ,
$$
with {\sl real\/} coefficients $a_{e1}, a_{e2}, a_{e3}$.

In this case, $d \phi = 0$, automatically. The curvature is then equal to
$\phi^2$ and its expression is quite simple, it reads
$$
   \rho = \left( a_{e1}a_{e3} - a_{e2}^2 \right)(1-q) \, dx \, dy \ .
$$
Notice that it is a singlet, the one-dimensional representation obtained
as a subrepresentation of the $3_{odd}$ of $\Omega_{WZ}^2$.

%%%%%%%%%%%%%%%%%%%%%%%%%%%%%%%%%%%%%%%%%%%%%%%%%%%%%%%%%%%%%%%%%%%%%%%%%%
%%%%%%%%%%%%%%%%%%%%%%%%%%%%%%%%%%%%%%%%%%%%%%%%%%%%%%%%%%%%%%%%%%%%%%%%%%

\section{Incorporation of Space-Time}
\label{sec:space-time}

%%%%%%%%%%%%%%%%%%%%%%%%%%%%%%%%%%%%%%%%%%%%%%%%%%%%%%%%%%%%%%%%%%%%%%%%%%

\subsection{Algebras of differential forms over $C^\infty(M) \otimes
            {\mathcal M}$}

Let $\Lambda$ be the algebra of usual differential forms over a space-time
manifold $M$ (the De Rham complex) and
$\Omega_{WZ} \doteq \Omega_{WZ}({\mathcal M})$,
the differential algebra over the reduced quantum plane introduced in
Section~\ref{sec:diff-calculus}. Remember that
$\Omega_{WZ}^0 = {\mathcal M}$,
$\Omega_{WZ}^1 = {\mathcal M} \: dx + {\mathcal M} \: dy$, and that
$\Omega_{WZ}^2 = {\mathcal M} \: dx \, dy$.
Calling $\Xi$ the graded tensor product of these two differential algebras,
$$
   \Xi \doteq \Lambda \otimes \Omega_{WZ} \ ,
$$
we see that:

\begin{itemize}
\item
   An element of $\Xi^0 = \Lambda^0 \otimes \Omega_{WZ}^0$ is a
   $3 \times 3$ matrix with elements in $C^\infty(M)$. It can be thought
   as an $M_3(\CC)$-valued scalar field.

\item
   A generic element of
   $\Xi^1 = \left( \Lambda^0 \otimes \Omega_{WZ}^1 \right) \oplus
            \left( \Lambda^1 \otimes \Omega_{WZ}^0 \right)$
   is given by a triplet $\omega = ( A_\mu, \phi_x, \phi_y )$, where
   $A_\mu$ determines a one-form (a vector field) on the manifold $M$
   with values in $M_3(\CC)$ (which can be considered as the Lie algebra
   of the Lie group $GL(3,\CC)$), and where $\phi_x$ and $\phi_y$ are
   $M_3(\CC)$-valued scalar fields. Indeed
   $\phi_x (x^{\mu}) \: dx + \phi_y (x^{\mu}) \: dy
            \in \Lambda^0 \otimes \Omega_{WZ}^1$.

\item
   An arbitrary element of
   $\Xi^2 = \left( \Lambda^0 \otimes \Omega_{WZ}^2 \right) \oplus
            \left( \Lambda^1 \otimes \Omega_{WZ}^1 \right) \oplus
            \left( \Lambda^2 \otimes \Omega_{WZ}^0 \right)$
   consists of
   \begin{itemize}
   \item
      $F_{\mu \nu} dx^\mu dx^\nu \in \Lambda^2 \otimes \Omega_{WZ}^0$,
      a matrix-valued $2$-form on the manifold $M$,

   \item
      a matrix-valued scalar field on $M$, \ie an element of
      $\Lambda^0 \otimes \Omega_{WZ}^2$,

   \item
      two matrix-valued vector fields on $M$, given by an element of
      $\Lambda^1 \otimes \Omega_{WZ}^1$.
   \end{itemize}
\end{itemize}

The algebra $\Xi$ is endowed with a differential (of square zero, of course,
and obeying the Leibniz rule) defined by
$d \doteq d \otimes \id \pm \id \otimes d$. Here $\pm$ is the (differential)
parity of the first factor of the tensor product upon which $d$ is
applied, and the two $d$'s appearing on the right hand side are the usual
De Rham differential on antisymmetric tensor fields and the differential
of the reduced Wess-Zumino complex, respectively.

If $G$ is a Lie group acting on the manifold $M$, it acts also (by
pull-back) on the functions on $M$ and, more generally, on the differential
algebra $\Lambda$. For instance, we may assume that $M$ is Minkowski space
and $G$ is the Lorentz group. The Lie algebra of $G$ and its enveloping
algebra $\mathcal U$ also act on $\Lambda$, by differential operators.
Intuitively, elements of $\Xi$ have an ``external'' part (\ie functions on
$M$) on which $\mathcal U$ act, and an ``internal'' part (\ie elements
belonging to $\mathcal M$) on which $\mathcal H$ acts. We saw that
$\mathcal H$ is a Hopf algebra (neither commutative nor cocommutative)
whereas $\mathcal U$, as it is well known, is a non-commutative but
cocommutative Hopf algebra. To conclude, we have an action of the Hopf
algebra ${\mathcal U} \otimes {\mathcal H}$ on the differential algebra
$\Xi$.

%%%%%%%%%%%%%%%%%%%%%%%%%%%%%%%%%%%%%%%%%%%%%%%%%%%%%%%%%%%%%%%%%%%%%%%%%%

\subsection{Generalized gauge fields}
\label{subsec:generalized-gauge-fields}

Since we have a differential algebra $\Xi$ on the associative algebra
$C^\infty(M) \otimes {\mathcal M}$ we can define, as usual,
``abelian''-like connections by choosing a module which is equal to the
associative algebra itself. A Yang-Mills potential $\omega$ is an arbitrary
element of $\Xi^1$ and the corresponding curvature, $d \omega + \omega^2$,
is an element of $\Xi^2$. As we have said above,
$\omega = (A_\mu, \phi_x, \phi_y)$ consists of a usual Yang-Mills field
$A_\mu$ and a pair $\phi_x, \phi_y$ of scalar fields, all of them valued
in $M_3(\CC)$. We have
$\omega = A + \Phi$, where $A = A_\mu \, dx^\mu$ and
$\Phi = \phi_x \, dx + \phi_y \, dy \in \Lambda^0 \otimes \Omega_{WZ}^1
                                    \subset \Xi^1$.
We can also decompose $A = A^\alpha \lambda_\alpha$, with $\lambda_\alpha$
denoting the usual Gell-Mann matrices (together with the unit matrix) and
$A^\alpha$ a set of complex valued one-forms on the manifold $M$. To make
the distinction clearer, let us call $\delta$ the differential on $\Xi$,
$\underline{d}$ the differential on $\Lambda$ and $d$ the differential on
$\Omega_{WZ}$ (as before). The curvature is then
$\delta \omega + \omega^2$. Explicitly,
$$
\delta A = (\underline{d} A^\alpha) \lambda_\alpha -
           A^\alpha d\lambda_\alpha
$$
and
$$
\delta\Phi = (\underline{d}\phi_x) dx + (\underline{d}\phi_y) dy +
             (d \phi_x) dx + (d \phi_y) dy \ .
$$

\noindent It is therefore clear that the corresponding curvature will have
several pieces:
\begin{itemize}
\item
   The Yang-Mills strength $F$ of $A$,
   $$
      F \doteq (\underline{d} A^\alpha) \lambda_\alpha + A^2 \quad
               \in \Lambda^2 \otimes \Omega_{WZ}^0
   $$

\item
   A kinetic term ${\mathcal D} \Phi$ for the scalars, consisting of a
   purely derivative term, a covariant coupling to the gauge field and a
   mass term for the Yang-Mills field (linear in the $A_\mu$'s),
   $$
      {\mathcal D \Phi} \doteq (\underline{d}\phi_x) dx +
                               (\underline{d}\phi_y) dy +
                               A \Phi + \Phi A - A^\alpha d\lambda_\alpha
                               \quad \in \Lambda^1 \otimes \Omega_{WZ}^1
   $$

\item
   Finally, a self interaction term for the scalars
   $$
      (d \phi_x) dx + (d \phi_y) dy + \Phi^2 \quad
         \in \Lambda^0 \otimes \Omega_{WZ}^2
   $$
\end{itemize}

\noindent We recover the usual ingredients of a Yang-Mills-Higgs system
(the mass term for the gauge field, linear in $A$, is usually obtained
from the ``$A \Phi$ interaction'' after shifting $\Phi$ by a constant).

By chosing an appropriate scalar product on the space $\Xi^2$, one obtains
therefore a quantity that is quadratic in the curvatures (quartic in $A$ and
$\Phi$) and could be a candidate for the Lagrangian of a theory of
Yang-Mills-Higgs type. However, if we do not make specific choices for the
connection (for instance by imposing reality constraints or by selecting one
or another representation of $\mathcal H$), the results are a bit too
general and, in any case, difficult to interpret physically. Regarding this
construction, there is also another caveat that will be explained in the
next section.

%%%%%%%%%%%%%%%%%%%%%%%%%%%%%%%%%%%%%%%%%%%%%%%%%%%%%%%%%%%%%%%%%%%%%%%%%%
%%%%%%%%%%%%%%%%%%%%%%%%%%%%%%%%%%%%%%%%%%%%%%%%%%%%%%%%%%%%%%%%%%%%%%%%%%

\section{Discussion: quantum group symmetry and noncommuting fields}
\label{sec:discussion}

As mentioned in Section~\ref{subsec:generalized-gauge-fields}, the
building of physical models of gauge type will involve the consideration
of one-forms. If we restrict ourselves to the ``internal space'' part of
these one-forms, we have to consider objects of the form
$$
   \Phi = \sum_i \varphi_i \omega_i \ .
$$
Here $\{ \omega_i \}$ is a basis of some non-trivial indecomposable
representation of $\mathcal H$ (or any other non-cocommutative quantum
group) on the space of $1$-forms, and $\varphi_i$ are functions over some
space-time manifold. What about the transformation properties of the fields
$\varphi_i$? This is a question of central importance, since, ultimately,
we will integrate out the internal space (whatever this means), and the only
relic of the quantum group action on the theory will be the transformations
of the $\varphi_i$'s. There are several possibilities: one of them, as
suggested from the results of Section~\ref{sec:space-time} is to consider
$\mathcal H$ as a discrete analogue of the Lorentz group (actually, of
the enveloping algebra $\mathcal U$ of its Lie algebra). In such a case,
``geometrical quantities'', like $\Phi$ should be $\mathcal H$-invariant
(and $\mathcal U$-invariant). This requirement obviously forces the
$\varphi_i$ to transform. Another possibility would be to assume that
$\Phi$ itself transforms according to some representation of this quantum
group (in the same spirit one can study, classically, the invariance of
particularly chosen connections under the action of a group acting also on
the base manifold). In any case, the $\varphi_i$ are going to span some
non-trivial representation space of $\mathcal H$.

Usually, the components $\phi_i$ of fields are real (or complex) numbers
and are, therefore, commuting quantities. However, this observation leads
to the following problem: If the components of the fields commute, so that
$\varphi_i \varphi_j = \varphi_j \varphi_i$, then we get
$h.(\varphi_i \varphi_j) = h.(\varphi_j \varphi_i)$, for any
$h \in {\mathcal H}$. This would imply (here $\Delta h = h_1 \otimes h_2$)
\begin{eqnarray*}
   (h_1.\varphi_i)(h_2.\varphi_j) &=& (h_1.\varphi_j)(h_2.\varphi_i) \\
                                  &=& (h_2.\varphi_i)(h_1.\varphi_j) \ .
\end{eqnarray*}
This equality cannot be true in general unless we have a cocommutative
coproduct. Hence we should generally have a noncommutative product for the
fields. In our specific case, there is only one abelian $\mathcal H$-module
algebra, the $3_{odd}$ one (which is nothing else than the group algebra of
the group $\ZZ_3$). Only fields transforming according to this
representation could have an abelian product. However, covariance strongly
restricts the allowable scalar products on each of the representation spaces
(for instance, the results for the quantum group $\mathcal H$ are shown in
\ref{app:metrics-on-H-representations}, where we get both indefinite and
degenerate metrics). This fact is particularly important as one should have
a positive definite metric on the physical degrees of freedom. To this end
one should disregard the non-physical (gauge) ones, and look for
representations such that only positive definite states survive. Thus we
see that the selection of the representation space upon which to build the
physical model is not easy.

The fact of having noncommuting fields has a certain resemblance with the
case of supersymmetry. As the superspace algebra is noncommutative, the
scalar superfield must have noncommutative component fields in order to
match its transformation properties. In this last situation, everything can
be casted in terms of Grassmann variables and fields. As a consequence,
instead of having ---on each space-time point--- just the Grassmann algebra
over the complex numbers, we see the appearance of an enlarged algebra
generated by both the variables and fields. Introducing a $q$-deformed
superspace leads to a more complicated algebraic structure \cite{Montani}.
Therefore, it is reasonable to expect that the addition of a non-trivial
quantum group as a symmetry of the space forces an even more constrained
algebra.

We should point out that the reasoning followed above is very general, and
is independent of the details of the fields. That is, the arguing leading
to such a conclusion relies only in the existence of a non-cocommutative
Hopf algebra acting in a nontrivial way on the fields.

For this reason, we did not plan, in this paper, to make specific choices
and discuss Lagrangian models. Actually, trying to write down such a
definite physical model would involve the making of very particular choices;
many of them are possible and we do not know, at the present level of our
analysis, which kind of constraint could give rise to interesting physics.

Our purpose was rather to investigate the structures involved and work out
(or present) the mathematical tools that would allow such physical
applications. We hope that the present exposition of our recent work will
trigger some interesting ideas in that direction.

%%%%%%%%%%%%%%%%%%%%%%%%%%%%%%%%%%%%%%%%%%%%%%%%%%%%%%%%%%%%%%%%%%%%%%%%%%
%%%%%%%%%%%%%%%%%%%%%%%%%%%%%%%%%%%%%%%%%%%%%%%%%%%%%%%%%%%%%%%%%%%%%%%%%%

\section*{Acknowledgements}

We are greatly indebted to Oleg Ogievetsky for his comments and
discussions.

%%%%%%%%%%%%%%%%%%%%%%%%%%%%%%%%%%%%%%%%%%%%%%%%%%%%%%%%%%%%%%%%%%%%%%%%%%
%%%%%%%%%%%%%%%%%%%%%%%%%%%%%%%%%%%%%%%%%%%%%%%%%%%%%%%%%%%%%%%%%%%%%%%%%%

\appendix

\section*{Appendices}

\setcounter{section}{0}
\setcounter{subsection}{0}
\setcounter{subsubsection}{0}
\def\thesection{}
\def\thesubsection{Appendix \Alph{subsection}.}

%%%%%%%%%%%%%%%%%%%%%%%%%%%%%%%%%%%%%%%%%%%%%%%%%%%%%%%%%%%%%%%%%%%%%%%%%%

\subsection{Expression of Gell-Mann matrices in terms of $x$ and $y$}
\label{app:lambda-matrices}

Usual Gell-Mann matrices $\{ \lambda_i \}_{i=1\ldots 8}$ (a base for the
Lie algebra of hermitian traceless $3 \times 3$ matrices) can be written in
terms of elementary matrices, and therefore also in terms of the generators
$x$ and $y$. Remember that

$$
\begin{tabular}{lll}
   $\lambda_1 = E_{12} + E_{21}$                                   &
      $\lambda_2 = i (-E_{12} + E_{21})$                           &
      $\lambda_3 = E_{11} - E_{22}$                                \\
   $\lambda_4 = E_{13} + E_{31}$                                   &
      $\lambda_5 = i (- E_{13} + E_{31})$                        & \\
   $\lambda_6 = E_{23} + E_{32}$                                   &
      $\lambda_7 = i (-E_{23} + E_{32})$                           &
      $\lambda_8 = {1 \over \sqrt 3} (E_{11} + E_{22} - 2 E_{33})$ \\
\end{tabular}
$$
hence we obtain
\begin{eqnarray*}
   \lambda_1 &=& (y + xy + y^2 + x^2y + q xy^2 + q^2 x^2y^2) / 3       \\
   \lambda_2 &=& -i(y + xy - y^2 + x^2y - q xy^2 - q^2 x^2y^2) / 3     \\
   \lambda_3 &=& ((1-q) x + (1-q^2) x^2)/3                             \\
   \lambda_4 &=& (y + q^2 xy + y^2 + q x^2y + xy^2 + x^2 y^2)/3        \\
   \lambda_5 &=& i(y + q^2 xy - y^2 + q x^2y - xy^2 - x^2y^2)/3        \\
   \lambda_6 &=& (y + q xy + y^2 + q^2 x^2y + q^2 xy^2 +q x^2y^2)/3    \\
   \lambda_7 &=& -i(y + q xy - y^2 + q^2 x^2y - q^2 xy^2 - q x^2y^2)/3 \\
   \lambda_8 &=& -(q^2 x + q x^2)/\sqrt 3
\end{eqnarray*}

%%%%%%%%%%%%%%%%%%%%%%%%%%%%%%%%%%%%%%%%%%%%%%%%%%%%%%%%%%%%%%%%%%%%%%%%%%

\subsection{A set of Gell-Mann matrices with coefficients in $\mathcal F$}
\label{app:F-lambda-matrices}

Here we replace $x$ and $y$ by $\Delta_L x$ and $\Delta_L y$ in the
expression of the usual Gell-Mann matrices given in
\ref{app:lambda-matrices} We obtain in this way a new set of matrices
with entries in the quantum group $\mathcal F$, denoted by $\lambda_i'$\/.
We only show the results for $\lambda_3$ and $\lambda_8$ since we do not
need these matrices explicitly in our work.

\begin{eqnarray*}
\lambda_3' &\doteq& {1 \over 3} \pmatrix{
   (1-q)\,a + (1-q^2)\,a^2 & (1-q)\,b + (q-q^2)\,ab  & (1-q^2)\,b^2 \cr
   (1-q^2)\,b^2            & (q^2-1)\,a + (q-1)\,a^2 &
                             (1-q)\,b + (1-q)\,ab                   \cr
   (1-q)\,b + (q^2-1)\,ab  & (1-q^2)\,b^2            &
                             (q-q^2)\,a + (q^2-q)\,a^2              \cr} \\
\lambda_8' &\doteq& {1 \over \sqrt{3}} \pmatrix{
   -q \, a - q \, a^2 & -q^2 b + ab  & -q \, b^2       \cr
   -q \, b^2          & -a - q^2 a^2 & -q^2 b + q^2 ab \cr
   -q^2 b + q \, ab   & -q \, b^2    & -q^2 a - a^2}
\end{eqnarray*}

Using the commutation relations for the reduced quantum group $\mathcal F$,
one can check that the usual commutation relations of $SU(3)$ are satisfied.
One can also check that $Tr(\lambda_i) = 0$ and that
$Tr(\lambda_i \lambda_j) = 2 \delta_{ij}$
(when $i \neq j$ the trace turns out to be proportional to $1+q+q^2 = 0$).

%%%%%%%%%%%%%%%%%%%%%%%%%%%%%%%%%%%%%%%%%%%%%%%%%%%%%%%%%%%%%%%%%%%%%%%%%%

\subsection{A faithful representation of $\mathcal F$}
\label{app:F-faithful-rep}

Let $\xi_1$ and $\xi_2$ be two {\sl commuting\/} symbols
($\xi_1 \xi_2 = \xi_2 \xi_1$) whose cube power vanishes
($\xi_1^3 = \xi_2^3 = 0$). Let us write

$$
\begin{tabular}{ccc}
$b = \xi_1 \pmatrix{0 & 1 & 0 \cr
                    0 & 0 & 1 \cr
                    1 & 0 & 0}$ &
$c = \xi_2 \pmatrix{0 & 1 & 0 \cr
                    0 & 0 & 1 \cr
                    1 & 0 & 0}$ &
$d = \pmatrix{1 & 0 & 0 \cr
              0 & q^{-1} & 0 \cr
              0 & 0 & q^{-2}}$
\end{tabular}
$$

\noindent and, given that $a \doteq (1 + q bc) d^2$,
$$
a = \pmatrix{1 & 0 & \xi_1 \xi_2 \cr
             \xi_1 \xi_2 q & q & 0 \cr
             0 & \xi_1 \xi_2 q^2 & q^2} \ .
$$
It can be checked that this is a faithful representation of the algebra
$\mathcal F$. It is a representation in terms of $3 \times 3$ matrices
with entries in the ring generated over $\CC$ by $1, \xi_1, \xi_2$. This
representation is due to \cite{Ogievetsky}.

%%%%%%%%%%%%%%%%%%%%%%%%%%%%%%%%%%%%%%%%%%%%%%%%%%%%%%%%%%%%%%%%%%%%%%%%%%

\newpage

\subsection{The lattice of submodules of $\mathcal H$}
\label{app:H-submodules}

\begin{figure}
\epsfxsize=7.0cm
$$
   \epsfbox{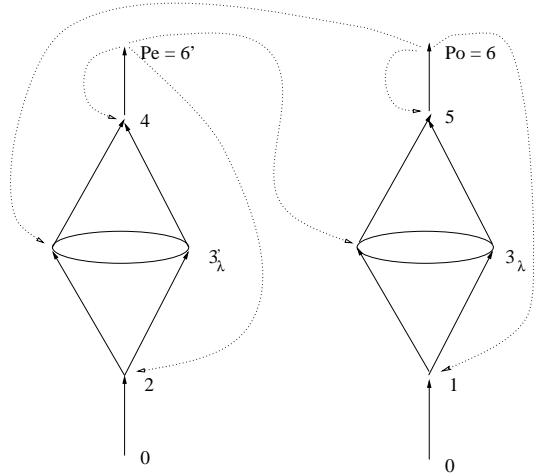}
$$
\caption{The lattices of submodules for the principal modules of
         $\mathcal H$.}
\label{fig:lattice-of-submodules}
\end{figure}

The theory of complex representations of quantum groups of type
$U_q(sl(2,\CC))$ at roots of unity has been investigated by a number of
people, but it is not yet in a satisfactory state. Here we are interested
in representation theory of the finite-dimensional quotient $\mathcal H$.
Notice that these representations can be considered as particular
representations $\rho$ of $U_q(sl(2,\CC))$, namely those for which
$\rho (X_+^3) = \rho (X_-^3) = 0$ and $\rho (K^3) = 1$. The structure of
the regular representation was studied in \cite{Alekseev} and the
representation theory ---in particular the lattice of submodules--- was
given in \cite{Coquereaux}; the general study, for arbitrary $N$, is to be
found in \cite{Ogievetsky}. For the reduced algebra $\mathcal H$ there are
three principal modules (\ie projective indecomposable modules). One is
three-dimensional and irreducible. The two others, $6_{eve}$ and $6_{odd}$
are six-dimensional. When $\mathcal H$ is written in terms of matrices with
entries in the Grassmann algebra with two generators, these two
representations can be written as
$6_{odd} \doteq (\gamma\theta_1 + \delta\theta_2,
                 \gamma'\theta_1 + \delta'\theta_2,
                 \alpha + \beta\theta_1\theta_2) $
and $6_{eve} \doteq (\alpha + \beta\theta_1\theta_2,
                     \alpha' + \beta'\theta_1\theta_2,
                     \gamma\theta_1 + \delta\theta_2)$.
Their lattices of submodules are obtained by requiring stability under the
$\mathcal H$ action. They are given in
Figure~\ref{fig:lattice-of-submodules}. The notation $P_o \equiv 6_{odd}$
and $P_e \equiv 6_{eve}$ refers to the fact that, when quotiented out by
their respective radicals, those two indecomposable modules give
irreducible representations of odd and even dimensions.

Remember that there exists a one-to-one correspondence between irreducible
representations of the algebra $\mathcal H$ and the principal modules
$3_{irr}$, $6_{eve}$ and $6_{odd}$. Irreducible representations are obtained
from these principal modules by factorizing their radical, which amounts to
kill the Grassmann ``$\theta$'' variables. As seen on the figure, the
radical of $6_{eve}$ is of dimension $4$ and the radical of $6_{odd}$, of
dimension $5$. This gives us three irreducible representations, of
dimensions $3$, $2 = 6-4$ and $1 = 6-5$. These are the three irreducible
representations corresponding to the quotient $\overline{\mathcal H}$ of
$\mathcal H$ by its Jacobson radical: namely
$\overline{\mathcal H} = \CC \oplus M_2(\CC) \oplus M_3(\CC)$. The dashed
lines in Figure~\ref{fig:lattice-of-submodules} refer to the projective
covers of the various representations.

The explicit definition given for $\mathcal H$ allows one to compute any
tensor product of representations and reduce them. Here we consider only the
tensor products of projective indecomposable representations
($\underline{6}_{odd}$, $\underline{6}_{eve}$ and $\underline{3}_{irr}$)
and the nontrivial irreducible ones ($\underline{2}$ and
$\underline{3}_{irr}$).

$$
\begin{tabular}{llllll}
$\underline{2} \times \underline{2}$             & $\equiv$ &
   $\underline{1} + \underline{3}_{irr}$             &
$\underline{6}_{eve} \times \underline{3}_{irr}$ & $\equiv$ &
   $2(\underline{6}_{eve}) + 2(\underline{3}_{irr})$ \\
$\underline{2} \times \underline{3}_{irr}$       & $\equiv$ &
   $\underline{6}_{eve}$                             &
$\underline{6}_{odd} \times \underline{3}_{irr}$ & $\equiv$ &
   $2(\underline{6}_{eve}) + 2(\underline{3}_{irr})$ \\
$\underline{3}_{irr} \times \underline{3}_{irr}$ & $\equiv$ &
   $\underline{6}_{odd} + \underline{3}_{irr}$       &
$\underline{6}_{eve} \times \underline{6}_{eve}$ & $\equiv$ &
   $2(\underline{6}_{eve}) + 2(\underline{6}_{odd})
                           + 4(\underline{3}_{irr})$ \\
$\underline{6}_{eve} \times \underline{2}$       & $\equiv$ &
   $\underline{6}_{odd} + 2(\underline{3}_{irr})$    &
$\underline{6}_{eve} \times \underline{6}_{odd}$ & $\equiv$ &
   $2(\underline{6}_{eve}) + 2(\underline{6}_{odd})
                           + 4(\underline{3}_{irr})$ \\
$\underline{6}_{odd} \times \underline{2}$       & $\equiv$ &
   $\underline{6}_{eve} + 2(\underline{3}_{irr})$    &
$\underline{6}_{odd} \times \underline{6}_{odd}$ & $\equiv$ &
   $2(\underline{6}_{eve}) + 2(\underline{6}_{odd})
                           + 4(\underline{3}_{irr})$ \\
\end{tabular}
$$

%%%%%%%%%%%%%%%%%%%%%%%%%%%%%%%%%%%%%%%%%%%%%%%%%%%%%%%%%%%%%%%%%%%%%%%%%%

\newpage

\subsection{Metrics on indecomposable representations of $\mathcal H$}
\label{app:metrics-on-H-representations}

Using the unique Hopf compatible star operation $*$ on $\mathcal H$, we
can calculate the most general metric on the vector spaces of each of the
indecomposable representations of $\mathcal H$. Obviously, as we did in
Section~\ref{subsec:scalar-product-on-M}, we restrict the inner product
to be a quantum group invariant one.

On each representation space we use the basis obtained from appropriate
restrictions of the natural basis of the column representations of
$\mathcal H$ given in Section~\ref{subsec:structure-of-H}. For each
indecomposable representation, we first write down the matrices of $X_+$,
$X_-$ and $K$ in the selected base, then we give an explicit expression of
the most general covariant metric, and finally we calculate its signature.

\medskip

\begin{itemize}

\item[$\bullet \: \mathbf 3_{irr}$]
   $$
      X_+ = \pmatrix{ 0& 1& 0\cr 0& 0& 1\cr 0& 0& 0} \quad
      X_- = \pmatrix{ 0& 0& 0\cr-1& 0& 0\cr 0&-1& 0} \quad
      K   = \pmatrix{q^2&0& 0\cr 0& 1& 0\cr 0& 0& q}
   $$
   We get for the metric, up to a real global normalization,
   $$
      G = \pmatrix{ 0& 0& -q^2\cr 0& 1& 0\cr -q& 0& 0} \ .
   $$
   Diagonalizing this matrix we get $G \sim \diag(1,1,-1)$, so the
   signature is
   $$
      \sigma = (++-) \ .
   $$
   \medskip

\item[$\bullet \: \mathbf 6_{odd}$]
   $$
      X_+ = \pmatrix{ 0& 0& 1& 0& 0& 0\cr
                      0& 0& 0& 1& 0& 0\cr
                      0& 0& 0& 0& 1& 0\cr
                      0& 0& 0& 0& 0& 0\cr
                      0& 0& 0& 0& 0& 0\cr
                      0& 1& 0& 0& 0& 0} \quad
      X_- = \pmatrix{ 0& 0& 0& 0& 0& 0\cr
                      0& 0& 0& 0& 1& 0\cr
                      1& 0& 0& 0& 0& 0\cr
                      0& 1& 0& 0& 0& 0\cr
                      0& 0& 0& 0& 0& 0\cr
                      0& 0& 1& 0& 0& 0}
   $$
   $$
      K   = \pmatrix{ q& 0&   0&   0& 0& 0\cr
                      0& q&   0&   0& 0& 0\cr
                      0& 0& q^2&   0& 0& 0\cr
                      0& 0&   0& q^2& 0& 0\cr
                      0& 0&   0&   0& 1& 0\cr
                      0& 0&   0&   0& 0& 1}
   $$
   Up to a normalization, the metric should be ($\beta \in \RR$)
   $$
      G = \pmatrix{   0&   0& 0& q&     0& 0\cr
                      0&   0&-q& 0&     0& 0\cr
                      0&-q^2& 0& 0&     0& 0\cr
                    q^2&   0& 0& 0&     0& 0\cr
                      0&   0& 0& 0& \beta& 1\cr
                      0&   0& 0& 0&     1& 0} \ .
   $$
   A change of basis tells us that
   $G \sim \diag(1,1,-1,-1,\lambda_+,\lambda_-)$, with
   $\lambda_+ > 0, \lambda_- < 0$. Thus, in this case, the signature is
   $$
      \sigma = (+++---) \ .
   $$
   \medskip

\item[$\bullet \: \mathbf 5_{odd}$]
   $$
      X_+ = \pmatrix{ 0& 0& 1& 0& 0\cr
                      0& 0& 0& 1& 0\cr
                      0& 0& 0& 0& 0\cr
                      0& 0& 0& 0& 0\cr
                      0& 1& 0& 0& 0} \quad
      X_- = \pmatrix{ 0& 0& 0& 0& 0\cr
                      0& 0& 0& 0& 0\cr
                      1& 0& 0& 0& 0\cr
                      0& 1& 0& 0& 0\cr
                      0& 0& 1& 0& 0} \quad
      K   = \pmatrix{ q& 0& 0& 0& 0\cr
                      0& q& 0& 0& 0\cr
                      0& 0& q^2& 0& 0\cr
                      0& 0& 0& q^2& 0\cr
                      0& 0& 0& 0& 1} \quad
   $$
   Up to a real factor, the metric we obtain is ($\beta,\gamma \in \RR$,
   $g \in \CC$)
   $$
      G = \pmatrix{ 0& 0&    iq\gamma&       g& 0\cr
                    0& 0& -q^2 \bar g& iq\beta& 0\cr
                    -i q^2 \gamma&         -qg& 0& 0& 0\cr
                           \bar g& -iq^2 \beta& 0& 0& 0\cr
                                0&           0& 0& 0& 0}
   $$
   Its signature is
   $$
      \sigma = (++--0) \ ,
   $$
   because
   $G \sim \diag(\lambda_+,-\lambda_+,\lambda_-,-\lambda_-,0)$.
   \medskip

\item[$\bullet \: \mathbf 3_{odd}$]
   $$
      X_+ = \pmatrix{ 0& 1& 0\cr 0& 0& 0\cr \lambda_2& 0& 0} \quad
      X_- = \pmatrix{ 0& 0& 0\cr 1& 0& 0\cr 0& \lambda_1& 0} \quad
      K   = \pmatrix{ q& 0& 0\cr 0& q^2& 0\cr 0& 0& 1}
   $$
   Up to a real factor,
   $$
      G = \pmatrix{ 0& iq& 0\cr -iq^2& 0& 0\cr 0& 0& 0}
   $$
   thus
   $$
      \sigma = (+-0) \ .
   $$
   \medskip

\item[$\bullet \: \mathbf 6_{eve}$]
   $$
      X_+ = \pmatrix{ 0& 0&    1& 0& 0& 0\cr
                      0& 0& -1/2& 1& 0& 0\cr
                      0& 0&    0& 0& 0& 0\cr
                      0& 0&    0& 0& 0& 1\cr
                      1& 0&    0& 0& 0& 0\cr
                      0& 0&    0& 0& 0& 0} \quad
      X_- = \pmatrix{    0& 0&  0& 0&  0& 0\cr
                         0& 0&  0& 0& -1& 0\cr
                         1& 0&  0& 0&  0& 0\cr
                      -1/2& 1&  0& 0&  0& 0\cr
                         0& 0&  0& 0&  0& 0\cr
                         0& 0& -1& 0&  0& 0}
   $$
   $$
      K   = \pmatrix{ q& 0&   0&   0& 0& 0\cr
                      0& q&   0&   0& 0& 0\cr
                      0& 0& q^2&   0& 0& 0\cr
                      0& 0&   0& q^2& 0& 0\cr
                      0& 0&   0&   0& 1& 0\cr
                      0& 0&   0&   0& 0& 1}
   $$
   Up to a normalization, the metric should be ($\beta \in \RR$)
   $$
      G = \pmatrix{ 0& 0& iq\beta& -iq& 0& 0\cr
                    0& 0&     -iq&   0& 0& 0\cr
                    -iq^2 \beta& iq^2& 0& 0& 0& 0\cr
                           iq^2&    0& 0& 0& 0& 0\cr
                    0& 0& 0& 0&  0& i\cr
                    0& 0& 0& 0& -i& 0}
   $$
   The signature is
   $$
      \sigma = (+++---) \ ,
   $$
   because diagonalizing $G$ we find
   $G \sim \diag(1,-1,\lambda_+,-\lambda_+,\lambda_-,-\lambda_-)$
   with $\lambda_+ > 0, \lambda_- < 0$.
   \medskip

\item[$\bullet \: \mathbf 4_{eve}$]
   $$
      X_+ = \pmatrix{ 0& 1& 0& 0\cr
                      0& 0& 0& 1\cr
                      0& 0& 0& 0\cr
                      0& 0& 0& 0} \quad
      X_- = \pmatrix{ 0& 0&-1& 0\cr
                      1& 0& 0& 0\cr
                      0& 0& 0& 0\cr
                      0& 0& 0& 0} \quad
      K   = \pmatrix{ q& 0& 0& 0\cr
                      0& q^2& 0& 0\cr
                      0& 0& 1& 0\cr
                      0& 0& 0& 1} \quad
   $$
   We obtain the metric ($\alpha, \beta \in \RR$, $g \in \CC$)
   $$
      G = \pmatrix{ 0& 0&      0&     0\cr
                    0& 0&      0&     0\cr
                    0& 0& \alpha&     g\cr
                    0& 0& \bar g& \beta}
   $$
   The signature is now dependant upon the parameters, because the non-null
   block is an arbitrary hermitian matrix.
   \medskip

\item[$\bullet \: \mathbf 3_{eve}$]
   $$
      X_+ = \pmatrix{ 0& 1& 0\cr 0& 0& \lambda_2\cr 0& 0& 0} \quad
      X_- = \pmatrix{ 0& 0&-\lambda_1\cr 1& 0& 0\cr 0& 0& 0} \quad
      K   = \pmatrix{ q& 0& 0\cr 0& q^2& 0\cr 0& 0& 1}
   $$
   In this case, we find simply
   $$
      G = \pmatrix{ 0& 0& 0\cr 0& 0& 0\cr 0& 0& 1}
   $$
   \medskip

\item[$\bullet \: \mathbf 2_{eve}$]
   $$
      X_+ = \pmatrix{ 0& 1\cr 0& 0} \quad
      X_- = \pmatrix{ 0& 0\cr 1& 0} \quad
      K   = \pmatrix{ q& 0\cr 0& q^2}
   $$
   Therefore
   $$
      G = \pmatrix{ 0& iq\cr -iq^2& 0} \sim \diag(1,-1) \ .
   $$

\end{itemize}

%%%%%%%%%%%%%%%%%%%%%%%%%%%%%%%%%%%%%%%%%%%%%%%%%%%%%%%%%%%%%%%%%%%%%%%%%%

\newpage

\subsection{The space of differential operators on $\mathcal M$}
\label{app:diff-operators-on-M}

We now look at the structure of differential operators on the reduced
quantum plane $\mathcal M$, \ie the algebra of $3 \times 3$ matrices.

%%%%%%%%%%%%%%%%%%%%%%%%%%%%%%%%%%%%%%%%%%%%%%%%%%%%%%%%%%%%%%%%%%%%%%%%%%

\subsubsection*{The operators $\partial_x$ and $\partial_y$}

We already know what the operator
$d: \Omega_{WZ}^0 = {\mathcal M} \longrightarrow \Omega_{WZ}^1$ is.

{\it A priori\/} we can set
$d f = dx \, \partial_x (f) + dy \, \partial_y (f)$, where
$\partial_x (f), \partial_y (f) \in {\mathcal M}$. This equation defines
$\partial_x$ and $\partial_y$ as (linear) operators on $\mathcal M$. We
shall see later that they are twisted derivations on the algebra
$\mathcal M$. This definition implies in particular (take $f = x$ or
$f = y$):
\begin{eqnarray*}
   \partial_x (x) = 1 \qquad & \partial_y (x) = 0 \cr
   \partial_x (y) = 0 \qquad & \partial_y (y) = 1
\end{eqnarray*}

%%%%%%%%%%%%%%%%%%%%%%%%%%%%%%%%%%%%%%%%%%%%%%%%%%%%%%%%%%%%%%%%%%%%%%%%%%

\subsubsection*{The space $\mathcal D$ of differential operators on
                $\mathcal M$}

Generally speaking, operators of the type $f(x,y) \partial_x$ or
$f(x,y) \partial_y$ are called differential operators of order $1$.
Composition of such operators gives rise to differential operators of order
higher than $1$. Multiplication by an element of $\mathcal M$ is considered
as a differential operator of degree $0$. The space of all these operators
is a vector space ${\mathcal D} = \oplus_{i=0}^4{\mathcal D}_i$, graded
by operators of
\begin{description}
\item[Order $0$:]
   $x^r y^s$. $\dim({\mathcal D}_0) = 9$.

\item[Order $1$:]
   $x^r y^s \partial_x$, $x^r y^s \partial_y$.
   $\dim({\mathcal D}_1) = 9 + 9 = 18$.

\item[Order $2$:]
   $x^r y^s \partial_x \partial_x$,
   $x^r y^s \partial_x \partial_y$,
   $x^r y^s \partial_y \partial_y$.
   $\dim({\mathcal D}_2) = 9 + 9 + 9 = 27$.

\item[Order $3$:]
   $x^r y^s \partial_x \partial_x \partial_y$,
   $x^r y^s \partial_x \partial_y \partial_y$.
   $\dim({\mathcal D}_3) = 9 + 9 = 18$.

\item[Order $4$:]
   $x^r y^s \partial_x \partial_x \partial_y \partial_y$.
   $\dim({\mathcal D}_4) = 9$

\end{description}

\noindent All these operators are linearly independent. Moreover, since
$\dim({\mathcal D}) = 9+18+27+18+9 = 81 = 9^2 = \dim(End({\mathcal M}))$,
we can identify $\mathcal D$ with $End({\mathcal M})$.

{\sl Warning\/}:
We have here a problem of notations: the reader should distinguish, for
instance, $\partial_x \, f$ which is a differential operator from
$\mathcal M$ to $\mathcal M$ (namely, when acting on something, it
multiplies what follows by $f$, and then acts with $\partial_x$ on the
result) from $\partial_x (f)$ which is the {\sl evaluation\/} of
$\partial_x$ on $f$, hence an element of $\mathcal M$ (which can, in turn,
be considered as a differential operator of order $0$). The presence, or
absence, of parenthesis should be enough to make this clear.

%%%%%%%%%%%%%%%%%%%%%%%%%%%%%%%%%%%%%%%%%%%%%%%%%%%%%%%%%%%%%%%%%%%%%%%%%%

\subsubsection*{The twisting automorphisms $\sigma$ and
                $\tau$\footnote{
                   Some properties of these automorphisms are discussed
                   in \cite{Manin-2}}.}

Since we know how to commute $x,y$ with $dx,dy$, we can write, for any
element $f \in {\mathcal M}$
\begin{eqnarray*}
   f dx &=& dx \, \sigma_x^x(f) + dy \, \sigma_y^x(f) \cr
   f dy &=& dx \, \sigma_x^y(f) + dy \, \sigma_y^y(f)
\end{eqnarray*}
where each matrix element of
$$
   \pmatrix{\sigma_x^x & \sigma_x^y \cr \sigma_y^x & \sigma_y^y}
$$
is an element of $End({\mathcal M})$ to be determined. In particular,
taking $f=x$ and $f=y$ in the above equations leads to
\begin{eqnarray*}
   \sigma_x^x(x) = q^2 x & \qquad & \sigma_y^x(x) = 0 \cr
   \sigma_x^x(y) = q y   & \qquad & \sigma_y^x(y) = 0 \cr
                      {} & {}     & \cr
   \sigma_x^y(x) = (q^2-1) y & \qquad & \sigma_y^y(x) = q x   \cr
   \sigma_x^y(y) = 0         & \qquad & \sigma_y^y(y) = q^2 y \cr
\end{eqnarray*}

\noindent Moreover, let $f$ and $g$ be elements of $\mathcal M$. Using the
associativity property $(fg) dx = f(g dx)$, and the same with $dy$, we find
$$
   \sigma_i^j(fg) = \sigma_i^k(f) \, \sigma_k^j(g) \ ,
$$
with a summation over the repeated index $k \in \{x,y\}$. Thus the map
$\sigma: f \in {\mathcal M} \rightarrow \sigma(f) \in M_2({\mathcal M})$
is an algebra homomorphism ($\sigma(fg) = \sigma(f) \sigma(g)$) from
${\mathcal M}$ to the algebra $M_2({\mathcal M})$ of $2 \times 2$ matrices
with elements in $\mathcal M$.

In the same way, we could have written for any element $f \in {\mathcal M}$
(note the transposed index convention)
\begin{eqnarray*}
   dx \, f &=& \tau_x^x(f) \, dx + \tau_x^y(f) \, dy \cr
   dy \, f &=& \tau_y^x(f) \, dx + \tau_y^y(f) \, dy
\end{eqnarray*}
hence defining another ${\mathcal M} \rightarrow M_2({\mathcal M})$
homomorphism $\tau$, since $\tau_i^j(fg) = \tau_i^k(f) \tau_k^j(g)$.

%%%%%%%%%%%%%%%%%%%%%%%%%%%%%%%%%%%%%%%%%%%%%%%%%%%%%%%%%%%%%%%%%%%%%%%%%%

\subsubsection*{$\partial_x $ and $\partial_y $ are twisted derivations}

The usual Leibniz rule for $d$, namely $d(fg) = d(f)g+fd(g)$, implies
$$
   \pmatrix{\partial_x(fg) \cr \partial_y(fg)} =
      \pmatrix{\partial_x(f)g \cr \partial_y(f)g} +
      \pmatrix{\sigma_x^x(f) & \sigma_x^y(f) \cr
               \sigma_y^x(f) & \sigma_y^y(f)}
      \pmatrix{\partial_x(g) \cr \partial_y(g)}
$$
or even
$$
   \partial_i(fg) = \partial_i(f) \, g + \sigma_i^j(f) \, \partial_j(g) \ .
$$

\noindent This shows that $\partial_x $ and $\partial_y $ are twisted
derivations (derivations twisted by an homomorphism). Actually, it is
conceptually interesting to notice that one can get rid of the (explicit)
twisting by introducing two distinct module structures on $\Omega^1_{WZ}$.
First of all, elements of $\Omega_{WZ}^1$ are of the kind
$dx \, g_1 + dy \, g_2 $ and can be written as a column vector
$\pmatrix{g_1 \cr g_2}$. We can then consider
$\Omega_{WZ}^1 \simeq {\mathcal M}^2$ as an $\mathcal M$-bimodule, taking
as right module structure the standard one:
$$
   \pmatrix{g_1 \cr g_2} \cdot f \doteq \pmatrix{g_1 f \cr g_2 f} \ ,
$$
and making use of the automorphism $\sigma$ for the left module structure:
$$
   f \cdot \pmatrix{g_1 \cr g_2} \doteq
      \pmatrix{\sigma_x^x(f) & \sigma_x^y(f) \cr
               \sigma_y^x(f) & \sigma_y^y(f)} \pmatrix{g_1 \cr g_2} \ .
$$

\noindent The operator
${\underline d} \doteq \pmatrix{\partial_x \cr \partial_y}$ is a
derivation on the algebra $\mathcal M$ with values in the bimodule
$\Omega_{WZ}^1 \simeq {\mathcal M}^2$. Indeed, by using the two
(distinct) left and right module structures just given, the twisted
derivation properties of the operators $\partial_x$ and $\partial_y$,
when expressed in terms of ${\underline d}$, read simply
$$
   {\underline d}(fg) = {\underline d}(f) \cdot g +
                        f \cdot {\underline d}(g) \ .
$$

%%%%%%%%%%%%%%%%%%%%%%%%%%%%%%%%%%%%%%%%%%%%%%%%%%%%%%%%%%%%%%%%%%%%%%%%%%

\subsubsection*{Relations in $\mathcal D$}

For calculational purposes, it is useful to know the commutation relations
between $x,y$ and $\partial_x,\partial_y$, those between
$\partial_x$ and $\partial_y$, and the relations between the $\sigma_i^j$.
Calculations are straightforward but slightly cumbersome\ldots \ here are
the results (see also \cite{Wess-Zumino,Manin}).

Calculating $\partial_x(xf)$, $\partial_x(yf)$, $\partial_y(xf)$,
$\partial_y(yf)$ gives the following relations (here we do not suppose that
$q^N = 1$):
\begin{eqnarray*}
   \partial_x\, x &=& 1 + q^2 x\, \partial_x + (q^2-1) y \, \partial_y \cr
   \partial_x\, y &=& q y \, \partial_x \cr
               {} &{}& \cr
   \partial_y\, x &=& q x \, \partial_y \cr
   \partial_y\, y &=& 1 + q^2 y \, \partial_y
\end{eqnarray*}
One can check that all these relations are compatible with the defining
relations of $\mathcal M$.

We then calculate
$\partial_x\partial_y (x)$, $\partial_x\partial_y (y)$,
$\partial_y\partial_x (x)$, $\partial_y\partial_x (y)$.
Compatibility of the results implies the commutation relation
$$
   \partial_y \, \partial_x = q \, \partial_x \, \partial_y \ .
$$

\noindent Moreover, the fact that $\partial_{x,y}^3$ are central elements
of $\mathcal{D}$ if $q^3=1$ leads to
$$
   \partial_x^3 = \partial_y^3 = 0 \ .
$$

The commutation relations between the $\sigma$'s can be obtained from
the values of the $\sigma^i_j(x)$. Taking into account that
$\sigma_y^x \equiv 0$, the remaining non-trivial relations are
\begin{eqnarray*}
   \sigma_x^x \sigma_x^y &=& q^2 \sigma_x^y \sigma_x^x \\
   \sigma_x^x \sigma_y^y &=& \sigma_y^y \sigma_x^x     \\
   \sigma_x^y \sigma_y^y &=& q^2 \sigma_y^y \sigma_x^y
\end{eqnarray*}

%%%%%%%%%%%%%%%%%%%%%%%%%%%%%%%%%%%%%%%%%%%%%%%%%%%%%%%%%%%%%%%%%%%%%%%%%%

\subsubsection*{Scaling operators}

It is useful to introduce the operators \cite{Ogievetsky-2}
\begin{eqnarray*}
   \mu_x &=& 1 + (q^2-1)(x \, \partial_x + y \, \partial_y) \cr
   \mu_y &=& 1 + (q^2-1) \, y \, \partial_y
\end{eqnarray*}
Indeed,
\begin{eqnarray*}
   \mu_x\, x = q^2 x\, \mu_x & \qquad & \mu_y\, x = x\, \mu_y \cr
   \mu_x\, y = q^2 y\, \mu_x & \qquad & \mu_y\, y = q^2 y \, \mu_y \cr
\end{eqnarray*}
Observe that $\mu_x$ rescales $x$ and $y$ in the same way. This is
not so for $\mu_y$. Their product satisfies
$$
   \mu_x \, \mu_y = \mu_y \, \mu_x
$$
and also
$$
   \mu_x^3 = \mu_y^3 = \one \ .
$$

\noindent It is sometimes handy to rewrite the commutation relations
between $x,y$ and the first order operators $\partial_x,\partial_y$ in
terms of the $\mu_x, \mu_y$. For instance,
\begin{eqnarray*}
   \partial_x\, x &=& \mu_x + x \, \partial_x \cr
   \partial_y\, y &=& \mu_y + y \, \partial_y \cr
\end{eqnarray*}

%%%%%%%%%%%%%%%%%%%%%%%%%%%%%%%%%%%%%%%%%%%%%%%%%%%%%%%%%%%%%%%%%%%%%%%%%%

\subsubsection*{The action of $\mathcal H$ in terms of differential
                operators}

The twisted derivations $\partial_x, \partial_y$ considered previously
constitute a $q$-analogue of the notion of vector fields. Their powers
(including zero) build up arbitrary differential operators. Elements of
$\mathcal H$ act also like powers of generalized vector fields (consider
for instance the left action generated by $X_\pm^L,K^L$), but, of course,
they are differential operators of a special kind: remember that
$\dim({\mathcal H}) = 27$ whereas $\dim({\mathcal D}) = 81$. One can say
that elements of $\mathcal H$ act on $\mathcal M$ as {\sl fundamental\/}
differential operators since they are associated with the action of a
(quantum) group on a (quantum) space.

A priori, the generators $X_\pm^L,K^L$ can be written in terms of
$x,y, \partial_x, \partial_y$. In order to find these expressions, it
helps to notice that $\partial_x$ and $\partial_y$ are respectively of
weight $-1/2,1/2$ in $\ZZ / 3 \ZZ$ (see also the discussion at the end of
Section~\ref{subsec:actions-of-H}). Writing the generators as arbitrary
differential operators of a fixed weight, one can determine the
coefficients by imposing that equations (\ref{H-products}),
(\ref{H-M-relations}) are satisfied. A rather cumbersome calculation
leads to the following unique solution:
\begin{eqnarray*}
   X_+^L &=& x\,\partial_y + (q-1) \,xy\,\partial_y^2    \cr
   X_-^L &=& y\,\partial_x + (q-q^2) \,xy\,\partial_x^2 +
             (1-q) \,y^2 \partial_x \partial_y           \cr
   K^L   &=& \one + (q-1) \,x\,\partial_x + (q^2-1) \,y\,\partial_y -
             3 q \,x^2 \,\partial_x^2 +
             3 (1-q) \,x^2 y \,\partial_x^2 \partial_y +
             9 \,x^2 y^2 \,\partial_x^2 \partial_y^2     \cr
   K_-^L &\equiv& (K^L)^2 = \one + (q^2-1) \,x\,\partial_x +
             (q-1) \,y\,\partial_y - 3 \,xy\,\partial_x \partial_y -
             3q \,y^2\,\partial_y^2                      \cr
\end{eqnarray*}

\noindent An alternative way of writing these, is to make use of the
scaling operators,
\begin{eqnarray*}
   K^L_- &=& \mu_x \mu_y                   \\
   K^L   &=& \mu_x^2 \mu_y^2               \\
   X^L_+ &=& \mu_y^2 \, x \, \partial_y    \\
   X^L_- &=& q \, \mu_x \, y \, \partial_x
\end{eqnarray*}

%%%%%%%%%%%%%%%%%%%%%%%%%%%%%%%%%%%%%%%%%%%%%%%%%%%%%%%%%%%%%%%%%%%%%%%%%%

\newpage

\subsection{The universal $R$ matrix of $\mathcal H$}
\label{app:R-matrix}

We did not discuss the $R$-matrix aspects of $\mathcal H$ in the main body
of this paper, however it can be seen that $\mathcal H$ is actually a
braided and quasi-triangular finite-dimensional Hopf algebra ---as it is
well known, the quantum enveloping algebra of $SL(2)$ does {\em not} posess
these properties when $q$ is a root of unity. The universal $R$ matrix can
be obtained directly from a general formula given in \cite{Rosso} but one
also obtain in a pedestrian way. In any case, the answer is simply the
following:
\begin{eqnarray*}
R &=& \frac{1}{3q} \left[
           \one \otimes \one + (\one \otimes K + K \otimes \one)
           + (\one \otimes K^2 + K^2 \otimes \one) \right. \nonumber \\
  & & \quad \left. + q^2 (K \otimes K^2 + K^2 \otimes K)
           + q K \otimes K + q K^2 \otimes K^2 \right]     \\
  & & \times \left[ \one \otimes \one + (q-q^{-1}) X_- \otimes X_+
           + 3q X_-^2 \otimes X_+^2 \right]                \nonumber
\label{R-matrix}
\end{eqnarray*}
By using the explicit formulae for the generators $X_+$, $X_-$ and $K$
given in \ref{app:metrics-on-H-representations}$\!\!,$ one can obtain the
expression of $R$ in any representation (including reducible indecomposable
ones). One can then check that the defining quadratic equations for the
quantum plane and its Manin dual are respectively recovered by writing
$A \, xx = 0$ (respectively $S \, xx = 0$). Here $S$ and $A$ are the two
projectors entering the spectral decomposition of the numerical
$\hat R = (flip). R$ matrix
\begin{eqnarray*}
   \hat R &=& q\, S - q^{-1} A                \\
   S + A  &=& \one \qquad \qquad (Tr S = 3\,,\; Tr A = 1) \ ,
\end{eqnarray*}
$R$ being this time the numerical $R$-matrix in the fundamental
representation.

%%%%%%%%%%%%%%%%%%%%%%%%%%%%%%%%%%%%%%%%%%%%%%%%%%%%%%%%%%%%%%%%%%%%%%%%%%
%%%%%%%%%%%%%%%%%%%%%%%%%%%%%%%%%%%%%%%%%%%%%%%%%%%%%%%%%%%%%%%%%%%%%%%%%%


\begin{thebibliography}{99}

\bibitem{Gluschenkov} D. V. Gluschenkov and A. V. Lyakhovskaya,
   {\em Regular representation of the quantum Heisenberg double ($q$ is a
   root of unity)}, Zapiski LOMI 215 (1994).

\bibitem{Dabrowski} L. D{\c{a}}browski, F. Nesti and P. Siniscalco,
   {\em A finite quantum symmetry of $M(3,\CC)$},
   Int. J. Mod. Phys. {\bf A 13} (1998), 4147.

\bibitem{Alekseev} A. Alekseev, D. Gluschenkov and A. Lyakhovskaya,
   {\em Regular representation of the quantum group $SL_q(2)$ ($q$ is a
   root of unity)}, St. Petersburg Math. J. {\bf 6} (1994), 88.

\bibitem{Coquereaux} R. Coquereaux, {\em On the finite dimensional quantum
   group $M_3 \oplus (M_{2\vert 1}(\Lambda^2))_0$},
   Lett. Math. Phys. {\bf 42} (1997), 309, {\tt hep-th/9610114}.

\bibitem{Pusz} W. Pusz and S. L. Woronowicz, {\em Twisted second
   quantization}, Rep. Math. Phys. {\bf 27} (1989), 231.

\bibitem{Wess-Zumino} J. Wess and B. Zumino,
   {\em Covariant differential calculus on the quantum hyperplane},
   Nucl. Phys. B (Proc. Suppl.) {\bf 18B} (1990), 302.

\bibitem{Weyl} H. Weyl, {\em The theory of groups and quantum mechanics}
   (Dover Publications, 1931).

\bibitem{Manin} Yu. I. Manin, {\em Quantum groups and non commutative
   geometry}, preprint Montreal Univ. CRM-1561 (1988).

\bibitem{Ogievetsky} O. Ogievetsky, {\em Matrix structure of $SL_q(2)$
   when $q$ is a root of unity}, preprint CPT-96/P.3390, to appear.

\bibitem{Woronowicz} S. L. Woronowicz, {\em Compact matrix pseudogroups},
   Comm. Math. Phys. {\bf 111} (1987), 613.

\bibitem{Coquereaux-Schieber} R. Coquereaux and G. Schieber,
   {\em Action of a finite quantum group on the algebra of $N \times N$
   matrices}, Proceedings of the Lodz Conference (1998).

\bibitem{Connes-2} A. Connes, {\em Gravity coupled with matter and the
   foundation of noncommutative geometry},
   Comm. Math. Phys. {\bf 182} (1996), 155.

\bibitem{Humbert} P. Humbert, {\em Sur les fonctions de Bessel de
   troisi\`eme ordre}, C. R. Acad. Sci. Paris (1930).

\bibitem{CoGaTr} R. Coquereaux, A. O. Garc\'{\i}a and R. Trinchero,
   {\em Finite dimensional quantum group covariant differential calculus
   on a complex matrix algebra}, Phys. Lett. {\bf B 443} (1998), 221.

\bibitem{Coquereaux-JGP} R. Coquereaux, {\em Noncommutative geometry and
   theoretical physics}, J. Geom. Phys. {\bf 6} (1989), 425.

\bibitem{Connes} A. Connes, {\em Noncommutative geometry and reality},
   preprint IHES M/95/52.

\bibitem{Dubois-Violette} M. Dubois-Violette, K. Kerner and J. Madore,
   {\em Classical bosons in a non-commutative geometry},
   Class. Quant. Grav. {\bf 6} (1989), 1709.

\bibitem{Coquereaux-Haussling-Scheck} R. Coquereaux, R. Haussling and
   F. Scheck, Int. J. Mod. Phys. {\bf A 7} (1992), 6555.

\bibitem{Montani} H. Montani and R. Trinchero,
   {\em Quantum mechanics over a $q$-deformed $(0+1)$-dimensional
   superspace}, Int. J. Mod. Phys. {\bf A 13} (1998), 4173.

\bibitem{Manin-2} Yu. I. Manin, {\em Notes on quantum groups and quantum
   De Rahm complexes}, Teor. i Matem. Fiz. {\bf Tom 92} N3 (1992), 425.

\bibitem{Ogievetsky-2} O. Ogievetsky, {\em Differential operators on
   quantum spaces for $GL_q(N)$ and $SO_q(N)$},
   Lett. Math. Phys. {\bf 24} (1992), 245.

\bibitem{Rosso} M. Rosso, {\em Quantum groups at root of unity and tangle
   invariants}, in Topological and Geometrical methods in Field Theory,
   J. Mickelsson and O. Pekonen (Eds.), World Scientific, 347 (1992).

\end{thebibliography}
\end{document}